\documentclass[aps,floats,twocolumn,prb,showpacs,10pt]{revtex4-1}
\usepackage{graphicx}
\usepackage{color}
\usepackage{upgreek}
\usepackage{setspace}
\usepackage{amssymb}
\usepackage{amsmath}
\usepackage{pstricks}
\usepackage{dsfont}
\usepackage{fancyhdr}
\usepackage{array}
\usepackage{multirow}

\begin{document}
\title{Local Electronic Correlation at the Two-Particle Level}
\author{G. Rohringer$^1$, A. Valli$^1$, A. Toschi$^1$}
\affiliation{$^1$Institute of Solid State Physics, Vienna University of Technology, 1040
Vienna, Austria}
\date{Version 1, \today}

\begin{abstract}
Electronic correlated systems are often well described by dynamical mean field theory (DMFT).  While DMFT studies have mainly focused hitherto on one-particle properties, valuable information is also enclosed into local two-particle Green's functions and vertices. They represent the main ingredient to compute momentum-dependent response functions at the DMFT level and to treat non-local spatial correlations at all length scales by means of diagrammatic extensions of DMFT. The aim of this paper is to present a DMFT analysis of the local reducible and irreducible two-particle vertex functions for the Hubbard model in the context of an unified diagrammatic formalism. An interpretation of the observed frequency structures is also given in terms of perturbation theory, of the comparison with the atomic limit, and of the mapping onto the attractive Hubbard model.
\end{abstract}

\maketitle

\let\n=\nu \let\o =\omega \let\s=\sigma


\section{Introduction}
\label{Sec:Intro}
Electronic correlations are responsible for some of the most fascinating phenomena occurring in condensed-matter physics, such as the colossal magnetoresistance of the manganites\cite{magnetoresistance}, the Mott-Hubbard metal-insulator transition (MIT)\cite{MH} in the vanadates\cite{mcwhan}, the high-temperature superconductivity of cuprates\cite{bm1986} and (possibly) of iron-pnictides\cite{kamihara2008}, and even for the appearance of quantum critical points in particular heavy fermion compounds\cite{QCP}.
While the exact treatment of electronic correlation is an impossible task in real materials as well as in model systems (e.g., the Hubbard model\cite{Hubbard}), a major step forward has been obtained since the early Nineties with the dynamical mean field theory (DMFT)\cite{DMFTrev,metzner1989}. 

DMFT represents the quantum extension of  classical mean field theory, and, hence, can be rigorously derived as the exact solution of a quantum many body Hamiltonian (such as the Hubbard Hamiltonian) in the limit of infinite coordination number or dimensions ($d \rightarrow \infty$)\cite{metzner1989}. While the average over infinite spatial dimensions implies neglecting non-local spatial correlations, DMFT provides for a very accurate treatment of local quantum (dynamical) fluctuations.  In the case of localized electrons, these fluctuations play a pivotal role, as they can drive, e.g., the MIT in several compounds. The most convincing proof of the accuracy of DMFT and/or its combination with ab-initio methods (LDA+DMFT)\cite{LDADMFT}, however, comes from its impressive success in treating some of the most challenging problems in condensed matter physics. We recall, among the most successful applications of DMFT, the description of the $\delta$ phase of Pu\cite{savrasov2001}, the MIT in V$_2$O$_3$\cite{held2001}, the correlation effect in Fe and Ni\cite{lichtenstein2001},  the volume collapse in Ce\cite{held2001_bis}, and of a possible uncoventional mechanism for the supercoductivity in doped fullerenes\cite{massimo2002}. Using DMFT has become almost standard for treating electronic correlated systems in the last decade and, as an example, recent DMFT calculations have been able to explain the  appearance of kinks in the spectral functions and the specific heat of particular vanadates (as SrVO$_3$\cite{byczuk2007}, LiV$_2$O$_4$\cite{PRLkinks}), the spin-polaron peak structures in photo-emission\cite{PRBafdmft} and optical spectroscopic data\cite{PRBafopt} of strongly coupled antiferromagnets, such as  V$_2$O$_3$ and LaSrMnO$_4$, the anomalies of the optical spectra and sum rules in the high-temperature superconducting cuprates\cite{sumrule}, as well as in V$_2$O$_3$\cite{PRBoptv2o3}, and some of the spectral\cite{spectpnictDMFT} and magnetic properties\cite{pnictideDMFT} of iron-based superconductors.

However, by looking at the existing DMFT or LDA+DMFT studies in more detail, one clearly sees some limitations. For instance, the theoretical calculations and the comparison with experiments  are mostly performed for one-particle quantities only, such as momentum-integrated or momentum-resolved spectral functions. The analysis of the two-particle quantities is usually restricted to the easiest cases of optical and thermal conductivities, for which --in DMFT-- one can safely consider\cite{DMFTrev} the ``joint-density of state'' (``bubble'') term only, i.e., in other words, it essentially remains at the  one-particle level. Calculations of spectral properties at the ``actual'' two-particle level are -with few exceptions\cite{kunes2011,park2011,DCAext}- done for local susceptibilities only. These, in turn, can be also directly approximated with the results of the self-consistently determined Anderson impurity model (AIM) associated with DMFT\cite{PRBpnict}. The reason for these restrictions is --evidently-- the higher difficulty and heavier workload of performing calculations at the two-particle level. In fact, the standard procedure (see Ref. \onlinecite{DMFTrev}) to calculate a (particle-hole, particle-particle) momentum- and frequency-dependent susceptibility $\chi(\mathbf{q},\omega)$ at the DMFT level requires the determination of the  local irreducible vertices ($\Gamma_r$, with $r=d,m,s,t$ ) of the AIM in the corresponding particle-hole ({\it d}ensity/{\it m}agnetic) or particle-particle  ({\it s}inglet/{\it t}riplet) channel, which serve as an input for the related Bethe-Salpeter equation. Only very recently, an alternative procedure, based on DMFT calculations in presence of time- and space-dependent perturbating fields, has been proposed\cite{nanli2012}.


The importance of determining the properties of reducible and irreducible two-particle quantities for the AIM, however, goes well beyond the calculation of the momentum- and frequency-dependent response functions $\chi(\mathbf{q},\omega)$, needed for the comparison of DMFT with other spectroscopic experiments than photo-emission or optics. Indeed, reducible and irreducible vertices of the AIM are  the basic ingredients of two important diagrammatic extensions of DMFT, such as the dynamical vertex approximation (D$\Gamma$A)\cite{DGA1, DGA1a, Kusunose} and the dual fermion (DF)\cite{DF} approach. In fact, both methods aim at the inclusion of spatial correlations beyond DMFT at all length scales, starting from a two-particle level (local) input of an associated AIM.    
Leaving aside the theoretical and numerical challenges of performing the calculation for the local two-particle vertices\cite{note_jan2011}, a thorough analysis of their general properties and of the physical interpretation has been lacking in the literature hitherto. The full frequency dependence of two-particle quantities indeed has been shown or discussed only in selected cases and for very specific problems (e.g., Refs. \onlinecite{DGA1,DGA1a,dolfen,kunes2011}). The main scope of the present paper, hence, is  to fill this gap.\\
We provide a detailed DMFT study of the two-particle reducible and irreducible local vertices of the Hubbard model within a unified derivation and formalism.  In this framework, the interpretation of the main structure of the vertices in (Matsubara) frequency space will be made easier by the comparison with perturbation theory and atomic limit results, and by the mapping onto the attractive Hubbard model. Both the formal derivation and the physical interpretations are potentially of high impact for future developments of many-body theoretical schemes, for possible improvements of the existing numerical schemes, and for calculations relying on an increased understanding of correlations beyond the one-particle level.  


The scheme of the paper is the following: In Sec. II, we first introduce the formal and diagrammatic definitions for treating the reducible and irreducible two-particle Green's and vertex functions of the AIM associated with the self-consistent solution of DMFT. At the end of this section, we also mention the general symmetry properties that are expected for such vertices (while their formal derivation is explicitly given in the Appendix). In Sec. III, our DMFT results for the reducible and irreducible local vertex functions are presented together with their interpretation in terms of the corresponding perturbative and atomic limit results. Furthermore, we analyze the effect of different approximations at the two-particle level on selected physical quantities, and, at the end of the section, we briefly discuss the possible relevance of our study for the improvement of numerical algorithms at the two-particle level. Subsequently in Sec. IV, the mapping onto the attractive Hubbard model is exploited to gain further insight into the main structures of the two particle vertex functions. Finally Sec. V is devoted to summarizing our theoretical and numerical results and conclusions.

\vskip 5mm
\section{Two-particle diagrams: formalism and general properties}
\label{Sec:2pdiag}

Starting point for our analysis is a rigorous and coherent definition of the relevant one- and two-particle quantities and of their general properties, which we will use throughout the present paper. While part of the derivations reported in this section (and in the corresponding appendixes) is already known\cite{abrikosov,bickers}, to the best of our knowledge, a systematic and unified discussion of the two-particle properties has been reported only partially or implicitly in the standard literature of quantum field theory of many particle systems. Hence, the explicit derivation of local vertex definitions and properties is helpful for  an easier reading of the following sections, where our numerical and analytical results are presented.
     
As mentioned in the Introduction, we consider one of the most fundamental models for electronic correlations, the Hubbard model on a simple cubic lattice
\begin{equation}
 \label{equ:defhubbard}
 \hat{\cal{H}}_{\text{Hubbard}}=-t\sum_{\langle ij \rangle,\sigma}\hat{c}^{\dagger}_{i\sigma}\hat{c}_{j\sigma}+U\sum_i \hat{n}_{i\uparrow}\hat{n}_{i\downarrow}.
\end{equation}
Here $t$ denotes the hopping amplitude between nearest neighbors, $U$ is the on-site Coulomb interaction, and $\hat{c}^{\dagger}_{i\sigma}(\hat{c}_{i\sigma})$ creates (annihilates) an electron with spin $\sigma$ on site $i$; $\hat{n}_{i\sigma}=\hat{c}^{\dagger}_{i\sigma}\hat{c}_{i\sigma}$. In the following, consistently with previous DMFT and D$\Gamma$A papers\cite{DGA1,DGA1a,DGA2}, we will express all energies in units of $D=2\sqrt{6}t$, which ensures that the standard deviation of the non-interacting density of states (DOS) is kept fixed to $0.5$\cite{noteDOS}.  
 As we are mainly interested in purely local quantities at the DMFT level, for indicating these we will omit the site-index $i$ in the following. Specifically, as in the DMFT-limit of infinite coordination number, the Hubbard-model can be mapped onto an effective (self-consistently determined) AIM. We will use the latter to calculate analytically or numerically the local observables
\begin{equation}
 \label{equ:defanderson}
\hat{\cal{H}}=\sum_{\ell \sigma}\varepsilon_{\ell}\hat{a}^{\dagger}_{\ell \sigma}\hat{a}_{\ell \sigma}+\sum_{\ell \sigma}V_{\ell}(\hat{c}^{\dagger}_{\sigma}\hat{a}_{\ell \sigma}+\hat{a}^{\dagger}_{\ell \sigma}\hat{c}_{\sigma})+U\hat{n}_{\uparrow}\hat{n}_{\downarrow},
\end{equation} 
where $\hat{a}^{\dagger}_{\ell \sigma}(\hat{a}_{\ell \sigma})$ creates (annihilates) an electron with spin $\sigma$ at the bath-level of energy $\varepsilon_\ell$, $\hat{c}^{\dagger}_{\sigma}(\hat{c}_{\sigma} )$ creates (annihilates) an electron at the impurity site ($\hat{n}_{\sigma}=\hat{c}^{\dagger}_{\sigma}\hat{c}_{\sigma}$),  $V_{\ell}$ describes the hybridization between the bath and the impurity, and $U$ is the on-site repulsion between two electrons at the impurity. 

\subsection{Definitions and general properties}

The general definition of the $n$-particle Green's function $G_n$ reads\cite{abrikosov}
\begin{equation}
 \label{equ:defngreenfunction}
 G_{n,\sigma_1\ldots\sigma_{2n}}(\tau_1,\ldots,\tau_{2n}):=\bigl\langle\text{T}
  \bigl(\hat{c}^{\dagger}_{\sigma_1}(\tau_1)\ldots
  \hat{c}_{\sigma_{2n}}(\tau_{2n})\bigr)\bigr\rangle,
\end{equation}
where an odd/even index always corresponds to an creation/annihilation operator $\hat{c}^{\dagger}_{\sigma}$/$\hat{c}_{\sigma}$. This means that the creation and annihilation operators appear in alternating order in Eq.\ (\ref{equ:defngreenfunction}), and $\langle\hat{\cal{O}}\rangle\!=\!\frac{1}{Z}\mbox{tr}(e^{-\beta \hat{\cal{H}}}\hat{\cal{O}})$ with $Z\!=\!\mbox{tr}(e^{-\beta \hat{\mathcal{H}}})$ denotes the thermal expectation value for the observable $\hat{\mathcal{O}}$. T denotes the time-ordering operator\cite{abrikosov}. \\
For $n\!=\!1$, one obviously recovers $G_{1,\sigma_1\sigma_2}(\tau_1,\tau_2)\equiv G_{\sigma}(\tau_1,\tau_2)\!\equiv\!G(\tau_1,\tau_2)$, i.e., the one-particle Green's function. In the two-particle case ($n\!=\!2$), one usually considers the so-called ``generalized susceptibility'', defined by the following combination of one- and two-particle Green's functions
\begin{equation}
 \label{equ:defsusceptibility}
 \begin{split}
 \chi_{\sigma_1\sigma_2\sigma_3\sigma_4}&(\tau_1,\tau_2,\tau_3,\tau_4):=G_{2,\sigma_1\ldots\sigma_4}(\tau_1,\ldots,\tau_4)-\\-&G_{1,\sigma_1\sigma_2}(\tau_1,\tau_2)G_{1,\sigma_3\sigma_4}(\tau_3,\tau_4).
 \end{split}
\end{equation}
Without any loss of generality, one can always limit the domain for the imaginary times $\tau_i$ to the interval $[0,\beta]$ (cf. Ref. [\onlinecite{martin_schwinger}] and Appendix \ref{app:boundaryconditions}). Furthermore, due to the time-translational-invariance of the Hamiltonian, one can restrict oneself to only three time-arguments $\tau_1$, $\tau_2$, $\tau_3$ in the interval  $[0,\beta]$ (cf. Appendix \ref{app:boundaryconditions}), i.e., we can set $\tau_4=0$
 \begin{equation}
 \label{equ:twopartdef3times}
 \begin{split}
 \chi_{\sigma_1\sigma_2\sigma_3\sigma_4}&(\tau_1,\tau_2,\tau_3):=G_{2,\sigma_1\ldots\sigma_4}(\tau_1,\tau_2,\tau_3,0)-\\-&G_{1,\sigma_1\sigma_2}(\tau_1,\tau_2)G_{1,\sigma_3\sigma_4}(\tau_3,0).
 \end{split}
\end{equation}
One should also recall that, for the SU(2)-symmetric case considered here, the spin-indexes $\sigma_1\ldots\sigma_4$ are not completely independent, as a results of the conservation of spin. In fact, among the $2^4\!=\!16$ possible combinations of spins, only the following $3\times 2= 6$ remain: (i) $\sigma_1\!=\!\sigma_2 \!=\!\sigma_3\!=\!\sigma_4$, with $\sigma_1\!=\!\uparrow,\downarrow$; (ii)  $(\sigma_1\!=\!\sigma_2)\!\ne\!(\sigma_3\!=\!\sigma_4)$, with $\sigma_1\!=\!\uparrow,\downarrow$; (iii) $(\sigma_1\!=\!\sigma_4)\!\ne\!(\sigma_2\!=\!\sigma_3)$, with $\sigma_1=\uparrow,\downarrow$. This suggests the following definitions
\begin{subequations}
 \label{equ:chispindefinitions}
 \begin{equation}
  \label{equ:chispindef}
   \chi_{\sigma\sigma'}(\tau_1,\tau_2,\tau_3):=\chi_{\sigma\sigma\sigma'\sigma'}(\tau_1,\tau_2,\tau_3)
 \end{equation}
 \begin{equation}
  \label{equ:chispindefbar}
   \chi_{\overline{\sigma\sigma'}}(\tau_1,\tau_2,\tau_3):=\chi_{\sigma\sigma'\sigma'\sigma}(\tau_1,\tau_2,\tau_3),
 \end{equation}
\end{subequations}
which cover all six cases mentioned above. Eventually, using the crossing symmetry\cite{bickers}, one can show that the quantity defined in Eq.\ (\ref{equ:chispindefbar}) can be obtained from the one given in Eq.\ (\ref{equ:chispindef}) by means of a mere frequency shift as it is explained in Appendix \ref{app:crossingsymmetry}. For this reason we will commit ourselves to Eq.\ (\ref{equ:chispindef}) and consider Eq.\ (\ref{equ:chispindefbar}) only later when dealing explicitly with the spin-structure of the irreducible vertices. \\
When switching to frequency space, it is convenient to define the Fourier transform of $\chi$ in two different ways, which we refer to as particle-hole ($ph$) and particle-particle ($pp$) notation, respectively
\begin{subequations}
\label{equ:fouriertransdef}
\begin{equation}
\label{equ:fouriertransdefph}
\begin{split}
 \chi_{ph,\sigma\sigma'}^{\nu\nu'\omega}:= 
 \chi\bigl(\underset{\text{outgoing electrons}}{\underbrace{\nu\sigma,(\nu'+\omega)\sigma}}; &
 \underset{\text{incoming electrons}}{\underbrace{\nu'\sigma',(\nu+\omega)\sigma'}}\bigr):= \\
 =\int_0^{\beta}\!{d\tau_1 d\tau_2 d\tau_3}\, \chi_{\sigma\sigma'}(\tau_1,
 \tau_2,\tau_3) & e^{-i\nu\tau_1}e^{i(\nu+\omega)\tau_2}e^{-i(\nu'+\omega)\tau_3},
\end{split}
\end{equation}
\begin{equation}
\label{equ:fouriertransdefpp}
\begin{split}
 \chi_{pp,\sigma\sigma'}^{\nu\nu'\omega}:=
 \chi\bigl(\underset{\text{outgoing electrons}}{\underbrace{\nu\sigma,(\omega-\nu)\sigma'}}; &
 \; \underset{\text{incoming electrons}}{\underbrace{(\omega-\nu')\sigma,\nu'\sigma'}}\bigr):=\\
 =\int_0^{\beta}\!{d\tau_1 d\tau_2 d\tau_3}\,\chi_{\sigma\sigma'}(\tau_1,
 \tau_2,\tau_3) & e^{-i\nu\tau_1}e^{i(\omega-\nu')\tau_2}e^{-i(\omega-\nu)\tau_3},
 \end{split}
\end{equation}
\end{subequations}
with $\nu$ and $\nu'$ being fermionic Matsubara frequencies (i.e., $\nu^{(\prime)}=\frac{\pi}{\beta}(2n^{(\prime)}+1), n^{(\prime)}\in
\mathbb{Z}$) and $\omega$ being a bosonic Matsubara frequency (i.e., $\omega=\frac{\pi}{\beta}(2m), m\in
\mathbb{Z}$).\\
The choice of the frequency convention for both cases has a clear physical motivation. (i) In the $ph$-case one considers the scattering process of a hole with energy $-\nu$ and an electron with energy $\nu+\omega$, i.e. the total energy of this process is $\omega$. 
 \begin{figure}[h!]
 \centering
 \includegraphics[width=5cm,height=14mm]{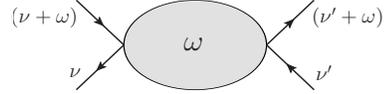}
 \caption{Particle-hole scattering.}
\end{figure}
\\
\noindent
(ii) In the $pp$-case we look at the scattering of two electrons with energies $\nu'$ and $\omega-\nu'$. Again the total energy of this process is $\omega$.
\begin{figure}[h!]
 \centering
 \includegraphics[width=5cm,height=14mm]{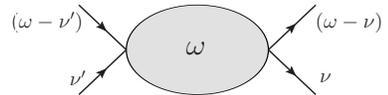}
 \caption{Particle-particle scattering.}
\end{figure}
\\Since in the full two-particle Green's function both processes are included, it is possible to express the $\chi_{pp}$ in terms of $\chi_{ph}$ and vice versa
\begin{equation}
\label{equ:phppnotation}
 \begin{split}
 &\chi^{\nu\nu'\omega}_{pp,\sigma\sigma'}=\chi^{\nu\nu'(\omega-\nu-\nu')}_{ph,\sigma\sigma'}\\
 &\chi^{\nu\nu'\omega}_{ph,\sigma\sigma'}=\chi^{\nu\nu'(\omega+\nu+\nu')}_{pp,\sigma\sigma'}.
 \end{split}
\end{equation}
In the following, we will constrict ourselves to $\chi_{ph}\equiv\chi$ and return to $\chi_{pp}$ only when explicitly needed (all the definitions, results etc. of the following section apply also to $\chi_{pp}$).\\
In the case of an interacting system ($U \neq 0$), the susceptibility $\chi$ can be decomposed into two parts, in order to divide the bubble terms (independent propagation of the two particles) from the vertex corrections, as it is illustrated in Fig. \ref{fig:chiph_eps}
\begin{figure}[t]
 \centering
 \includegraphics[width=0.4\textwidth]{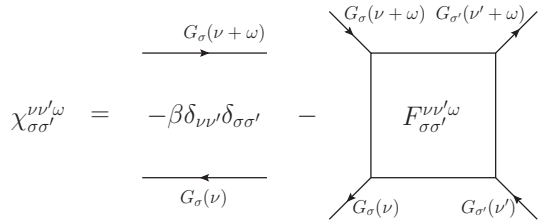}
\caption{Diagrammatic representation of the generalized susceptibility $\chi_{\sigma\sigma'}^{\nu\nu'\omega}$, as defined in Eqs.\ (\ref{equ:fouriertransdefph}) and (\ref{equ:chi_decomposition}). In the interacting case $\chi_{\sigma\sigma'}^{\nu\nu'\omega}$ is naturally decomposed into a bubble term ($\chi_0$, see Eq.\ (\ref{equ:chi0def})) and vertex correction terms ($F$).}
 \label{fig:chiph_eps}
\end{figure}
\begin{equation}
 \label{equ:chi_decomposition}
 \begin{split}
 \chi^{\nu\nu'\omega}_{\sigma\sigma'} &= 
 -\beta G_{\sigma}(\nu)G_{\sigma}(\nu+\omega)\delta_{\nu\nu'}\delta_{\sigma\sigma'} \\ &- 
 G_{\sigma}(\nu)G_{\sigma}(\nu+\omega)
 F^{\nu\nu'\omega}_{\sigma\sigma'}
 G_{\sigma'}(\nu')G_{\sigma'}(\nu'+\omega).
 \end{split}
\end{equation}
The  full vertex function $F$ appearing on the r.h.s. of Eq. (\ref{equ:chi_decomposition}) includes all possible vertex corrections, or in other words, all possible scattering events between the two propagating fermions, and can be hence interpreted in terms of the amplitude of a scattering process between two quasi-particles\cite{abrikosov,bickers}, at least in the Fermi-liquid regime, where the one-particle excitations are unambiguously defined. Eq. (\ref{equ:chi_decomposition}) can be also more compactly written in terms of the ``one-particle''-like bubble part of $\chi$, defined as
\begin{equation}
 \label{equ:chi0def}
 \chi_0^{\nu\nu'\omega}=-\beta G_{\sigma}(\nu)G_{\sigma}(\nu+\omega)\delta_{\nu\nu'},
\end{equation}
where  the spin-indexes on the l.h.s. can be omitted by restricting oneself to the paramagnetic case
\begin{equation}
 \label{equ:chi_chi0_def}
 \chi_{\sigma\sigma'}^{\nu\nu'\omega}=\chi_0^{\nu\nu'\omega}\delta_{\sigma\sigma'}-\frac{1}{\beta^2}\sum_{\nu_1\nu_2}
  \chi_0^{\nu\nu_1\omega}
  F^{\nu_1\nu_2\omega}_{\sigma\sigma'}
  \chi_0^{\nu_2\nu'\omega}.
\end{equation}
Analogous definitions can be introduced for the particle-particle notation. 

\subsection{Diagrammatics and mutual relations}
\label{subsec:digrammatics}
\begin{figure}[t]
 \centering
\includegraphics[width=0.35\textwidth]{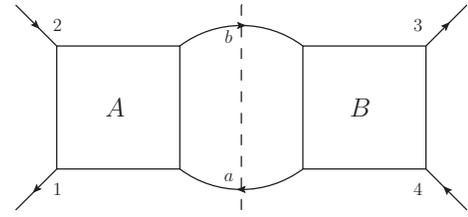}
\caption{Schematic representation of a generic particle-hole reducible diagram  contributing to the (full) scattering amplitude $F$.}
 \label{fig:phreduciblegeneric}
\end{figure}

The full vertex-function $F$ defined in Eq. (\ref{equ:chi_decomposition}) is the connected part of the complete four-point function. From a diagrammatic point of view $F$ consists of all ``fully connected'' two-particle diagrams, i.e., all diagrams which are not separated into two parts. These diagrams, in turn,  can be classified with respect to the way how they can be split into two parts by cutting {\sl two} internal Green's function lines.\\
\noindent
(i)  {\sl fully irreducible:} Diagrams of $F$, which cannot be split into two parts by cutting two internal Green's function lines. They represent the two-particle ``counter-part'' of the self-energy diagrams at the one-particle level.\\
\noindent
(ii) {\sl reducible:} Diagrams of $F$, which can be split by cutting two fermionic lines. At the two-particle level, however, the concept of {\sl reducibility} is more articulated than at the one-particle level. In fact, there are more possibilities of cutting lines than in the one-particle case, and, therefore, the concept of reducibility has to be referred to a specific channel: This specifies in which way two of the four outer legs of a given diagram can be separated from the other two. Labeling the outer legs of the two-particle diagrams with $1,2,3,4$, it is clear that three different possibilities exist: If the outer legs 1 and 3 denote outgoing particles (and 2 and 4 the incoming ones) than the diagrams where (13) can be separated from (24) are called particle-particle reducible, while the two other cases, i.e., (12) from (34) and (14) from (23), correspond to particle-hole longitudinal($ph$) and transverse($\overline{ph}$) reducible diagrams, respectively. One example for a (longitudinal) particle-hole reducible diagram is shown in Fig. \ref{fig:phreduciblegeneric} where (12) can be separated from (34) by cutting the internal lines $a$ and $b$. 

It is worth  recalling that each diagram is either fully irreducible or reducible in exactly one channel, i.e., there are no diagrams that are reducible in two or more channels \cite{note_reduc}. As a consequence, the complete vertex function $F$ can be decomposed into four parts - a fully irreducible part $\Lambda$ and the reducible contributions $\Phi_r$ in the three different channels (one particle-particle and two particle-hole)
\begin{equation}
F=\Lambda+\Phi_{pp}+\Phi_{ph}+\Phi_{\overline{ph}},
\label{equ:parquet}
\end{equation}
which have been written by now with a schematic notation, omitting spin- and frequency arguments (they will be explicitly introduced in the next sub-section).

\begin{figure*}[t]
 \centering
 \caption{Parquet equation.}
 \vspace{0.3cm}
 \includegraphics[width=1.0\textwidth]{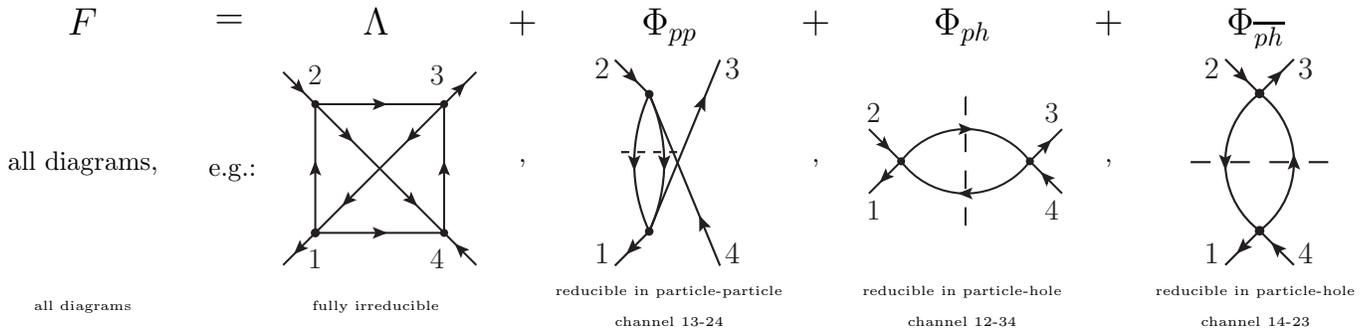}
 \label{fig:parquettable}
\end{figure*}

Such a decomposition of $F$ is known as {\sl parquet equation}\cite{bickers,janis} and it is schematically illustrated in Fig. \ref{fig:parquettable} with one low-order diagram shown for each of the four contributions. Note that the parquet equation represents just a ``classification'' of {\sl all} connected two-particle diagrams in four classes, and, therefore, does not imply in itself any kind of approximation.  This is analog to the one-particle case, where {\sl all} connected one-particle diagrams can be divided into a set of reducible and irreducible ones (defining the self-energy).

The full vertex function $F$ appearing in the parquet equation can be calculated from the complete four-point matrix element $\chi$ via Eq. (\ref{equ:chi_decomposition}). In order to work with the parquet equation one needs additional relations connecting $F$ and the reducible vertices $\Phi_r$. This can be achieved by defining new quantities $\Gamma_r$ 
 \begin{equation}
  \label{equ:phigamma}
   F=\Phi_r+\Gamma_r,\quad r=pp,ph,\overline{ph}.
 \end{equation}
Since $F$ contains all diagrams and $\Phi_r$ contains all the diagrams which are reducible in the given channel $r$, $\Gamma_r$ is the set of all diagrams, which are irreducible in a given channel $r$. Since each diagram is either fully irreducible or reducible in a given channel, we have that  $\Gamma_r=\Lambda+\Phi_{j_1}+\Phi_{j_2},\;j_1,j_2\ne r$. \\
The $\Gamma_r$ vertices, in turn, can be calculated from $F$ by means of an integral-equation, the so called {\sl Bethe-Salpeter equation}
\begin{equation}
 \label{equ:bethesalpeter}
 F=\Gamma_r+\int\Gamma_r GG F,
\end{equation}
where the integral-symbol denotes an integration/summation over all internal degrees of freedom (e.g.: frequencies, spin, $\ldots$). 
The interpretation of this equation is very simple: $F$ is the sum of all connected diagrams which are irreducible in the given channel $r$ (i.e., $\Gamma_r$) and the diagrams that are reducible in this channel (i.e., $\Phi_r$). The latter can be easily expressed by connecting the corresponding irreducible vertex $\Gamma_r$ to the full vertex function $F$, via two Green's function lines. Such a ``decomposition'' procedure, which avoids any possible double-counting of diagrams, is obviously not unique, as it can be performed independently for all channels.

The two particle-hole channels are connected by the crossing relations\cite{bickers}
\begin{equation}
 \label{equ:crossinggammageneralph}
 \Gamma_{ph}(1234)=-\Gamma_{\overline{ph}}(1432),
\end{equation}
which corresponds to interchanging the two incoming particles. In contrast the particle-particle channel fulfills a crossing relation on its own, namely
\begin{equation}
 \label{equ:crossinggammageneralpp}
 \Gamma_{pp}(1234)=-\Gamma_{pp}(1432)
\end{equation}
which is identical to the crossing relation for the full vertex $F$ (cf. Appendix \ref{app:crossingsymmetry}). \\

In order to clarify the meaning of the different reducible and irreducible vertex functions ($\mathcal{V}_{(r)}\!=\!F$, $\Gamma_r$ or $\Lambda$), defined in this section, it is important to discuss how some well-known approximation schemes correspond  diagrammatically to different levels of approximation of the two-particle vertex functions.
\begin{table}[t]
\caption{Approximations at different vertex levels. ${\cal V}_{(r)}=F,\Gamma_r$ or $\Lambda$, respectively.}
\label{tab:approximations}
\begin{center} 
\hspace{-0.7cm}
\begin{tabular}{m{0.03\textwidth}|m{0.03\textwidth}|m{0.21\textwidth}|m{0.21\textwidth}|}
\cline{2-4}
 &\begin{center} $\mathcal{V}_{(r)}$\end{center} & \begin{center}static $\left({\cal V}_{(r)}=U\right)$\end{center} & \begin{center}local, dynamic $\left({\cal V}_{(r)}={\cal V}_{(r),loc}^{\nu\nu'\omega}\right)$\end{center}\\
\cline{2-4}
\multirow{3}*{
\fcolorbox{white}{white}{
  \begin{picture}(7,100) (0,0)
  \end{picture}}
}
 & 
 \begin{center}$F$\end{center}
 & \begin{center}2$^{\text{nd}}$-order perturbation theory\end{center} & 
\begin{center}$-$\end{center}\\ 
\cline{2-4}
 & 
\begin{center}$\Gamma_{\text{r}}$\end{center}
& \begin{center}RPA, FLEX, 
pseudopotential parquet\end{center} & \begin{center} Moriyasque D$\Gamma$A\end{center}\\ 
\cline{2-4}
& 
\begin{center}$\Lambda$\end{center}
& \begin{center}parquet approximation\end{center} & \begin{center}D$\Gamma$A \end{center}\\
\cline{2-4}
\end{tabular}
\end{center}
\end{table}
An overview over approximations adopted for the Hubbard model at different vertex levels is given in Tab. \ref{tab:approximations}. 

The simplest schemes are obtained, obviously, by remaining at the ``surface'' of the two-particle diagrammatic complexity, i.e., when making approximations directly at the level of the full vertex function $F$. For instance,  replacing $F$ with the bare Hubbard interaction $U$ leads directly to second order perturbation theory for the self-energy, as it can be easily seen by making such a replacement ($F\!=\!U$) in the Schwinger-Dyson equation of motion (cf. Eq. (\ref{equ:dysonschwinger}) and Fig. \ref{fig:dysonschwinger} below). This method is used to approximate the self-energy of the Hubbard model in the asymptotic cases of weak and strong coupling (see also, at half-filling, the iterated perturbation theory\cite{DMFTrev,itp}). \\

Making a step further in the diagrammatics means to apply an approximation at the level of the irreducible vertices $\Gamma_r$ (third row of Tab. \ref{tab:approximations}).  For example, one can calculate the full scattering amplitude $F$ by simply replacing $\Gamma_{ph}$ (or $\Gamma_{pp}$) with the bare interaction $U$, which corresponds to the well-known random phase approximation (RPA)\cite{mahan}. Adopting this substitution for all three irreducible channels and the fully irreducible one leads to the Fluctuation Exchange (FLEX) approximation\cite{FLEX,FLEXSE}. If one chooses a different ``effective'' constant $\Gamma_r=U^{\text{eff}}_r$ for each of the three channels, where the $U^{\text{eff}}_r$ is determined by some additional condition, one ends up with the so-called pseudopotential parquet approximation\cite{bickers,note_park}. All these methods represent reasonable approximations in the case of small-intermediate $U$, i.e., in the weak-to-intermediate coupling regime, and improve systematically the second order perturbation theory results. \\

Remaining at the same level of the diagrammatics, a more complex approach, which aims to include non-perturbatively the physics of the MIT, is to replace $\Gamma_r$ by its purely local counterpart,  instead of  the lowest order (and frequency-independent) contribution $U$ only. This is done in the ladder (or Moriyasque) version of the D$\Gamma$A\cite{DGA2}. There, the reducible vertex $\Phi_r$ (typically in one or two of the different channels) is allowed to be non-local. Such an approximation is justified, when one of the channels dominates the physics (e.g., the spin channel for the half-filled Hubbard model at intermediate-to-strong coupling): In that case, by neglecting the interference between different channels, one constructs a non-local $\Phi_r$ from the local $\Gamma_r$ only in selected channels, while the remaining ladders can be considered as purely local quantities.  \\

Finally, going to the deepest level of the two-particle diagrammatics, one may apply approximations directly to the fully irreducible vertex $\Lambda$ (fourth row in Tab. \ref{tab:approximations}). Again, for the Hubbard model the simplest approximation of this class is obtained by replacing $\Lambda$ with the bare Hubbard interaction $U$. This approach is called parquet approximation\cite{bickers,janis,fotso}. Similarly as for $\Gamma_r$, however, one can also replace $\Lambda$ with its purely local (but frequency dependent) counter-part, which corresponds to the D$\Gamma$A\cite{DGA1}. While the approximations at the level of $\Lambda$ are usually very expensive computationally, and particular numerical tricks\cite{tam} have to be used, they may be necessary to capture the complicate physics of the Hubbard model in situations where none of the channels really dominates over the others (e.g., for the doped case, which is relevant for the physics of the high-temperature superconducting cuprates). It is also  worth recalling that the fully irreducible diagrams $\Lambda$ have a very compact structure, which cannot --per definition-- include any ladder diagrams. Therefore, for the case of the Hubbard model (where the naked interaction is completely local in space) it is reasonable to expect a weak spatial dependence, as it seems to be confirmed\cite{maier} by a dynamical cluster approximation\cite{DCA,DCA3d} study for the two-dimensional Hubbard model. \\  
To conclude the diagrammatic classification of several known approximation schemes, it is worth to mention also the cases of DMFT\cite{DMFTrev} and DF\cite{DF}, which -in a strict sense- do not belong directly to any of the specific levels discussed above. In fact, DMFT is an exact theory in the limit $d\rightarrow\infty$, and {\sl all} local vertices ($F$, $\Gamma_r$, and $\Lambda$) are included in its diagrammatics. However, the internal Green's function lines are also local in DMFT and non-local correlations are totally neglected in contrast to  the methods discussed previously (from perturbation theory to D$\Gamma$A). On the other hand, DF does include spatial correlations beyond DMFT, via an expansion in a dual fermion space, defined via a Hubbard-Stratonovic decoupling of the non-local degrees of freedom in Eq.\ (\ref{equ:defhubbard}). The coefficients of the DF expansion are given by the  local generalized susceptibility of the AIM, which would correspond to a dynamical approximation at the level of  the full vertex function $F$. However, a classification of DF in Tab. \ref{tab:approximations} would be not easy, as the local but dynamical vertex $F$ represents the ``naked interaction'' in the dual fermion space. Hence, different diagrammatic degrees of accuracy are obtained by applying specific approximations (perturbation theory/ladder/Parquet) for the dual fermions. While this makes a classification of the DF in Tab. \ref{tab:approximations} difficult at the moment, our discussion calls for future investigations of the correspondence between a given approximation in DF and the diagrammatics of the real electrons.   

\subsection{Spin-dependence: Definition of the different channels}
\label{sec:spin-structure}
In this section, the spin-structure of the irreducible vertex in the three different channels is explicitly discussed. As mentioned before, for the SU(2)-symmetric case there are three independent spin-combinations, i.e., $\uparrow\uparrow$, $\uparrow\downarrow$ and $\overline{\uparrow\downarrow}$, see  Eq. (\ref{equ:chispindefinitions}). On the level of the full vertex $F$ the $\uparrow\downarrow$- and the $\overline{\uparrow\downarrow}$-spin-combination are connected by the crossing relation Eq. (\ref{equ:crossingphnotationF}) given in Appendix \ref{app:crossingsymmetry}. However, since at the level of $\Gamma$ the crossing relation connects two different channels (see Eq.\ (\ref{equ:crossinggammageneralph})) at least in the particle-hole case we will for now consider the $\Gamma_{r,\overline{\uparrow\downarrow}}$ as an independent quantity. Hence, we have three different spin-combinations for each of the three channels, which would lead to nine different $\Gamma$'s. Using crossing- and SU(2)-symmetry, however, one can show that only four of them are independent, which corresponds  to the definition of the so called density ($d$), magnetic ($m$), singlet ($s$) and triplet ($t$) ``channels'', given as follows
\begin{align}
\label{equ:channeldefdensity}
&\Gamma_{d}^{\nu\nu'\omega}=\Gamma_{ph,\uparrow\uparrow}^{\nu\nu'\omega}+\Gamma_{ph,\uparrow\downarrow}^{\nu\nu'\omega}\\
\label{equ:channeldefmagnetic}
&\Gamma_{m}^{\nu\nu'\omega}=\Gamma_{ph,\uparrow\uparrow}^{\nu\nu'\omega}-\Gamma_{ph,\uparrow\downarrow}^{\nu\nu'\omega}\\
\label{equ:channeldefsinglet}
&\Gamma_{s}^{\nu\nu'\omega}=\Gamma_{pp,\uparrow\downarrow}^{\nu\nu'\omega}-\Gamma_{pp,
\overline{\uparrow\downarrow}}^{\nu\nu'\omega}\\
\label{equ:channeldeftriplet}
&\Gamma_{t}^{\nu\nu'\omega}=\Gamma_{pp,\uparrow\downarrow}^{\nu\nu'\omega}+\Gamma_{pp,
\overline{\uparrow\downarrow}}^{\nu\nu'\omega},
\end{align}
The same definitions are valid at the level of $F$ and $\Lambda$ as well. However, since neither $F$ nor $\Lambda$ can be divided into different channels, only two of them are actually independent. \\
A more detailed discussion of the different irreducible channels can be found in Appendix \ref{app:spindiag} and in Ref. \onlinecite{bickers}.


\section{DMFT results}
\label{Sec:DMFTpos}

In this section we present our DMFT results for all local two-particle vertex-functions, i.e., $F$ (full vertex),  $\Gamma_r$ (irreducible in channel $r$) and  $\Lambda$ (fully irreducible vertex) of the half-filled Hubbard model on a cubic lattice. The frequency-dependent local vertex functions have been obtained by solving the AIM associated to the DMFT solution by means of exact diagonalization (ED).
Specifically, the DMFT(ED) algorithm used to compute the local two-particle vertex functions exploits the Lehmann representation for the generalized local susceptibilities $\chi_{ph}$, $\chi_{pp}$ (Eqs. (\ref{equ:fouriertransdefph})-(\ref{equ:fouriertransdefpp})) of the AIM, whose analytic expression has been derived and reported, e.g., in Refs. \onlinecite{DGA1,dolfen,hartmut}. From $\chi_{ph}$ and $\chi_{pp}$, the full (connected) two
particle vertex ($F$) is easily computed via Eq. (\ref{equ:chi_decomposition}). Then, all the two-particle vertices irreducible in one channel ($\Gamma_r$) are obtained via
inversion of the corresponding Bethe Salpeter equations (see Eq. (\ref{equ:bethesalpeter}), and Eqs. (\ref{equ:invertchi}), (\ref{equ:invertchipp}), (\ref{equ:invertchippupdown}) in Appendix B). Eventually,
the knowledge of the $\Gamma_r$ in all channels ($r=d,m,s,t$) allows
to determine the fully irreducible vertex ($\Lambda$) via the (inverse) parquet equation(s) (Fig. 5, Eqs. (\ref{equ:parquet})-(\ref{equ:phigamma}), and Eqs. (\ref{equ:defreducible})-(\ref{equ:parquetchannels4}) in Appendix C).  

The present ED-calculations have been performed with $N_s=5$ sites in the AIM, keeping (at least)  $160$ (positive) fermionic and bosonic Matsubara frequencies, which has required, for each determination of the generalized susceptibility, a parallel calculation of  about $100.000$ CPU-hours on the Vienna Scientific Cluster (VSC). This allowed for a precise calculation of the (Matsubara) frequency structures of the two-particle vertex functions at all levels of the diagrammatics, down to the fully irreducible objects. The accuracy of the calculations has been directly tested by checking the asymptotic behavior and the symmetry properties (see Appendix \ref{app:symmetries}) of the different vertex functions, as well as by comparing them to the corresponding atomic limit results.
 Furthermore, the numerical robustness of our DMFT(ED) results for reducible and irreducible local vertices has been also successfully  verified by comparing  with corresponding results obtained with a Hirsch-Fye quantum Monte Carlo algorithm\cite{hirschfye} as impurity solver, in a slightly higher temperature regime ($\beta\!=\!20.0$) than that considered here.

For presenting our DMFT results we will follow the thread underlying the discussion of Tab. \ref{tab:approximations} at the end of Sec. \ref{subsec:digrammatics}: we will start analyzing the most conventional (and easiest to compute) among the vertex functions, i.e., the full vertex $F$, in the next subsection (Sec. \ref{sec:fullvertex}). Subsequently, in Sec. \ref{subsec:irreduciblechannel} we will make a step deeper in the diagrammatics, presenting our DMFT results for the vertices irreducible in one specific channel ($\Gamma_r$), and, finally, in Sec. \ref{subsec:fullyirreducible}, results for the most fundamental block of the two-particle diagrammatics,  the fully irreducible vertex function $\Lambda$, will be presented and discussed.

In all cases, the frequency structure of the local vertices will be first examined  at small values of $U$ (e.g., $U=0.5$), which allows for a direct comparison with perturbation theory. Deviations from the perturbation theory predictions will be also discussed, and in Sec. \ref{subsec:effects}, their effects on more conventional physical and thermodynamical quantities will be eventually addressed. Finally, the impact of our analysis on possible improvements of numerical calculations of two-particle vertex functions is briefly discussed in Sec. \ref{practuse}.





\subsection{Full vertex functions}
\begin{figure}[t]
 \centering
 \includegraphics[width=0.4\textwidth]{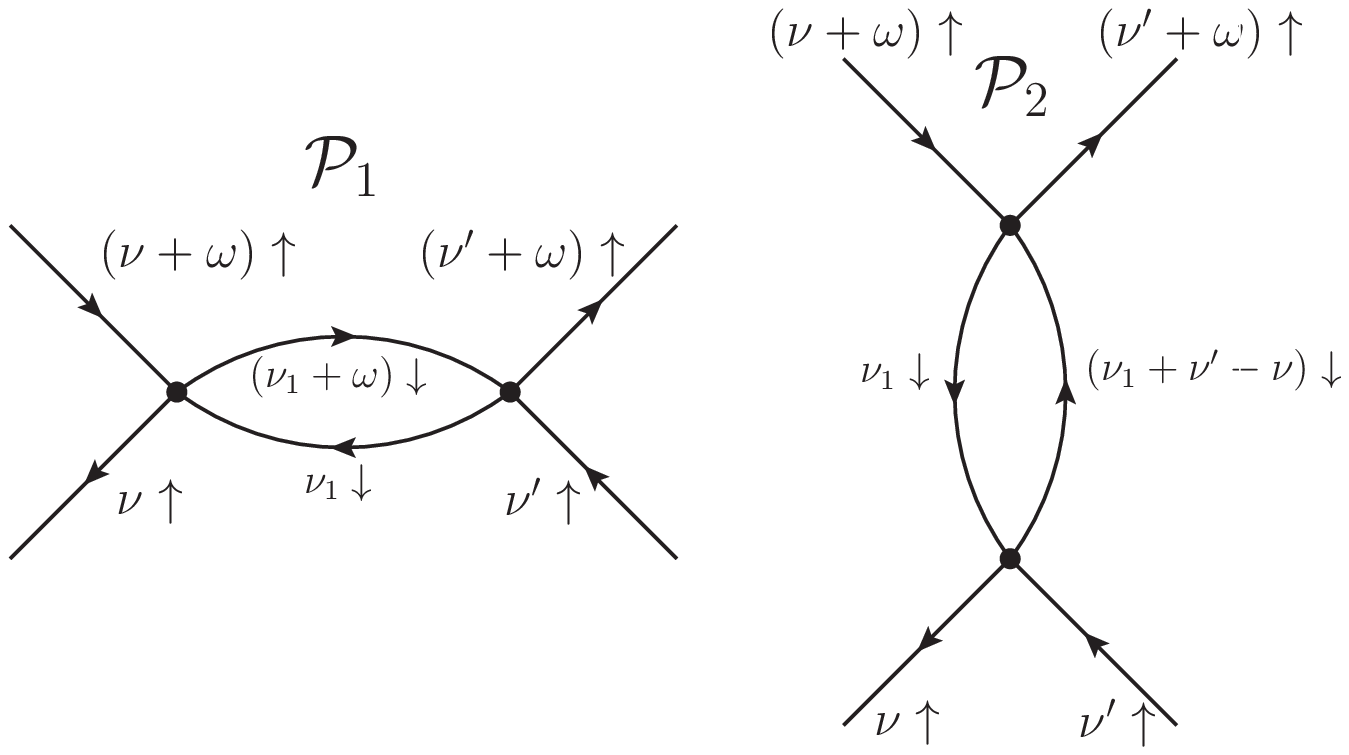}
 \includegraphics[width=0.5\textwidth]{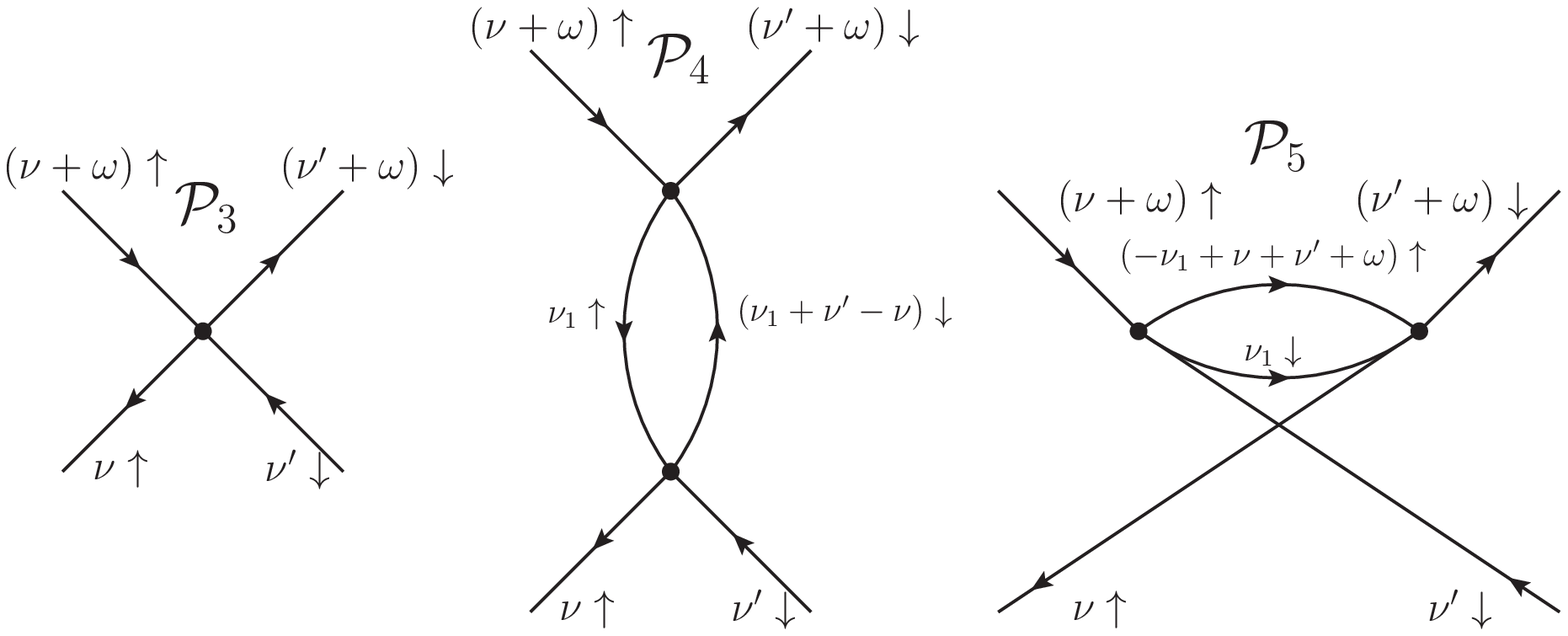}
 \caption{Upper row: lowest order (perturbative) diagrams for $F_{\uparrow\uparrow}$, Lower row: the same for $F_{\uparrow\downarrow}$}
 \label{fig:vertex_perturb}
\end{figure}

The full vertex $F$ contains all connected diagrams with two particles coming in and two particles going out. In Fig. \ref{fig:vertex_perturb} the lowest order diagrams for the two possible spin combinations are shown in the particle-hole frequency convention (the corresponding results in the  particle-particle notation can be simply obtained via the transformation $\omega\rightarrow\omega-\nu-\nu'$, see also Eq. (\ref{equ:phppnotation})). We recall, moreover, that on the level of $F$, the singlet- and the triplet-channel are just linear combinations of $F_d$ and $F_m$ (see discussion in Sec. \ref{sec:spin-structure}).  
\label{sec:fullvertex}
In terms of Green's functions the lowest order contributions for $F$ read as follows
\begin{subequations}
 \label{equ:secondorderupup}
 \begin{equation}
  \label{equ:secondorderupup1}
  {\cal P}_1 = +\frac{U^2}{\beta}\sum_{\nu_1}G(\nu_1)G(\nu_1+\omega),
  \end{equation}
  \begin{equation}
   \label{equ:secondorderupup2}
   {\cal P}_2 = -\frac{U^2}{\beta}\sum_{\nu_1}G(\nu_1)G(\nu_1+\nu'-\nu),
  \end{equation}
\end{subequations}
for the $\uparrow\uparrow$-case and
\begin{subequations}
 \label{equ:secondorderupdown}
 \begin{equation}
  \label{equ:secondorderupdown1}
   {\cal P}_3= U,
 \end{equation}
 \begin{equation}
  \label{equ:secondorderupdown2}
  {\cal P}_4= -\frac{U^2}{\beta}\sum_{\nu_1}G(\nu_1)G(\nu_1+\nu'-\nu),
 \end{equation}
 \begin{equation}
  \label{equ:secondorderupdown3}
  {\cal P}_5 = -\frac{U^2}{\beta}\sum_{\nu_1}G(\nu_1)G(-\nu_1+\nu+\nu'+\omega),
 \end{equation}
\end{subequations}
for the $\uparrow\downarrow$-case. The lowest order contributions for the four different channels, as defined for the $\Gamma$'s in Eqs. (\ref{equ:channeldefdensity})-(\ref{equ:channeldeftriplet}), hence, are given by
\begin{align}
\label{equ:channellowestorder}
&F_{d}^{\nu\nu'\omega}=F_{ph,\uparrow\uparrow}^{\nu\nu'\omega}+F_{ph,\uparrow\downarrow}^{\nu\nu'\omega}=U+O(U^2)\\
&F_{m}^{\nu\nu'\omega}=F_{ph,\uparrow\uparrow}^{\nu\nu'\omega}-F_{ph,\uparrow\downarrow}^{\nu\nu'\omega}=-U+O(U^2)\\
&F_{s}^{\nu\nu'\omega}=F_{pp,\uparrow\downarrow}^{\nu\nu'\omega}-F_{pp,\overline{\uparrow\downarrow}}^{\nu\nu'\omega}=2U+O(U^2)\\
&F_{t}^{\nu\nu'\omega}=F_{pp,\uparrow\downarrow}^{\nu\nu'\omega}+F_{pp,\overline{\uparrow\downarrow}}^{\nu\nu'\omega}=0+O(U^2).
\end{align}
\begin{figure}[t]
 \centering
 \includegraphics[width=0.5\textwidth]{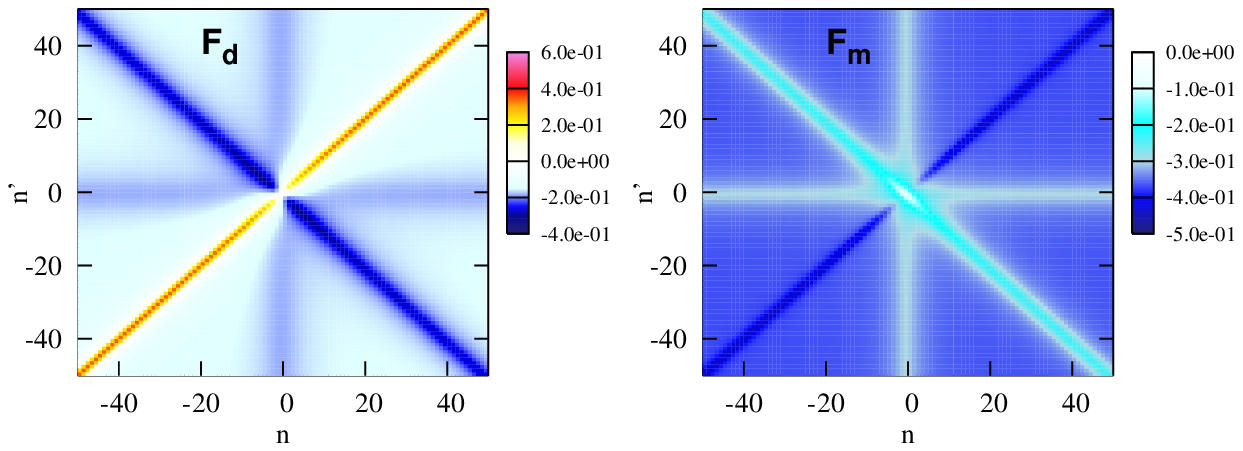}\\[0.3cm]
 \includegraphics[width=0.5\textwidth]{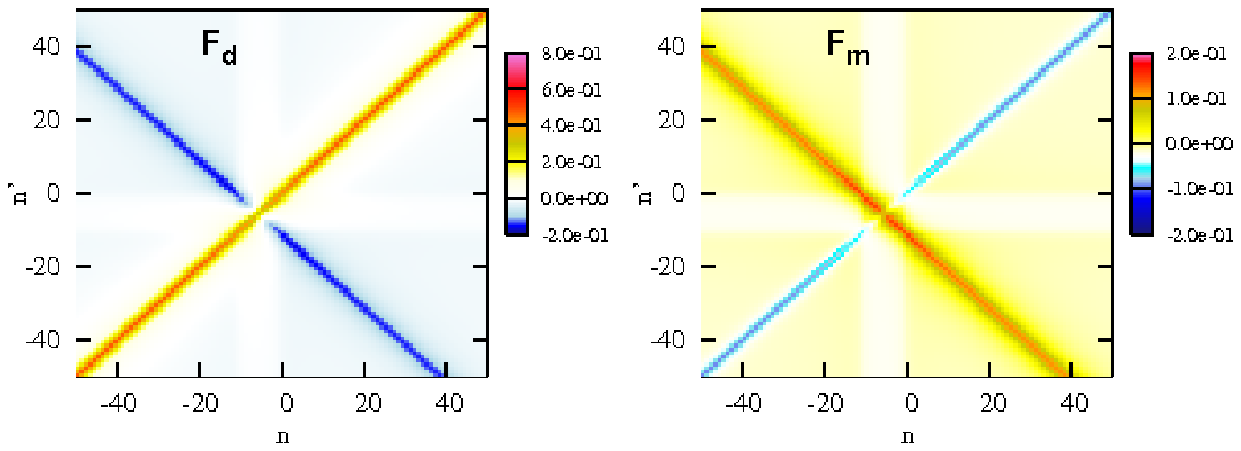}
 \caption{(color online). Vertex functions vs. the two fermionic frequencies $\nu\!=\!\frac{\pi}{\beta}(2n\!+\!1)$ and $\nu'\!=\!\frac{\pi}{\beta}(2n'\!+\!1)$ ($n,n'\in\mathds{Z}$): density part $F_d^{\nu\nu'\omega}\!-\!U$ (left) and magnetic part $F_m^{\nu\nu'\omega}\!+\!U$ (right) for $U\!=\!0.5$  at half-filling ($\beta=26.0$) for fixed $\omega\!=\!\frac{\pi}{\beta}(2m)$ ($m\in\mathds{Z}$); Upper row: $\omega\!=\!0$ ($m\!=\!0$), Lower row: $\omega\!=\!20\frac{\pi}{\beta}$ ($m\!=\!10$).}
 \label{fig:f_densityplot}
\end{figure}
\begin{figure*}[ht]
 \centering
 \includegraphics[width=\textwidth]{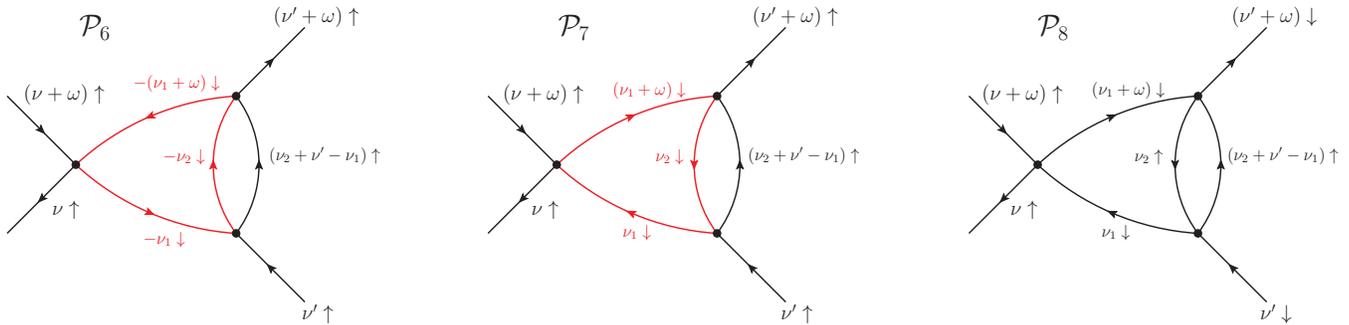}
 \caption{Third order (perturbative) diagrams for $F_{\uparrow\uparrow}$ and $F_{\uparrow\downarrow}$}
 \label{fig:pertthirdorder}
\end{figure*}
The full vertex functions $F$ in the density ($F_d\!=\!F_{\uparrow\uparrow}+F_{\uparrow\downarrow}$) and magnetic ($F_m\!=\!F_{\uparrow\uparrow}\!-\!F_{\uparrow\downarrow}$) channel calculated by means of DMFT are shown in Fig. \ref{fig:f_densityplot} for the case $\omega\!=\!0$ (upper row) and $\omega\!\ne\!0$ (lower row). The $x$-axis corresponds to $\nu$ while the $y$-axis is assigned to $\nu'$.  Note that, for the sake of readability of the figure, instead of the absolute values of the Matsubara-frequencies just the corresponding indexes are given. The vertex functions $F$ are calculated for $U=0.5$  at half-filling, at a temperature value ($\beta=26.0$) close to the critical end-point of the MIT in DMFT\cite{DMFTrev}. It should be recalled that for the half-filled system all vertex functions are purely real (see Eq. (\ref{equ:particleholesuscupupfourier}) in Appendix \ref{app:particleholesymmetry}). Furthermore, here as in the following, the (constant) contribution of the first order diagram, namely the Hubbard $U$, is subtracted in order to better highlight the frequency structure of the two-particle vertices beyond the standard lowest order perturbative results. \\
One can now trace the different features of the two-dimensional plot of $F$ back to different types of diagrams.\\
First of all, let us note that a constant background is still present, despite the subtraction of the lowest order term. This constant background stems from higher order diagrams that are independent of $\nu$ and $\nu'$. An example in second order perturbation theory is given in Fig. \ref{fig:vertex_perturb}: The left diagram in the upper row (${\cal P}_1$) has no $\nu$- or $\nu'$-dependence, as it also follows from Eq. (\ref{equ:secondorderupup1}). The same holds also for diagrams of higher order with all possible vertex corrections inside  the bubble of ${\cal P}_1$: The sum of all diagrams of this family yields the constant background observed in the upper row of Fig. \ref{fig:f_densityplot}. However, this feature is reduced with an increasing value of $\omega$ as one can observe in the lower row of Fig. \ref{fig:f_densityplot}.\\
Secondly, the evident structure along the main diagonal (i.e., the region around the line $\nu\!=\!\nu'$) stems from diagrams like the second ones (${\cal P}_2$, ${\cal P}_4$) in the upper or lower row of Fig. \ref{fig:vertex_perturb} (see also Eqs. (\ref{equ:secondorderupup2}) and (\ref{equ:secondorderupdown2})), which describe (at the order considered) scattering processes {\sl reducible} in the transverse particle-hole channel. More specifically, these diagrams, as well as similar diagrams of the same type but with vertex corrections included, depend only on $(\nu\!-\!\nu')$, which means that they give a constant contribution along the lines $\nu\!-\!\nu'\!=\!const$. The largest contribution, however, is expected for the case $\nu-\nu'=0$ when the scattering between the particle and the hole occurs at the Fermi surface. One can easily identify these structures in Fig. \ref{fig:f_densityplot}. For the density-case one obviously has to add the diagrams of the $\uparrow\uparrow$- and the $\uparrow\downarrow$-channel which leads to twice the contribution of such diagrams in second order perturbation theory. For the magnetic vertex, instead, these second-order contributions cancel exactly each other, and only higher order contributions to this diagonal line remain, which explains the difference between the two channels.\\

Furthermore, one also observes an enhanced scattering rate along the secondary diagonal $\nu'\!=\!-\nu$. 
The origin of this structure stems from diagrams like ${\cal P}_5$ in the lower row of Fig. \ref{fig:vertex_perturb} (see also Eq. (\ref{equ:secondorderupdown3})), which build up scattering processes {\sl reducible} in the particle-particle channel. In fact, such diagrams (with and without vertex corrections in the bubble) describe the scattering of two particles with energies $(\nu+\omega)$ and $\nu'$. Hence, the corresponding scattering amplitude is enhanced for total energies at the Fermi level, i.e., for $\nu'=-\nu-\omega$. If $\omega=0$ this yields the secondary diagonal in the plots shown in the upper row of Fig. \ref{fig:f_densityplot}. However, for a finite $\omega$ this line is expected to be shifted to $\nu'=-\nu-\omega$. This behavior is shown for case of the  tenth bosonic Matsubara frequency, i.e., for $\omega=\frac{\pi}{\beta}(2\times 10)=20\frac{\pi}{\beta}$, in the lower row of Fig. \ref{fig:f_densityplot}. The main diagonal remains unchanged, as it stems from $\omega$-independent diagrams, while the secondary diagonal is shifted compared to the upper row. \\
Finally, one can also note a cross-structure (shaped as a ``+``) in the upper row of Fig. \ref{fig:f_densityplot},
i.e., one observes an enhanced scattering amplitude compared to the constant background along the lines $\nu=0$ and $\nu'=0$. In order to explain the origin of these structures one has to go at least to third order perturbation theory. The contribution of the diagrams shown in Fig. \ref{fig:pertthirdorder} reads as 
\begin{equation}
\label{equ:3order}
 {\cal P}_{7,8}= -\frac{U^3}{\beta^2}\sum_{\nu_1\nu_2}G(\nu_1)G(\nu_1+\omega)G(\nu_2)G(\nu_2+\nu'-\nu_1).
\end{equation}
One sees that it is independent of $\nu$ and therefore it gives a constant contribution along the horizontal line $\nu'=const.$ in Fig. \ref{fig:f_densityplot}, with a maximum for $\nu' \sim 0$, as only in this situation one has the possibility to have all Green's functions appearing in Eq. (\ref{equ:3order}) simultaneously at the Fermi level. In complete analogy one can construct diagrams that do not depend on $\nu'$ (in this case the vertical bubble should be on the left side). At $\omega=0$ these result in a maximum at $\nu=0$,  which  explains  the cross-structure  observed in $F$. A more quantitative understanding of the ''+''-shaped cross-structure in $F$ requires a closer look at the spin-dependence of the third-order diagrams shown in Fig. \ref{fig:pertthirdorder}: While the only contribution to $F_{\uparrow\downarrow}$ is given by the last diagram in this figure (${\cal P}_8$), for $F_{\uparrow\uparrow}$ one has to include two topologically non-equivalent diagrams (${\cal P}_6$ and ${\cal P}_7$). The explicit expression for ${\cal P}_7$ is completely equivalent to that for the $\uparrow\downarrow$-vertex (i.e., ${\cal P}_8$). As for ${\cal P}_6$, its expression one can be obtained from ${\cal P}_{7}$ by simply inverting the (internal) $\downarrow$-lines (and the corresponding frequencies). This leads to
\begin{equation}
\label{equ:3order_upup}
 {\cal P}_{6}= -\frac{U^3}{\beta^2}\sum_{\nu_1\nu_2}G(-\nu_1)G(-\nu_1-\omega)G(-\nu_2)G(\nu_2+\nu'-\nu_1).
\end{equation}
At half-filling one has $G(-\nu)\!=\!-G(\nu)$ due to particle-hole symmetry (see Eq. (\ref{equ:particleholegreenfourier}) in Appendix \ref{app:particleholesymmetry}), which implies ${\cal P}_6\!=\!-{\cal P}_7$. Hence, the diagrams ${\cal P}_6$ and ${\cal P}_7$ cancel each other, and only the contribution ${\cal P}_8$ to the $\uparrow\downarrow$-vertex remains in this order of perturbation theory.
This can be viewed as a manifestation of the so-called Furry's theorem\cite{mandel-shaw} of quantum electrodynamics, which states that -as a results of the electron-positron symmetry- {\sl a closed fermionic loop containing an odd number of Fermions always vanishes}. This also explains the different signs of the cross-structure originated by the diagrams shown in Fig. \ref{fig:pertthirdorder}: In the density/magnetic channel $F_{\uparrow\downarrow}$ enters with a plus/minus-sign which leads to a negative/positive contribution of diagram ${\cal P}_8$ w.r.t. the negative background of $F$ as one can directly see by the colored features of the plots in the first row of Fig. \ref{fig:f_densityplot}. 
\begin{figure}[t]
 \centering
 \includegraphics[width=0.5\textwidth]{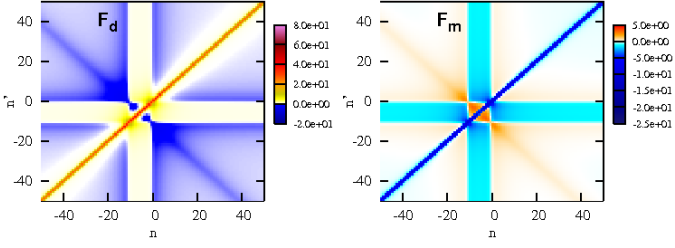}\\[0.3cm]
 \caption{(color online). Vertex functions vs. the two fermionic frequencies $\nu\!=\!\frac{\pi}{\beta}(2n\!+\!1)$ and $\nu'\!=\!\frac{\pi}{\beta}(2n'\!+\!1)$: density part $F_d^{\nu\nu'\omega}\!-\!U$ (left) and magnetic part $F_m^{\nu\nu'\omega}\!+\!U$ (right) for $U\!=\!2.0$  at half-filling ($\beta\!=\!26.0$) for $\omega\!=\!20\frac{\pi}{\beta}$.}
 \label{fig:f_densityplot_U2}
\end{figure}

Extending our analysis to the finite-$\omega$-case, we observe a broadening of the cross-structure of $F$, with the formation of an horizontal and a vertical band, both extended from $ - \omega$ to $0$, see Figs. \ref{fig:f_densityplot} (lower row) and \ref{fig:f_densityplot_U2}. This more general feature can be traced to the sign changes (i.e., their jumps at zero frequencies) of the four Green's functions in the Eq. (\ref{equ:3order}). 
It is important to notice, here, that the combination of this broadened cross-structure with the (shifted) diagonal maxima/minima of the vertex function $F$ generates the appearance of a sort of square-like feature in the frequency plots, whose importance will be further discussed in Sec. \ref{practuse}.

In the final part of this section, we investigate how our DMFT-results change, upon increasing the value of the Hubbard interaction beyond the weak-coupling regime. 
While the quantitative comparison with perturbation theory is obviously deteriorating when increasing $U$, it is interesting to note that at least the ``topology'' of the main frequency structure of the vertex $F$ survives qualitatively unchanged also for higher values of $U$, and -to a good extent- even in the atomic limit.

For the sake of conciseness, we focus here on a generic case at finite bosonic frequency ($\omega =20 \frac{\pi}{\beta}$), which allows for better identifying, separately, the main features of the vertex functions  due to the different frequency shifts of the several structures discussed above. 

In Fig. \ref{fig:f_densityplot_U2} we report our results for $U\!=\!2.0$, i.e., four times larger than the interaction value considered before. 
Notice that this value of $U$ lies well beyond the perturbative regime, and corresponds, e.g., to the $U$ for which the maximum of the N\'eel-Temperature of the antiferromagnetic instability is predicted by DMFT at half-filling\cite{DMFTrev,kent}. 

From a first visual inspection of the plots, it emerges clearly that the main frequency structures of the vertex functions correspond well to those we have just discussed for the perturbative case: One can easily identify  similar structures as in the plots of Fig. 7, lower panels, in the same position as before, and even with the same sign for the deviation w.r.t. the lowest order constant contribution.
Remarkably, a similar situation can be observed even in the extreme case of the atomic limit ($D=0$), where an analytic expression of the full vertex functions $F_d, F_m$ can be derived\cite{nota_hartmut} directly from the Lehmann representation:
\begin{widetext}
\begin{align}
\label{equ:fdmatomic}
F_{d,m}^{\nu\nu'\omega} &  = F_{\uparrow\uparrow}^{\nu\nu'\omega} \pm F_{\uparrow\downarrow}^{\nu\nu'\omega}\\
\label{equ:defatomicup}
F_{\uparrow\uparrow}^{\nu\nu'\omega} & =-\beta\frac{U^2}{4}\frac{\delta_{\nu\nu'}-\delta_{\omega0}}{\nu^2(\nu'+\omega)^2}\left(\nu^2+\frac{U^2}{4}\right)\left((\nu'+\omega)^2+\frac{U^2}{4}\right)  \\
\label{equ:defatomicdown}
F_{\uparrow\downarrow}^{\nu\nu'\omega} & = 
-U+\frac{U^3}{8}\frac{\nu^2+(\nu+\omega)^2+(\nu'+\omega)^2+(\nu')^2}{\nu(\nu+\omega)(\nu'+\omega)\nu'}+
\frac{3U^5}{16}\frac{1}{\nu(\nu+\omega)(\nu'+\omega)\nu'}+\nonumber\\ &\phantom{=}+
\beta\frac{U^2}{4}\frac{1}{1+e^{\beta U/2}}\frac{2\delta_{\nu(-\nu'-\omega)}+\delta_{\omega0}}{(\nu+\omega)^2(\nu'+\omega)^2}\left((\nu+\omega)^2+\frac{U^2}{4}\right)\left((\nu'+\omega)^2+\frac{U^2}{4}\right)+\nonumber\\ &\phantom{=}-
\beta\frac{U^2}{4}\frac{1}{1+e^{-\beta U/2}}\frac{2\delta_{\nu\nu'}+\delta_{\omega0}}{\nu^2(\nu'+\omega)^2}\left(\nu^2+\frac{U^2}{4}\right)\left((\nu'+\omega)^2+\frac{U^2}{4}\right)
\end{align}
\end{widetext}
From the analytical expression of the atomic limit Eqs. (\ref{equ:fdmatomic})-(\ref{equ:defatomicdown}), in fact, one can easily recognize the same main frequency features of the vertex $F$, appearing in the cases studied with DMFT ($U=0.5$, $U=2.0$). For instance, when considering -for the sake of generality- the case at finite $\omega$, one immediately identifies the $ph$  and the $pp$ diagonal structures in the terms proportional to $\delta_{\nu\nu'}$ and to $\delta_{\nu(-\nu'-\omega)}$ in Eqs. (\ref{equ:fdmatomic}), (\ref{equ:defatomicup}) and (\ref{equ:defatomicdown}). Not surprisingly for the atomic limit of repulsive models, however, the magnitude of the $pp$ structure is exponentially suppressed when $T \ll U$, as it can be inferred from the corresponding prefactor.  
Interestingly, beyond the two diagonal structures, one can recognize in the atomic limit formulas also the broadened ``+``-shaped cross-structure, which was generated at small $U$  by the third order diagrams, Eq. (\ref{equ:3order}). In fact, this structure corresponds --even in the atomic limit-- to the term proportional to $U^3$ in Eq. (\ref{equ:defatomicdown}).    
\begin{figure*}[t]
 \centering
 \includegraphics[width=1.0\textwidth]{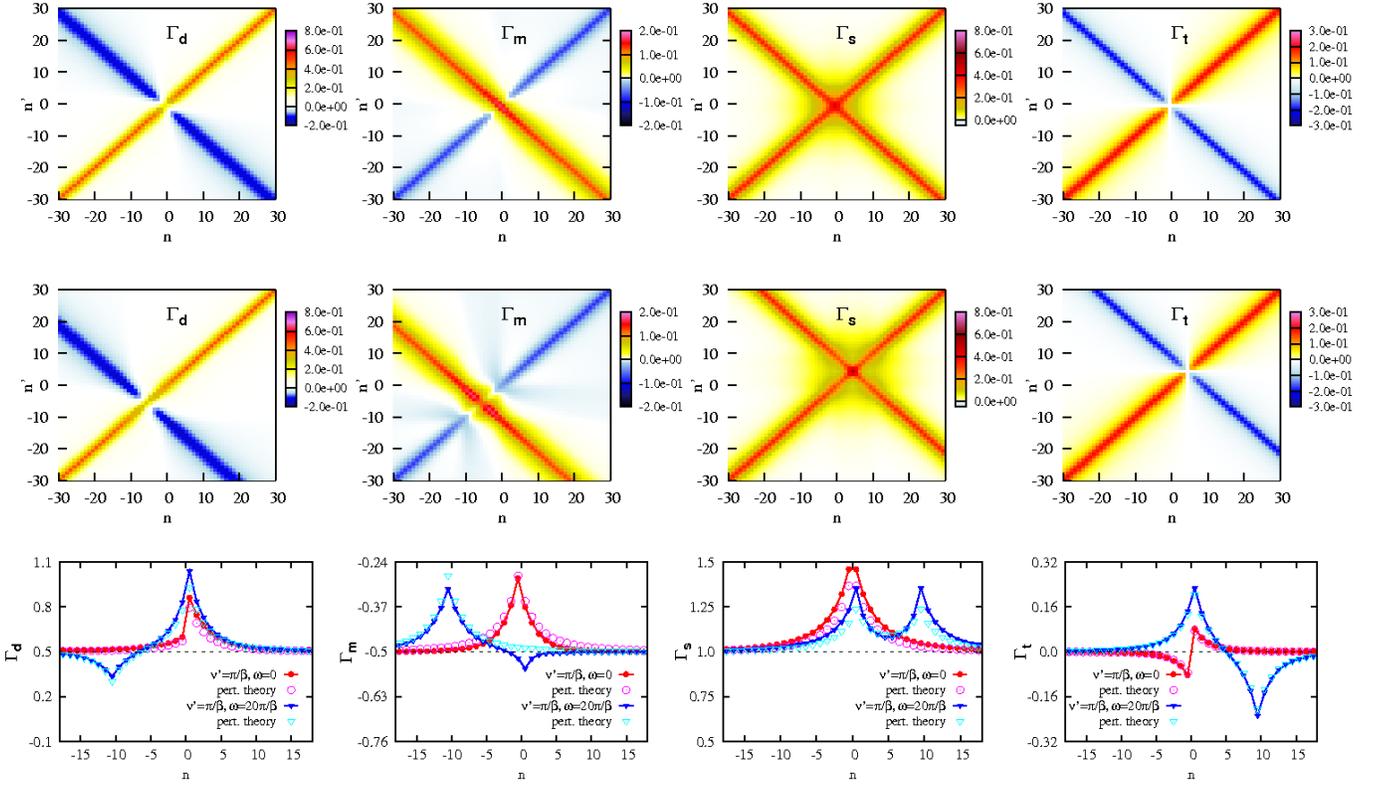}
 \caption{(color online). Vertex functions irreducible the different channels. First and second row:  $\Gamma_d^{\nu\nu'\omega}\!-\!U$, $\Gamma_m^{\nu\nu'\omega}\!+\!U$, $\Gamma_s^{\nu\nu'\omega}\!-\!2U$, $\Gamma_t^{\nu\nu'\omega}$ for $U\!=\!0.5$ at half-filling ($\beta\!=\!26.0$) for $\omega\!=\!0$ (first row) and $\omega\!=\!20\frac{\pi}{\beta}$ (second row) vs. the two fermionic frequencies $\nu$ and $\nu'$. For singlet- and triplet-channel particle-particle notation was adopted.
Third row: one-dimensional snapshot of the same vertex functions for  $\nu'\!=\!\frac{\pi}{\beta}$ ($n'\!=\!0$, fixed) and the two values of $\omega$, compared to the corresponding (lowest order) perturbative results.}
\label{fig:gamma_density}
\end{figure*}

\subsection{Irreducible vertices in one selected channel}
\label{subsec:irreduciblechannel}
At the level of irreducible vertices one has necessarily to consider four independent quantities: the density and the magnetic vertex correspond to the two possible spin-combinations in the longitudinal channel ($\Gamma_{ph}$) while the singlet and the triplet vertex are linear combinations of the two different spin-directions in the particle-particle channel ($\Gamma_{pp}$). The transverse channel is not independent since it can be obtained from the longitudinal one by means of the crossing symmetry (see previous section and Appendix \ref{app:crossingsymmetry}).\\
\begin{figure}[t]
 \centering
 \includegraphics[width=0.50\textwidth]{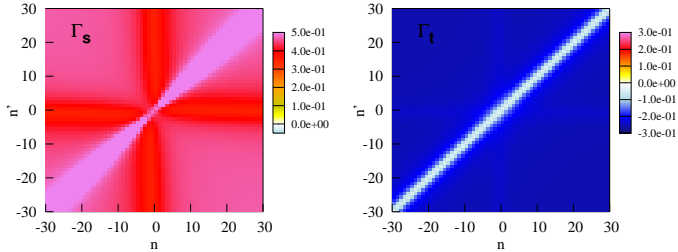}
 \caption{(color online). Irreducible particle-particle vertices in particle-hole notation: $\Gamma_s^{\nu\nu'(\nu+\nu'+\omega)}\!-\!2U$ (left) and $\Gamma_t^{\nu\nu'(\nu+\nu'+\omega)}$ (right) vs. $\nu$ and $\nu'$ for the same parameters as in Fig. \ref{fig:gamma_density} (for $\omega\!=\!0$).}
 \label{fig:gamma_density_ph_notation}
\end{figure}
We start with the discussion of the two-dimensional density-plots for the four channels for $U\!=\!0.5$ and two different values of $\omega$ ($\omega\!=\!0$ and $\omega\!=\!20\frac{\pi}{\beta}$, Fig. \ref{fig:gamma_density}). It is important to recall that for the two particle-particle channels, i.e., the singlet- and the triplet-channel, the particle-particle notation is adopted. 
\begin{figure*}[t]
 \centering
 \includegraphics[width=1.0\textwidth]{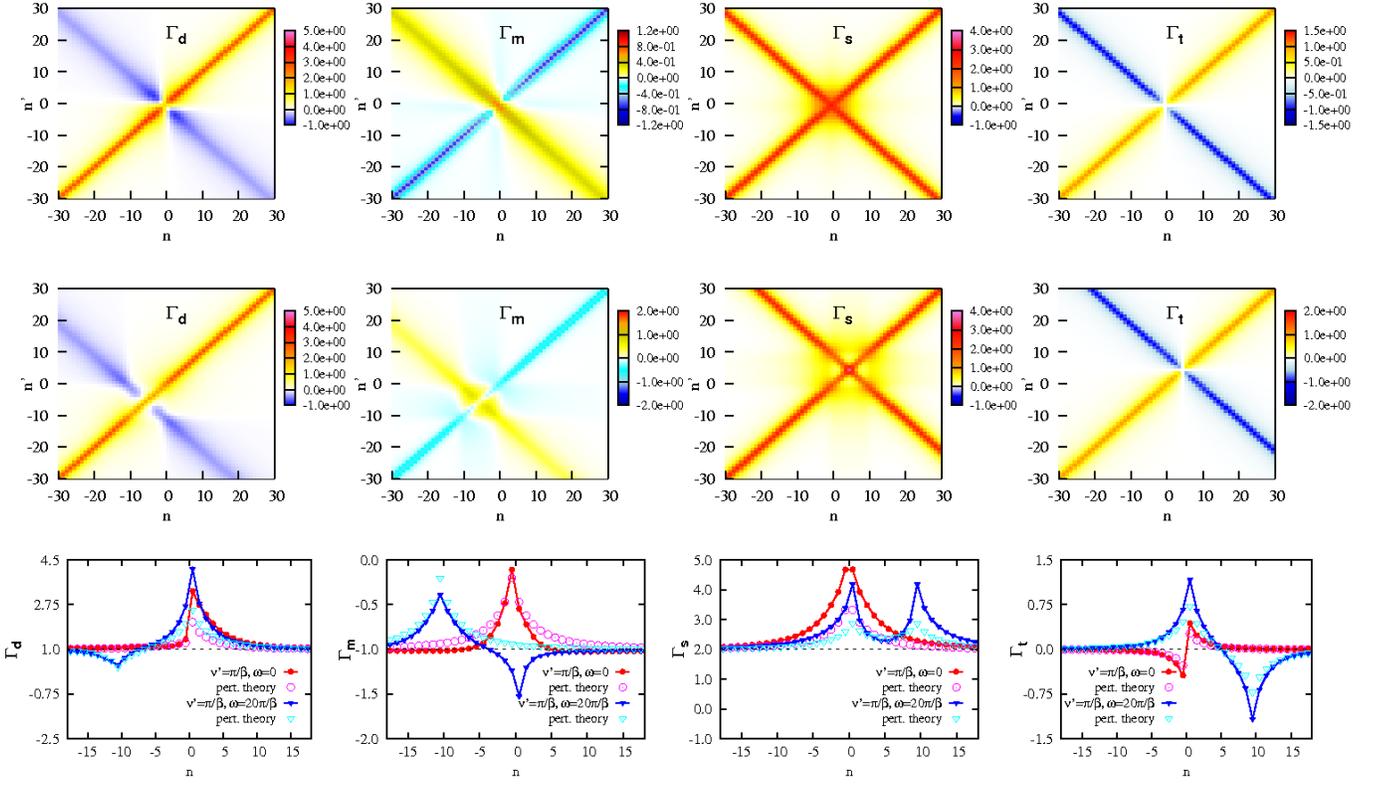}
 \caption{(color online). Same as in Fig. \ref{fig:gamma_density} but for U=1.0. The comparison with perturbation theory in the third row shows that, while the most important features of the vertex structures are located exactly in the same position as for U=0.5, the values of the irreducible vertices $\Gamma$ (with the exception of the triplet channel)  deviate already by more than a factor two from perturbation theory in the most significant points of the frequency space, e.g., in the proximity to the maxima and the minima of the vertex functions. At the same time, perturbation theory appears to describe reasonably the region in between the main structures as well as the asymptotic behavior of the vertex.}
 \label{fig:gamma_densityU1.0}
\end{figure*}
For the density- and the magnetic channel (first two plots in each row of Fig. \ref{fig:gamma_density}) one identifies the main and the secondary diagonal as it was the case for the full vertex function. However, the constant background and the cross-structure in the center are missing. In fact,  such features originate from diagrams like ${\cal P}_1$ in Fig. \ref{fig:vertex_perturb} and ${\cal P}_8$ in Fig. \ref{fig:pertthirdorder} which are reducible in the longitudinal channel and therefore do not contribute to $\Gamma_d$ and $\Gamma_m$. \\
For the particle-particle channel in particle-hole notation one would expect again a constant background as well as the cross-structure (see Fig. \ref{fig:gamma_density_ph_notation}) but no secondary diagonal since the diagram ${\cal P}_5$ in Fig. \ref{fig:vertex_perturb} does not contribute (it is particle-particle reducible). \\
The situation is, however, different when adopting particle-particle notation, i.e., $\omega\rightarrow\omega-\nu-\nu'$ for the particle-particle irreducible channels: In this case, the first diagram $\mathcal{P}_1$ in the upper row of Fig. \ref{fig:vertex_perturb}, in fact, depends on $\omega-\nu-\nu'$ (instead of being independent of $\nu$ and $\nu'$ at all) and therefore yields a constant contribution along the lines $\omega=\nu+\nu'$. For the case $\omega=0$ this contribution reaches a maximum yielding the secondary diagonal structure, as it appears in the density-plots for $\Gamma_s$ and $\Gamma_t$ (last two plots in each row of Fig. \ref{fig:gamma_density}). On the other hand, in the particle-particle notation, the diagram ${\cal P}_5$ in Fig. \ref{fig:vertex_perturb} becomes independent from $\nu$ and $\nu'$ and, hence, would lead to a constant background. Since this diagram is particle-particle reducible such a contributions is missing in $\Gamma_s$ and $\Gamma_t$ which, therefore do not exhibit a constant background as it can be observed in Fig. \ref{fig:gamma_density}. 

Let us mention an interesting feature characterizing the triplet vertex $\Gamma_t$ in the particle-hole notation: The triplet vertex $\Gamma_t$ coincides with the $\uparrow\uparrow$-vertex. Hence, (in the particle-hole notation) it describes the effective interaction between two electrons with spin $\uparrow$ and energies $\nu+\omega$ and $\nu'$, respectively (these are the energies associated with the two annihilation operators in the particle-hole notation). However, for $\nu'=\nu+\omega$ both electrons would be in the same state, which is forbidden by the Pauli-principle. Therefore the triplet-vertex is expected to be strongly suppressed along this line, as it can be actually observed (for $\omega=0$) in Fig. \ref{fig:gamma_density_ph_notation} (right panel).\\ 
In the lowest row of Fig. \ref{fig:gamma_density} one-dimensional slices of the four irreducible vertex functions are shown: $\nu'$ is kept fixed to the first fermionic Matsubara frequency in that case, and $\Gamma$ is plotted for two different values of $\omega$ as a function of $\nu$. One observes a good agreement with perturbation theory obtaining deviations of the order $ U^3 \sim 0.1$. This is to be expected since third order diagrams have not been considered in the perturbation expansion.
\begin{figure*}[t]
 \centering
 \includegraphics[width=1.0\textwidth]{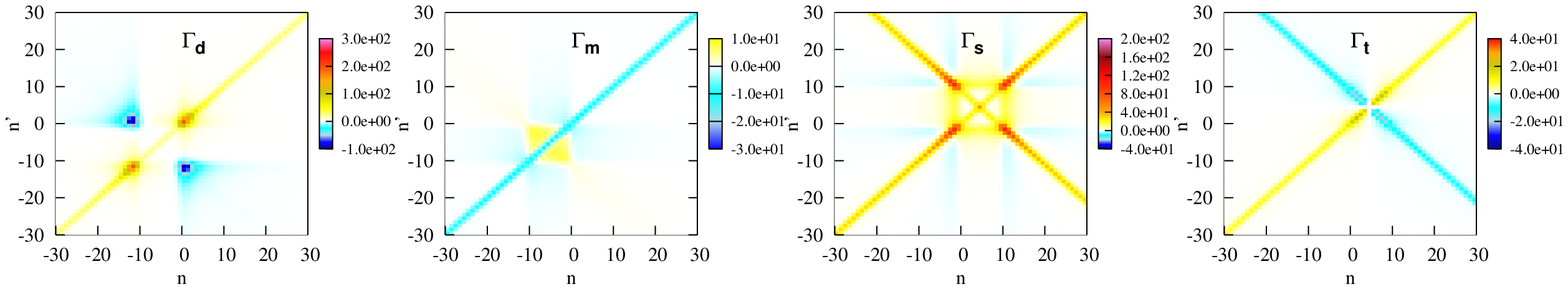}\\[0.3cm]
 \includegraphics[width=1.0\textwidth]{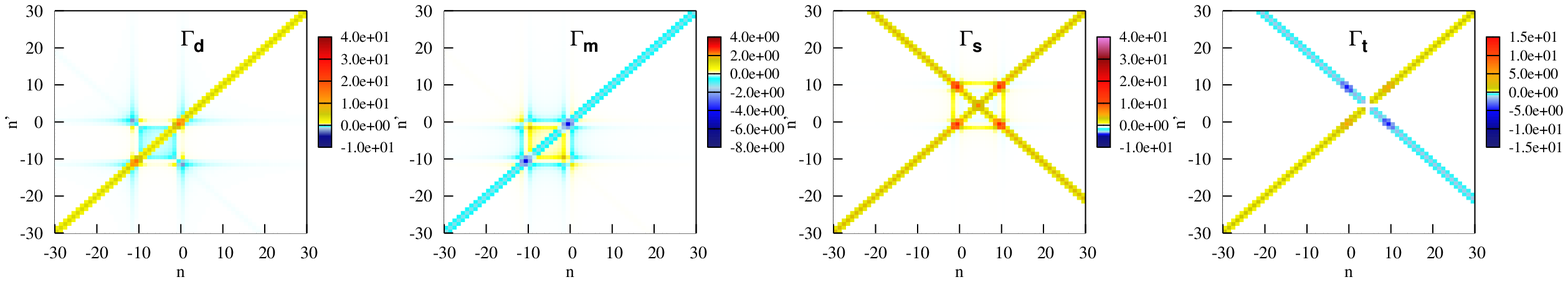}
\caption{(color online). Upper row: Same as in Fig. \ref{fig:gamma_density} but for $U\!=\!2.0$ and at finite $\omega\!=\!20\frac{\pi}{\beta}$. Lower row: atomic limit ($D\!=\!0.0$, $U\beta\!=\!10.0$) calculation.
For singlet- and triplet-channel particle-particle notation was adopted.}
 \label{fig:gamma_density_U2.0}
\end{figure*}

At $U=1.0$ quantitative deviations from perturbation theory results become gradually visible in the ``low-frequency'' (small $\nu,\nu'$, if $\omega\!=\!0$) region, see Fig. \ref{fig:gamma_densityU1.0}, with the possible exception of  the triplet channel. To define more generally such ``low-frequency'' region, however, one should consider the data at {\sl finite} bosonic frequency ($\omega=20\frac{\pi}{\beta}$), whose quantitative comparison  with perturbation theory is shown in the lowest row of Fig.\ref{fig:gamma_densityU1.0}.
Here, one observes that the largest deviations are found  in correspondence to the main structures of the vertex functions, i.e., in the proximity of maxima/minima and saddle points of the $\Gamma_r$ functions (where the exact values deviate already by more than a factor two from perturbation theory). Similarly as for $F$, however, the position of these frequency structures is unchanged w.r.t. perturbation theory. Moreover, from the quantitative point of view, perturbation theory still works reasonably well, not only for the asymptotics, but also for the region in between the main vertex structures.

This trend is preserved -to some extent- when increasing $U$ further, as it is immediately understandable  from the plots for the most general case (i.e., for finite bosonic frequency) Fig. \ref{fig:gamma_density_U2.0}. At $U\!=\!2.0$ (upper row in Fig. \ref{fig:gamma_density_U2.0}), the main frequency structures of the four $\Gamma_r$ are located in the same position as for lower values of $U$, where perturbation theory was still applicable for their understanding and classification. 
By a closer inspection of the $U\!=\!2.0$ results, however, some general trends emerge. First, one observes a weakening of the secondary diagonal ($\nu\!=\!-\nu'\!-\!\omega$) in $\Gamma_d$ and $\Gamma_m$ in the upper row of Fig. \ref{fig:gamma_density_U2.0}. This is a consequence of the suppression of particle-particle scattering events in repulsive models as it was already discussed at the end of section \ref{sec:fullvertex} for the the full vertex-function $F$ in the atomic limit. Secondly, one can see that, for large values of $U$, the triplet vertex (last plot in the first row of Fig. \ref{fig:gamma_density_U2.0}) consists almost completely of the double diagonals. This ``$\times$``-structure can be understood in terms of the atomic limit where one can find an exact parametrization of the triplet vertex in terms of these two diagonals, i.e., 
\begin{equation}
 \label{equ:gammatripletatomic}
\Gamma_t^{\nu\nu'\omega}=f(\nu,\omega)\delta_{\nu\nu'}+g(\nu,\omega)\delta_{\nu(-\nu'+\omega)} 
\end{equation}
as it can be deduced from Eq. (\ref{equ:defatomicup}) and the plot for the triplet vertex in the atomic limit (last plot in the lower row of Fig. \ref{fig:gamma_density_U2.0}).\\
Noteworthy, the main features of the vertex functions are preserved also when considering the extreme case of zero bandwidth, i.e, the atomic limit ($D\!=\!0$), shown in the lower row of Fig. \ref{fig:gamma_density_U2.0} for a generic choice of $U\beta\!=\!10.0$. 
One can clearly identify the dominant diagonal- and square-like features at $\omega\!\ne\!0$ whose locations are unchanged w.r.t. to the case of finite $D$.

The principal message of our analysis of the vertices $F$ and $\Gamma_r$ can be, hence, summarized as follows:
The ``topology'' of the vertex functions is to a large extent preserved when increasing $U$. This, in turn, may be used to build up approximated schemes for parameterizing the vertex functions as well as for simplifying two-particle calculations based on DMFT input (see also Sec. \ref{practuse}). One example for such a parametrization of the triplet vertex function is provided by the same Eq. (\ref{equ:gammatripletatomic}), obtained for the atomic limit.\\ 
At the same time, the numerical values of the vertex functions for $U \gtrsim 1$ are completely unrelated to those of perturbation theory, and their knowledge requires, therefore, an explicit calculation by means of DMFT (or of its extensions). In this respect, our results are also suggestive that the non-perturbative nature of the Mott transition at the two-particle level is to some extent 'enclosed' precisely in this part of the vertex functions, i.e., in their (large) values at the extremal points. 

In fact, while it is not easy to find any general interpretation scheme for the ``low-energy'' features of the vertex functions to be applicable to the non-perturbative regime,  some clear hallmarks of  the physics of the Mott MIT are readily identified in the full vertex $F$ as well as in the irreducible one $\Gamma_r$. Specifically, all these vertex functions contain {\sl reducible} processes in the magnetic channel\cite{spinreducibility}.
Hence, as the local magnetic spin susceptibility $\chi_m(\omega=0)$ {\sl diverges} at the MIT (at $T\!=\!0$), the effects of this divergence will be visible in all the frequency structures, which contain such diverging diagrammatic contributions. According to our previous discussion, in order to individuate their location, we can still resort to the perturbative analysis of the previous subsection. Indeed, although the {\sl divergence} of the local spin susceptibility at the MIT is ultimately a non-perturbative phenomenon driven by the vertex corrections of the local spin correlation function, the 'labeling' of the vertex structures performed by means of perturbation theory does not lose its validity at higher $U$. \\
In practice, this means that all the peak structures we have previously identified via second order diagrams with a bubble in the $ph$ magnetic channel, will become strongly enhanced by increasing $U$ (as one can see for example in Fig. \ref{fig:gamma_density_U2.0} for $U\!=\!2.0$), with diverging maxima/minima exactly at the MIT. This happens because with increasing $U$ the bubble contributions of the perturbation theory are dressed by all higher order contributions, including an infinite resummation of the internal vertex corrections. In the non-perturbative regime, these vertex corrections modify the values of the corresponding structures (with the bubble replaced by the corresponding full susceptibility $\chi_r$), without changing their positions.  In the actual cases we have considered here, this situation occurs for the $ph$ main diagonal ($\nu=\nu'$) structures of the $F$ vertex function  and of the $\Gamma_r$, which include, in different ways, the local spin (magnetic) bubble in the perturbative regime (and, hence, the full enhanced local susceptibility at larger $U$).  

\subsection{Fully irreducible vertices}
\label{subsec:fullyirreducible}
In this subsection we present results for the fully irreducible vertex $\Lambda$. The formulas used for the actual calculations are given in Appendix \ref{app:parquet}. As mentioned before, the fully irreducible vertex is the most fundamental ``brick'' among the two-particle vertex functions, representing the diagrammatic analog of the self-energy at the two-particle level. Hence, approximations based on this level of the diagrammatics, such as the parquet approximation or the D$\Gamma$A are extremely appealing from a theoretical point of view. At the same time, the calculation of fully irreducible vertex functions is quite challenging, so that the few calculations\cite{janis,fotso,tam} based on approximations for $\Lambda$, simply replace the latter with its lowest-order contribution ($U$). Motivated by the lack of studies on the frequency dependence of the fully irreducible local vertices, even at the level of perturbation theory, we will present our  numerical and analytical results with more details than in the previous subsections and we also explicitly consider the effects of the frequency dependence of $\Lambda$ in selected physical and thermodynamical quantities as a function of the Hubbard interaction $U$.   

By definition no channel-dependence of the fully irreducible vertex function $\Lambda$ can exist, since it is irreducible in {\sl all} channels. Hence, as in Sec. III A, here we also restrict ourselves to the DMFT result for the density and the magnetic vertices, which represent the two possible spin combinations. Diagrammatically, the lowest order contribution to the fully irreducible vertex is the bare Hubbard interaction $U$ (diagram ${\cal P}_6$ in Fig. \ref{fig:vertex_perturb}). The next terms in the perturbation expansion are already of 4$^{\text{th}}$ order: These diagrams have the form of an envelope, and, hence,  are usually referred to as ``envelope''-diagrams. 
\begin{figure}[t]
 \centering
 \includegraphics[height=0.9\textwidth]{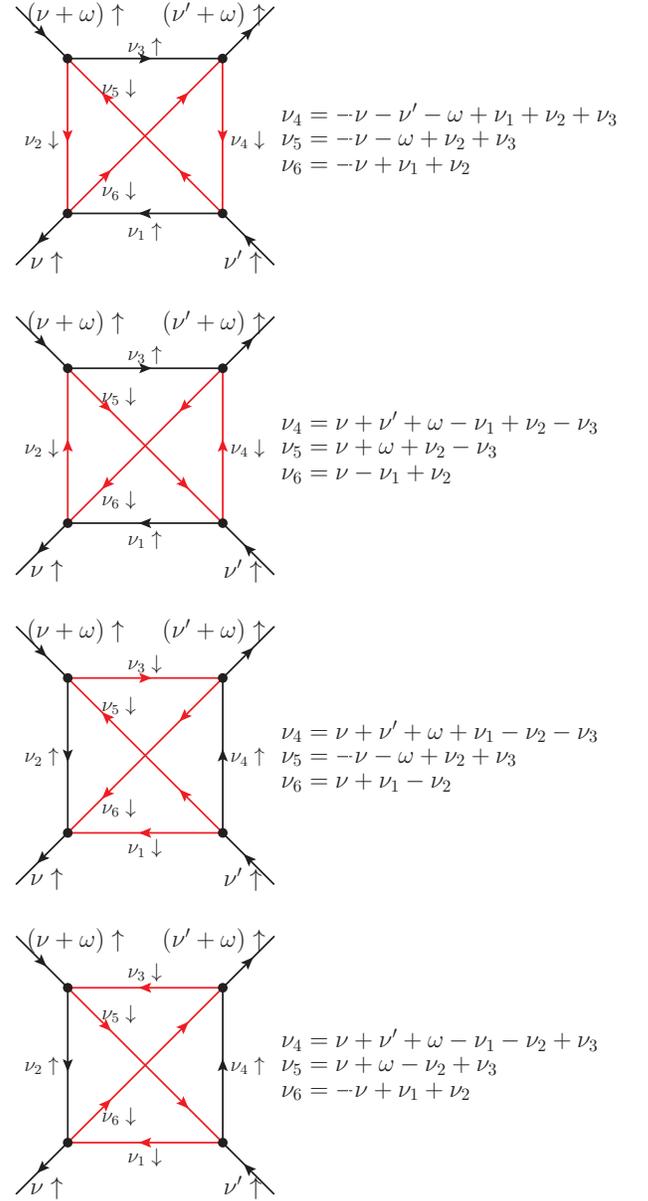}
 \caption{(color online). $U^4$-contributions to the perturbative expansion of the fully irreducible vertex $\Lambda^{\nu\nu'\omega}_{\uparrow\uparrow}$ (``envelope''-diagrams) in particle-hole notation. }
 \label{fig:envelope_up_up}
\end{figure}
\begin{figure}[t]
 \centering
 \includegraphics[height=0.45\textwidth]{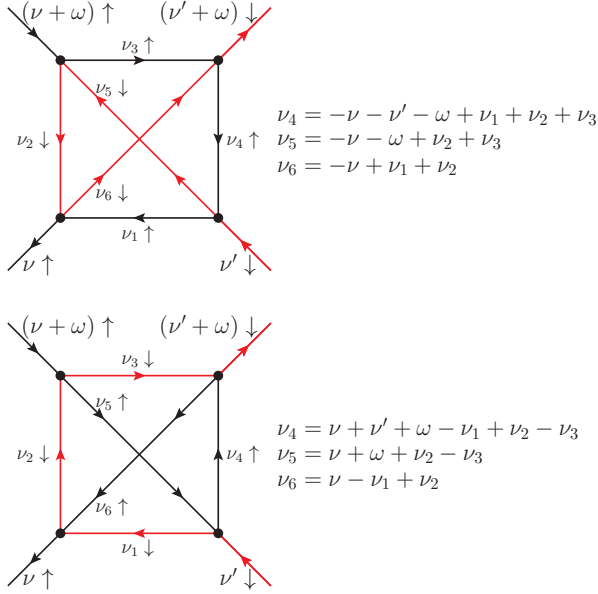}
 \caption{(color online). $U^4$-contributions to the perturbative expansion for the fully irreducible vertex $\Lambda^{\nu\nu'\omega}_{\uparrow\downarrow}$ (``envelope''-diagrams) in particle-hole notation.}
 \label{fig:envelope_up_down}
\end{figure}
The envelope-diagrams for the $\uparrow\uparrow$- and the $\uparrow\downarrow$-case are shown in Figs. (\ref{fig:envelope_up_up}) and (\ref{fig:envelope_up_down}), respectively. \\
Let us just mention one interesting feature for the $\uparrow\uparrow$ diagrams, which is relevant for the particle-hole symmetric case discussed here: At half-filling the contributions of the first and the second (as well as of the third and the fourth) diagrams become exactly the same, i.e., one can take only the first and the third diagrams and assign a factor two to them. This happens because all these diagrams differ only for the direction of the propagators in their closed fermion loops containing four internal electron $\downarrow$-lines. This is again analog to Furry's theorem\cite{mandel-shaw} in quantum electrodynamics, which was already discussed in Sec. \ref{sec:fullvertex} below Eq. (\ref{equ:3order}) for a third-order contribution to the full vertex function $F$. In contrast to the situation explained there, where an odd number of fermion lines in a fermionic loop led to a cancellation of diagrams (see also Fig. \ref{fig:pertthirdorder}), we are dealing here with a closed loop containing an even number of fermions. This leads to a factor 2 for the diagram under consideration rather than to a cancellation.\\
\begin{figure}[t]
 \centering
 \vspace{-0.1cm}
 \includegraphics[width=0.5\textwidth]{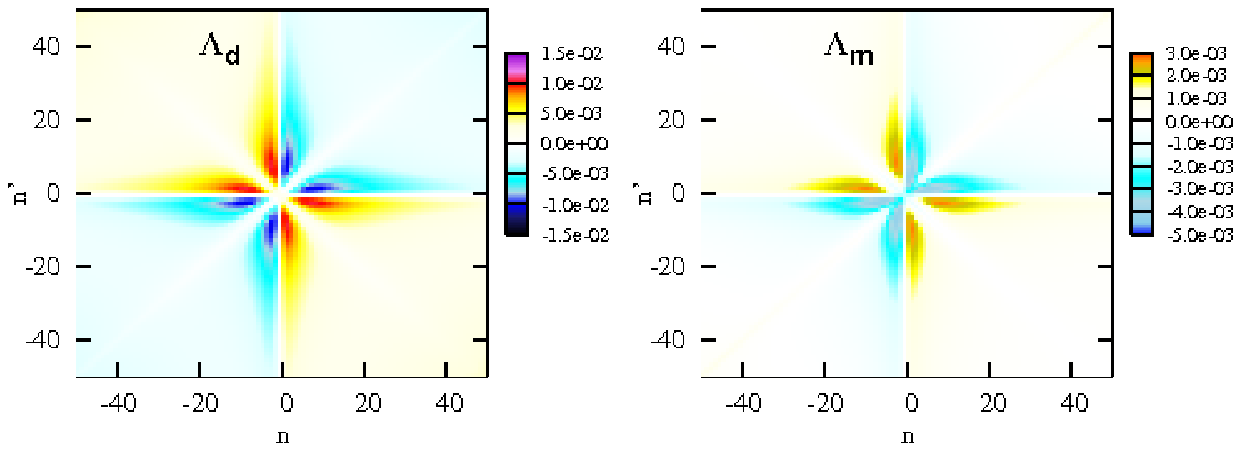}\\[0.3cm]
 \includegraphics[width=0.5\textwidth]{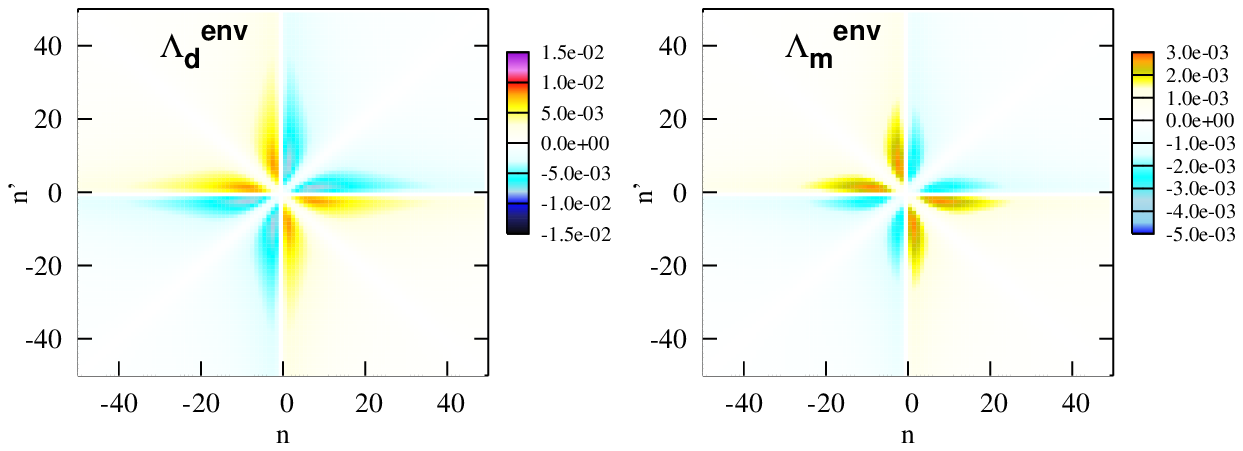}
 \caption{(color online). Upper row: fully irreducible vertex functions vs. the two fermionic frequencies $\nu$ and $\nu'$: density part $\Lambda_d^{\nu\nu'\omega}\!-\!U$ (left) and magnetic part $\Lambda_m^{\nu\nu'\omega}\!+\!U$ (right) for $U\!=\!0.5$ at half-filling ($\beta\!=\!26.0$) for $\omega\!=\!0$. Lower row: 4$^{\text{th}}$-order perturbation theory results (``envelope'' diagrams in Figs. \ref{fig:envelope_up_up} and \ref{fig:envelope_up_down}).}
 \label{fig:lambda_plot}
\end{figure}
In Fig. \ref{fig:lambda_plot}, eventually, our DMFT results for $\Lambda_d^{\nu\nu'\omega}$ and $\Lambda_m^{\nu\nu'\omega}$ are compared with the the $U^4$-contributions from perturbation theory, given by the envelope diagrams in Figs. \ref{fig:envelope_up_up} and \ref{fig:envelope_up_down}. Algebraically, the contribution stemming from such a diagram is
\begin{equation}
 \label{equ:envelopealgebraic}
 \Lambda^{env} = (\pm)  \frac{U^4}{\beta^3}\sum_{\nu_1\nu_2\nu_3}G(\nu_1)G(\nu_2)G(\nu_3)G(\nu_4)G(\nu_5)G(\nu_6),
\end{equation}
where $\nu_4$, $\nu_5$ and $\nu_6$ are functions of $\nu_1$, $\nu_2$ and $\nu_3$ (reported in Figs. \ref{fig:envelope_up_up} and \ref{fig:envelope_up_down}) rather than independent summation variables.
We recall, that the lowest order diagram, which is simply given by the bare Hubbard interaction $U$ (diagram ${\cal P}_3$ in Fig. \ref{fig:vertex_perturb}), is subtracted in both cases, i.e., only deviations from this constant contribution are plotted. 

\begin{figure}[t]
 \centering
 \includegraphics[width=0.5\textwidth]{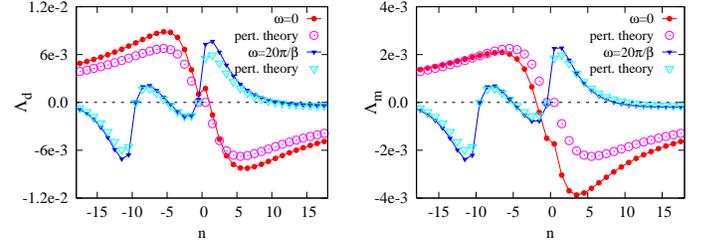}
 \caption{(color online). $\Lambda_d^{\nu\nu'\omega}\!-\!U$ (left) and $\Lambda_m^{\nu\nu'\omega}\!+\!U$ (right) for $U\!=\!0.5$ at half-filling ($\beta\!=\!26.0$) for selected one-dimensional snapshot at fixed $\omega\!=\!0$ or $\omega\!=\!20\frac{\pi}{\beta}$, $\nu'\!=\!\frac{\pi}{\beta}$, as a function of $\nu$.}
 \label{fig:lambda1d_plot}
\end{figure}

From Fig. \ref{fig:lambda_plot} one can see that the structure of the DMFT results for $\Lambda_{d,m}^{\nu\nu'\omega}$ resembles very much that of the envelope diagram. This is expected for a relatively small $U=0.5$ and it is also demonstrated in Fig. \ref{fig:lambda1d_plot}. There a one-dimensional slice of $\Lambda$ is plotted ($\omega$ {\sl and} $\nu'$ are fixed) in comparison with perturbation theory, i.e., the envelope-diagrams.
The deviations from the constant term $\pm U$ for $\Lambda_d^{\nu\nu'\omega}$ and $\Lambda_m^{\nu\nu'\omega}$, respectively, are of the order $U^4\!\sim\!10^{-2}\!-\!10^{-3}$ which is perfectly consistent with our numerical data. Our results demonstrate also that, contrary to the case of $F$ and $\Gamma_r$, due to the complete absence of reducible contributions, the high-frequency asymptotic value of $\Lambda_{d,m}$ is {\sl always} given by the lowest order terms ($\pm U$). As this asymptotic property is intimately connected with the intrinsic fully irreducible nature of $\Lambda$ it holds evidently independently from the value of $U$. A consequence for numerical calculations based on this finding will be discussed in Sec. \ref{practuse}.\\

\subsection{Effects on physical quantities}
\label{subsec:effects}
In this section, we  want to establish a connection between the results for two-particle quantities, we have presented so far, and the more familiar results at the one-particle level, i.e., those for the self-energy of the system. In a second step, the connection with some selected physical quantities, which are typically analyzed in the context of the Hubbard model, will also be illustrated.

\begin{figure}[t]
 \centering
 \includegraphics[width=0.5\textwidth]{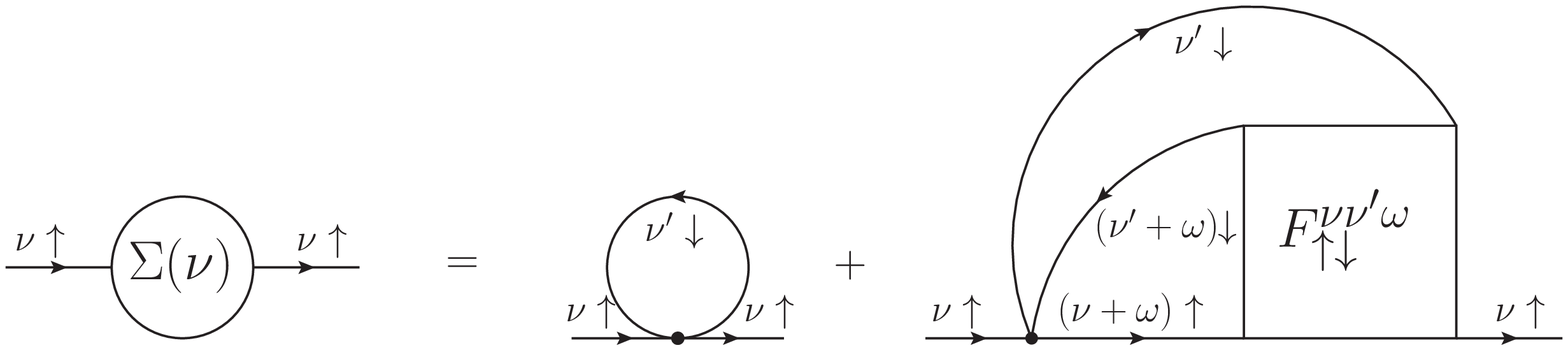}
 \caption{Schwinger-Dyson equation of motion.}
 \label{fig:dysonschwinger}
\end{figure}

As for the self-energy, this goal can be easily achieved by exploiting the Heisenberg (or Schwinger-Dyson) equation of motion
\begin{equation}
 \label{equ:dysonschwinger}
 \Sigma(\nu)=\frac{Un}{2}-\frac{U}{\beta^2}\sum_{\nu'\omega}F_{\uparrow\downarrow}^{\nu\nu'\omega}G(\nu')G(\nu'+\omega)G(\nu+\omega).
\end{equation}
whose diagrammatic representation is given in Fig. \ref{fig:dysonschwinger}.

Inserting the parquet Eq. (\ref{equ:parquet}) in Eq. (\ref{equ:dysonschwinger}), i.e., splitting up $F_{\uparrow\downarrow}^{\nu\nu'\omega}$ into a fully irreducible and the three reducible parts allows us to identify four different contributions to the self-energy stemming from the irreducible and the reducible part of the full vertex function $F$.
\begin{figure}[t]
 \centering
 \includegraphics[width=0.5\textwidth]{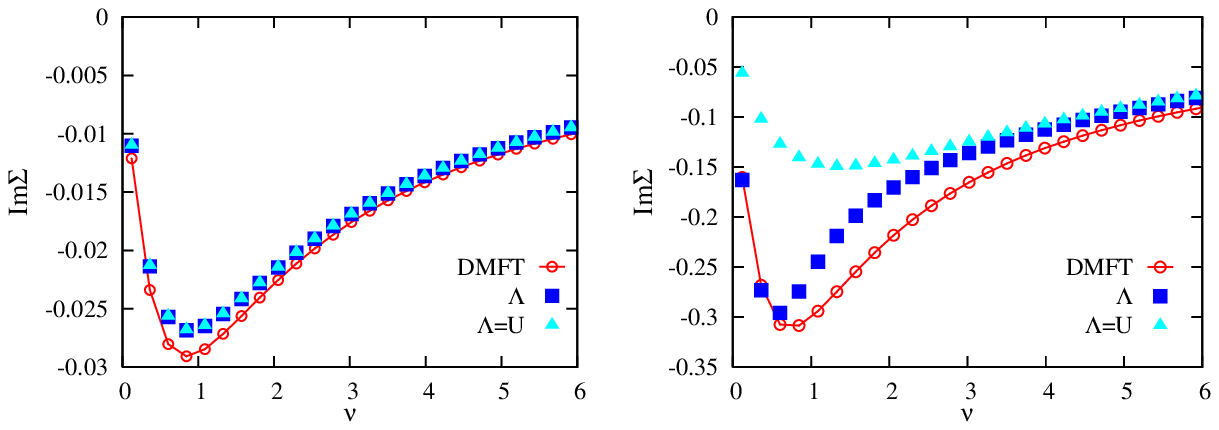}\\[0.3cm]
 \includegraphics[width=0.5\textwidth]{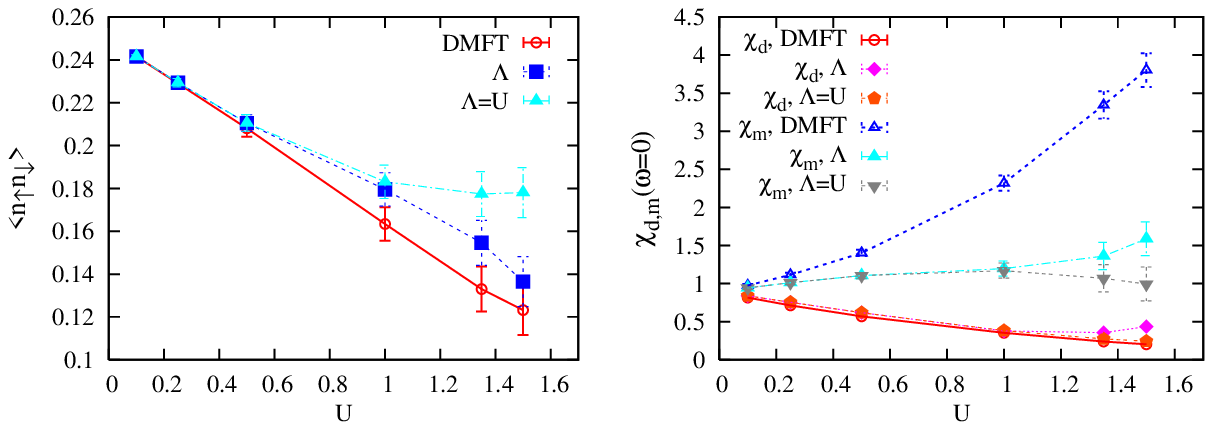}
 \caption{(color online). Upper row: DMFT self-energy (red circles) compared to the contributions stemming from $\Lambda$ (blue squares) and from $U$ (light-blue triangles) only, respectively, for $U\!=\!0.5$ (left) and $U\!=\!1.5$ (right); Lower row: double occupancy (left) and susceptibilities (right). The error bars refer to the finite frequency range adopted for the fermionic frequency summations over $\nu,\nu'$\cite{note_error}.}
 \label{fig:sigmalambda}
\end{figure}
Specifically, in Fig. \ref{fig:sigmalambda} we compare the DMFT self-energy $\Sigma(\nu)$ with its contribution which originate from the fully irreducible vertex $\Lambda$ only and from its lowest order contribution ($U$), respectively. For this purpose we used Eq. (\ref{equ:dysonschwinger}) and replaced $F_{\uparrow\downarrow}^{\nu\nu'\omega}$ by $\Lambda_{\uparrow\downarrow}^{\nu\nu'\omega}$ and $U$, respectively (where the second case simply yields diagrammatic contributions similar to those of the 2$^{\text{nd}}$-order perturbation theory).\\
For the relatively small $U=0.5$ (left panel) there is no visible difference between the self-energies calculated with the full $\Lambda$ and $U$. This is to be expected here, since in the perturbative regime the relative difference between the full $\Lambda$ and $U$ is extremely small, as we have already noticed in Fig. \ref{fig:lambda1d_plot}. \\

On the other hand, for a larger value of the Hubbard interaction ($U=1.5$, upper right panel in Fig. \ref{fig:sigmalambda}) one can see that, while the major part of the self-energy is still coming from the fully irreducible part of $F$, the frequency dependence of $\Lambda$ becomes essential for the calculation of the self-energy. Hence, setting $\Lambda=U$, as it is done, e.g., in the parquet approximation, would yield results quite far from the correct structure of the one-particle local self-energy, which appears to be determined to a major extent by frequency-dependent high-order terms of $\Lambda$. 

Very similar conclusions can be drawn by analyzing the contribution of $\Lambda$ to the value of the double occupancy  $n_\uparrow n_\downarrow = \frac{1}{\beta U} \sum_{\nu} \Sigma(\nu) G(\nu)$ as a function of $U$, which is shown in the lower-left panel of Fig. \ref{fig:sigmalambda}. By comparing the results obtained with the full DMFT self-energy, with those obtained considering $\Lambda$ only, we observe that also in this case the irreducible vertex gives a significant contribution to the well-known reduction of the double-occupancy value w.r.t. its not interacting value of $n_\uparrow n_ \downarrow =n_\uparrow \times n_\downarrow = 0.25$ with increasing $U$. Also in this case, however, for $U > 1$ the results calculated with the approximation $\Lambda = U$ deteriorate very quickly, so that at $U \sim 1.4$ a very incorrect estimate for $n_{\uparrow}n_{\downarrow}$ would be obtained by neglecting the frequency dependence of $\Lambda$.
The situation appears more articulated, however, when analyzing the case of two-particle local response functions, such as the density $\chi_d^{loc}$ and magnetic $\chi_m^{loc}$ local susceptibilities at zero bosonic frequency ($\omega=0$). Such thermodynamic quantities contain a very important piece of information for the physics of the Hubbard model: Approaching the MIT is marked by a constant enhancement of $\chi_m^{loc}$  with  increasing $U$. In fact, a  divergence of  $\chi_m^{loc}$ actually signalizes the transition line, as it corresponds to the formation of a stable local magnetic moment in the Mott phase. At the same time, the reduced mobility of the electrons with increasing value of the local Coulomb interaction $U$ is mirrored in a gradual suppression of the local charge  fluctuations, and, hence, in a monotonous decrease of $\chi_d^{loc}$, with $U$. Such trends are naturally well captured by our DMFT calculations, performed via a summation of both Matsubara fermionic frequencies $\nu, \nu'$ of the generalized susceptibility $\chi^{\nu \nu' \omega=0}$, defined as  in Eq. (\ref{equ:chi_decomposition}).  Following the same procedure described above, we have extracted the contribution to $\chi_d^{loc}$  and  $\chi_m^{loc}$ originated by the fully irreducible vertex $\Lambda$ and its lowest-order term ($U$). While only limited information can be extracted from the  $\chi_d^{loc}$, as it is becoming very small in the non-perturbative region, by analyzing the data for  $\chi_m^{loc}$ some relevant difference with the previous cases can be noted. The contribution to  $\chi_m^{loc}$ stemming from the irreducible vertices $\Lambda$ and reducible diagrams in $F$ are comparable. The latter contributions appear to become the predominant ones in the region $U > 1$ where a stronger enhancement of $\chi_m^{loc}$ is observed. We also note here that the relative error made by replacing $\Lambda=U$ is naturally increasing with $U$ but remains  weaker than in the previous cases.

\subsection{Possible algorithmic developments and improvements}
\label{practuse}

Before turning to the subject of the attractive Hubbard model, we close this section by discussing some possible practical applications of our results in improving or developing algorithms for computing the two-particle properties in DMFT, as well as in other many-body methods.

Let us briefly recall that the computation of two-particle vertex functions in ED, as well as in QMC, over a large number of frequencies poses evidently significant practical problems (from the stability of the results in the high-energy regime, to the storing of increasingly larger arrays). In this respect, numerical schemes capable to limit the frequency region of the actual calculations of the generalized susceptibility are very useful. A relevant example is the algorithm illustrated in Ref. [\onlinecite{kunes2011}], which allows for  a considerable reduction of the size of the frequency region for the numerical calculation of $\chi(\nu,\nu',\omega=0)$, aiming at a much faster computation of the {\boldmath $q$}-dependent susceptibilities in DMFT at {\sl zero frequency}.  
This algorithm is based on the replacement of the calculated high-frequency values of the irreducible vertices $\Gamma_r$, with their corresponding asymptotics. In this respect, our results demonstrate that the high-frequency asymptotics of the fully irreducible vertex $\Lambda$ always reduces to the lowest order perturbative contribution ($U$). This provides (i) an independent confirmation of the assumptions behind the analytical derivation of the high-frequency behavior of $\Gamma_r$ of Ref. [\onlinecite{kunes2011}] and (ii) useful information for its possible extension to the {\sl  finite frequency ($\omega\!\ne\!0$)} case. 
Specifically, the analysis of the frequency structure of the irreducible vertices $\Gamma_r$ at finite bosonic frequency (see, e.g.,  Figs. \ref{fig:gamma_density}, second row, and \ref{fig:gamma_density_U2.0}) suggests the way to generalize the results of  Ref. [\onlinecite{kunes2011}] to an arbitrary (bosonic) frequency case: One can easily note that the simple double diagonal (``$\times$'') structure of $\Gamma_r$ vertices at zero frequency is replaced by a square-like structure. Hence, the frequency region which one has to calculate exactly will be no longer the low frequency one, but one should rather keep all the vertex values for frequencies belonging to or in the proximity of the square-structure, replacing the remaining ones with the corresponding asymptotics.

Finally, let us stress here, that many calculations (not only DMFT based) of the Hubbard model aiming to include dynamical vertex corrections may greatly benefit from (approximated) simplifications or parameterizations of the vertex structures, e.g., considering only the most important features of $F$, $\Gamma$ or $\Lambda$ (see, e.g., the proposed parameterization schemes of the vertex function $F$ for functional renormalization group (fRG) calculations on the AIM\cite{frgAIM,karraschthesis}, the Hubbard model\cite{husemann} or even spin-only models\cite{goettel}). In this light, our results may either guide the construction of such approximations, or --at least-- provide a very precise reference for evaluating the correctness of the approximations already in use\cite{frgAIM,bauer,ayral}.
\begin{figure}[t]
 \centering
 \includegraphics[height=0.4\textwidth]{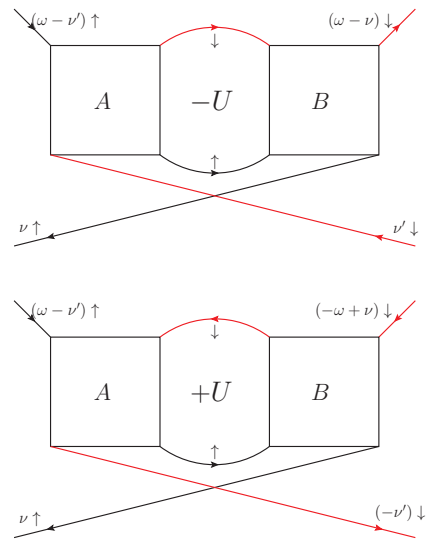}
 \caption{(color online). Upper diagram: reducible in particle-particle channel, Lower diagram: reducible in the transverse particle-hole channel, $-U$ denotes the attractive and $+U$ the repulsive model}
 \label{fig:particle_hole_transform}
\end{figure}

\section{DMFT Results for the attractive model}
\label{Sec:DMFTneg}

We will analyze in this section the two-particle vertex functions for the case of a local {\sl attractive} interaction ($U < 0$), i.e., for the attractive Hubbard model.

\begin{figure*}[t]
 \centering
 \includegraphics[width=1.0\textwidth]{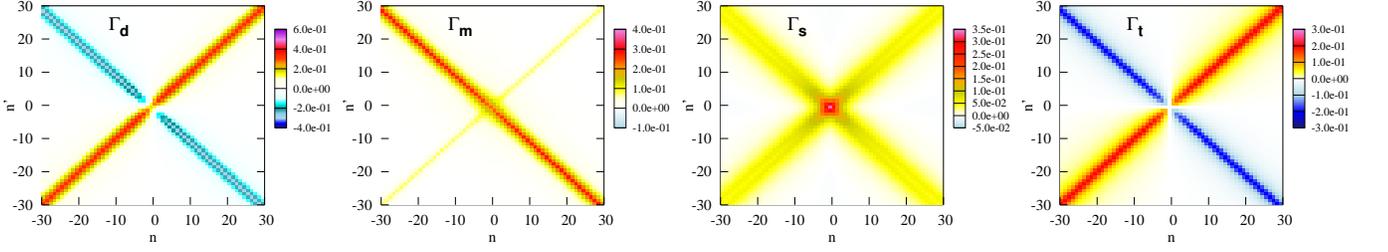}
 \caption{(color online). Irreducible vertices for the attractive model: $\Gamma_d^{\nu\nu'\omega} -U $, $\Gamma_m^{\nu\nu'\omega} + U$, $\Gamma_s^{\nu\nu'\omega} -2U $ and $\Gamma_t^{\nu\nu'\omega}$  for $U=-0.5$ at half-filling ($\beta=26.0$) for $\omega=0$.}
 \label{fig:gammaU-0.5}
\end{figure*}

Obviously an attractive interaction among electrons can represent at most  an {\sl ``effective''} description of more complex microscopic phenomena in condensed matter. However, the physics described by the attractive Hubbard model is far from being merely academic. In fact, the latter represents an ideal playground to investigate the physics of the superconducting and charge-density wave ordered phases in the intermediate-to-strong-coupling regime and, more generally, the important problem of the BCS-Bose Einstein crossover\cite{BCSBE}. 
These issues have raised a remarkable interest also because of their possible relevance to the physics of high-temperature superconductivity. Let us recall, e.g., the analyses of the actual role played by the fluctuations of the phase of the superconducting order parameter in the underdoped cuprates\cite{phasefluct}, and the possibility to derive phase-only effective theories\cite{phaseonly} to capture a part of the underlying physics of these materials, going beyond\cite{phaselen} the standard BCS assumptions. As for DMFT, its application to the attractive case was very useful to identify\cite{PRBattractive} the hallmarks of the BCS-Bose Einstein crossover in several thermodynamic and optical properties of correlated systems\cite{BCSBE2}.
We should also mention here the novel perspectives opened by the ``actual''  experimental realization of quantum models with tunable attractive or repulsive interaction, when confining ultra-cold atoms in the interference pattern of laser sources\cite{optlattices}. This exciting new physics is already stimulating novel DMFT studies\cite{antonio} of the attractive Hubbard model.
\begin{figure}[t]
 \centering
 \includegraphics[width=0.5\textwidth]{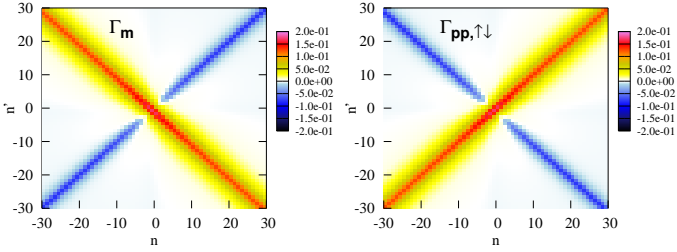}
 \caption{(color online). $\Gamma_m^{\nu\nu'\omega}$ for $U=+0.5$ (left) and $\Gamma_{pp,\uparrow\downarrow}^{\nu\nu'\omega}=\frac{1}{2}(\Gamma_s^{\nu\nu'\omega}+\Gamma_t^{\nu\nu'\omega})$ for $U=-0.5$ (right), both at half-filling ($\beta=26.0$) for $\omega=0$.}
 \label{fig:gammaspinupdown}
\end{figure}

Similarly as for the repulsive case, also for the attractive model, previous DMFT studies focused mainly on the one-particle properties. Even the well-known mapping\cite{particle-hole-transform} between the repulsive and the attractive Hubbard model (see below), to the best of our knowledge, has never been explicitly applied to investigate the connections between the repulsive and attractive models at the  level of the two-particle vertex functions. Hence, our aim is to extend our theoretical analysis of Sec. III also to the $U<0$ case, identifying and interpreting the observed frequency structures in the vertex functions in terms of the mapping onto the corresponding repulsive case and of the discussions of the previous Section.
  
In this respect, let us recall here, that the mapping between the repulsive ($U> 0$)  and the attractive ($U< 0$) Hubbard model is obtained via a partial particle-hole transformation of the half-filled Hubbard model. On the local level, which we are interested in, this mapping is performed by a unitary operator\cite{particle-hole-transform}, which transforms particles with a given spin (e.g., the $\downarrow$-spins) into holes and vice versa, i.e., an operator $\hat{c}^{\dagger}_{\downarrow}$ becomes $\hat{c}_{\downarrow}$ (and vice versa) under this transformation. Particles (holes) with the other spin-direction (i.e., $\uparrow$ in this case) are invariant. The Hamiltonian is also unchanged except for a change in sign of the Hubbard interaction parameter $U$. The same holds also for the purely local model, i.e., the AIM associated with the DMFT solution (see also Appendix \ref{app:particleholesymmetry}).\\ 
One of the consequences of this symmetry is, e.g., that the Green's function and the self-energy for systems which differ only in the sign of the Hubbard interaction $U$ are identical. Furthermore, for the local model, one can show that these one-particle functions are purely imaginary. \\
At the two-particle level the situation is logically more complicated. A detailed calculation (see Appendix \ref{app:particleholesymmetry}) shows that the $\uparrow\downarrow$-susceptibility calculated with $U < 0$ can be mapped onto to the magnetic susceptibility for $U > 0$ (Eq. (\ref{equ:particlehole2ndfrequencyshift})). Physically, this can be understood as follows: Fluctuations of the $x-$ and $y-$ spin component (i.e., the order parameter of the antiferromagnetic phase transition) at positive $U$ are equivalent to fluctuations of the ``cooper-pair density`` $\hat{c}^{\dagger}_{\uparrow}\hat{c}^{\dagger}_{\downarrow}$ (i.e., to the superconducting order parameter) at negative $U$. For the lattice model this means that for an antiferromagnetic instability at a given point $(U>0,T)$ in the phase diagram there exists a superconducting instability in the corresponding attractive model at $(-U,T)$.\\
While a complete algebraic derivation is given in Appendix \ref{app:particleholesymmetry}, we provide here a brief diagrammatic illustration of the relationship between the two channels (i.e., the magnetic and the particle-particle $\uparrow\downarrow$ ones). We start with an arbitrary $\uparrow\downarrow$ diagram which is reducible in the particle-particle channel, i.e., it contributes to $\Phi_{pp,\uparrow\downarrow,(-U)}^{\nu\nu'\omega}$ (see upper panel of Fig. \ref{fig:particle_hole_transform}). The $\downarrow$-Green's functions (plotted in red) are reversed under the particle hole transformation. Naturally the corresponding frequency arguments also change their sign. The diagram we obtain after the particle-hole transformation is a diagram which is reducible in the transverse (or vertical) particle-hole channel. Since this relation holds for all reducible diagrams we can formulate the following equation for the $\Phi$'s
\begin{equation}
 \label{equ:particleholediagrammatic}
 \Phi_{pp,\uparrow\downarrow,(-U)}^{\nu\nu'\omega}=-\Phi_{\overline{ph},\uparrow\downarrow,(+U)}^{\nu(\nu-\omega)(\omega-\nu-\nu')},
\end{equation}
where the minus-sign stems from the exchange of the two fermions. Furthermore SU(2)-symmetry states that $\Phi_{\overline{ph},\uparrow\downarrow}^{\nu\nu'\omega}=-\Phi_m^{\nu(\nu+\omega)(\nu'-\nu)}$ (see Eq. (\ref{equ:chiFrotation})). Using this relation in Eq.\ (\ref{equ:particleholediagrammatic}) yields
\begin{equation}
 \label{equ:particleholediagramsu2}
 \Phi_{pp,\uparrow\downarrow,(-U)}^{\nu\nu'\omega}=\Phi_{m,\uparrow\downarrow,(+U)}^{\nu(\omega-\nu')(-\omega)}.
\end{equation}
Finally the additional transformation $\nu'\rightarrow\omega-\nu'$ (see Appendix \ref{app:particleholesymmetry} and \ref{app:spindiag}) gives the mapping. Evidently this relation holds for the $\Gamma$ vertices as well. 

We can verify our analytical results and gain further insight of the vertex structures for $U\!<\!0$, by looking at the corresponding DMFT data. Our DMFT results for the $\Gamma$'s in the four different channels are shown in Fig. \ref{fig:gammaU-0.5} for the case $U\!=\!-0.5$. Comparing it with Fig. \ref{fig:gamma_density}, i.e., the $\Gamma$'s for the corresponding repulsive case $U\!=\!+0.5$, one observes that the triplet-channel $\Gamma_t^{\nu\nu'\omega}$ is unchanged. This is expected because the triplet channel is identical to the $\uparrow\uparrow$ particle-particle channel $\Gamma_{pp,\uparrow\uparrow}^{\nu\nu'\omega}$, and the $\uparrow$-creation- and annihilation-operators are not affected by the particle-hole transformation.\\
Furthermore, following Eqs. (\ref{equ:particleholediagramsu2}), (\ref{equ:particlehole2ndfrequencyshift}) and (\ref{equ:particleholegamma}), which state that the magnetic-channel is mapped onto the particle-particle $\uparrow\downarrow$-channel (plus an additional frequency shift), we compare these two channels in Fig. \ref{fig:gammaspinupdown}. Performing the additional transformation $\nu'\rightarrow(\omega=0)-\nu'$ in the plot for $\Gamma_{pp,\uparrow\downarrow}^{\nu\nu'\omega}$ (i.e., reflecting the plot along the $x=\nu$-axis) one sees that the two plots are identical. 

Our analytical and numerical results for the case $U\!<\!0$ can be, hence, summarized as follows. The main features of the vertices $\Gamma_r$ (and logically of the corresponding $F$) appear also for $U\!<\!0$ along the diagonals and originated from reducible processes. As a consequence, also in the attractive case, the ''topology`` of the vertex functions remains essentially preserved upon increasing $U$. In contrast to the repulsive case, however, as suggested by Eq. (\ref{equ:particleholediagramsu2}) and Fig. \ref{fig:gammaspinupdown}, the strong enhancement of the main diagonal structure, identified as an hallmark of the MIT, will be now visible in the {\sl secondary} diagonals ($\nu=-\nu'$) in some of the channels (e.g. magnetic, singlet). Physically, this reflects simply that for $U\!<\!0$, $\chi_{pp,\uparrow\downarrow}(0)$ (instead of $\chi_{m}(0)$) is diverging at the MIT, since the ''insulating`` phase is now consisting of a collection of preformed local Cooper pairs.

\section{Conclusions}
\label{Sec:Conclusion}


In this paper, we have presented a focused analysis of the general properties and the frequency structures of the two-particle local vertex functions by means of DMFT, applied to the half-filled Hubbard model on a cubic lattice. Starting point of our study is the DMFT calculation of the full scattering rate amplitude ($F$) and of the vertices irreducible in a given channel ($\Gamma_r$). The DMFT vertex functions have been then interpreted in terms of the corresponding perturbation theory/atomic limit results and of the mapping onto the attractive Hubbard model. These comparisons have allowed for a clear understanding of the main frequency structures of the local vertices in the weak coupling limit. Furthermore, we have also observed that, while the numerical values of the vertices start to deviate significantly from perturbation theory already at $U\!=\!1.0$, the main structures of $F$ and $\Gamma_r$ survive also for larger values of the Hubbard interaction, and -to a good extent- even in the atomic limit. This constitutes an important piece of information for possible improvements of the parametrization of the vertex functions such as those used, e.g., in fRG and of the numerical algorithms to treat more accurately the high-frequency part of the two-particle quantities. 

Finally, for the first time --to the best of our knowledge-- we have also presented  DMFT results for the {\sl fully irreducible} local vertices $\Lambda$.  In this respect, it has been shown how the frequency dependent part of the fully irreducible vertex function, crucially affects the one-particle self-energy at intermediate values of the Hubbard interaction. This would imply that approximations at the level of the fully irreducible vertex, which rely on the parquet formalism, should necessarily include its frequency dependence, as, e.g., in the D$\Gamma$A, in order to go beyond the weakly-correlated regime. Whether, and to what extent, it is possible to neglect the spatial (or momentum) dependence of the fully irreducible vertices  of the Hubbard model remains to be investigated, though a specific set of data\cite{maier} obtained in dynamical cluster approximation for a two-dimensional Hubbard model appears rather promising in this respect.

\vskip 6mm

\textit{Acknowledgments.} We are indebted to G. Sangiovanni for his constant support during the preparation of this paper, and for illuminating advices. We also thank K. Held, A. Katanin, T. Sch\"afer, M. Capone, E. Gull, C. Castellani, S. Ciuchi, and O. Gunnarsson for discussions and exchanging of ideas. We acknowledge financial support from Austria Science Fund (FWF) through Research Unit FOR: 1346 and Project No. I597-N16 (AT) and through the Austria-Russia joint Project No. I610-N16 (AV, GR). The numerical calculations have been performed on the Vienna Scientific Cluster (VSC).

\clearpage

\appendix

\section{(Imaginary) Time translational invariance - Boundary conditions}
\label{app:boundaryconditions}

We summarize here the so called Kubo-Martin-Schwinger (KMS) boundary conditions for the $n$-particle Green's function, which follow from the time-translational invariance and from the cyclic property of the trace (see Ref. [\onlinecite{dolfen}] and Ref. [\onlinecite{martin_schwinger}]).\\
\begin{figure}[t]
 \centering
 \includegraphics[width=7cm]{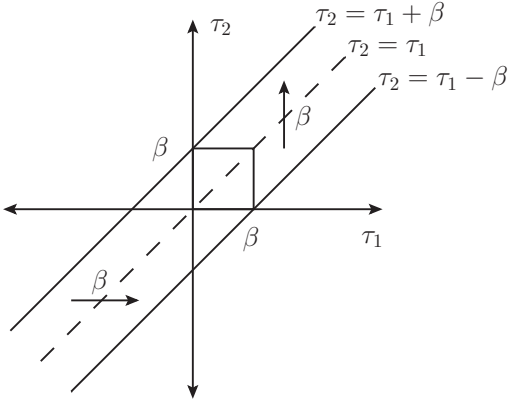}
 \caption{Domain of definition for $G(\tau_1,\tau_2)$}
 \label{fig:timeorder}
\end{figure}
We start from the $n$-particle Green's function defined in Eq. (\ref{equ:defngreenfunction}) omitting the spin-indexes for this section, since the subsequent considerations are independent of the spin. Note also that the results discussed here are valid for models with arbitrary degrees of freedom (like spin, lattice site, etc.) and not only for the local AIM Hamiltonian: The only requirement is time-translational invariance of the system, i.e., that the Hamiltonian $\hat{\mathcal{H}}$ is independent of $\tau$. \\
Assuming that $\tau_1$ is the largest and $\tau_{2n}$ is the smallest time argument of the $n$-particle Green's function $G_n(\tau_1,\ldots,\tau_{2n})$ one gets the following condition\cite{dolfen} for the $2n$ time-variables:
\begin{equation}
 \label{equ:ngreenboundary}
 \tau_{2n}+\beta>\tau_1>\ldots>\tau_i>\ldots>\tau_{2n},
\end{equation}
i.e., all time-arguments have to be located within an interval of the length $\beta$. Otherwise the term $e^{-(\beta+\tau_{2n}-\tau_1)\hat{\mathcal{H}}}$ in the definition of $G_n$ would lead to exponentially increasing contributions with growing eigenvalues $E_n$ of the system, and the trace occurring in Eq. (\ref{equ:defngreenfunction}) diverges. On the contrary, if condition (\ref{equ:ngreenboundary}) is fulfilled, the above-mentioned exponential factor suppresses the contribution to the trace for large eigenvalues $E_n$ of $\hat{\mathcal{H}}$ and hence, the trace converges and the $n$-particle Green's function is well defined. As an example, the domain of definition for the one-particle Green's function $G_1(\tau_1,\tau_2)=G(\tau_1,\tau_2)$ is shown in Fig. \ref{fig:timeorder} (region between the two solid diagonal lines). \\  
Due to the time-invariance of the Hamiltonian the $n$-particle Green's function $G_n$ does not depend on all $2n$ times explicitly but rather on time-differences, e.g., of the form $\tau_i-\tau_{2n}$, yielding 
\begin{equation}
 \label{equ:ngreentimetranslation}
 G_n(\tau_1,\ldots,\tau_{2n})=G_n(\tau_1-\tau_{2n},\ldots,\tau_{2n-1}-\tau_{2n},0).
\end{equation}
As a result, the one-particle Green's function is constant along diagonals of the form $\tau_2=\tau_1+\alpha, \alpha\in[-\beta,\beta]$ in Fig. \ref{fig:timeorder}. \\
Furthermore, the cyclic property of the trace leads to anti-periodicity of the $n$-particle Green's function which reads
\begin{equation}
 \label{equ:ngreenantiperiodic}
 G_n(\tau_1,\ldots,\tau_{2n})=-G_n(\tau_1-\beta,\ldots,\tau_{2n}),
\end{equation}
if we assume that $\tau_1>\ldots>\tau_{2n}>\tau_1-\beta$. \\
All imaginary time-variables can be restricted to the interval $[0,\beta]$, since the value of $G_n$ for all other combinations of time-arguments (that are allowed according to Eq. \ref{equ:ngreenboundary}) can be constructed by means of Eqs. (\ref{equ:ngreentimetranslation}) and (\ref{equ:ngreenantiperiodic}).\\ 
Considering the anti-periodicity condition (\ref{equ:ngreenantiperiodic}) one can express the $n$-particle Green's function $G_n$ as a Fourier-expansion
\begin{equation}
 \label{equ:nparticleinversefourier}
 \begin{split}
 &G_n(\tau_1,\ldots,\tau_{2n})=\frac{1}{\beta^n}\sum_{\{\nu_i\}}e^{-i(\nu_1\tau_1+\ldots-\nu_{2n}\tau_{2n})}\tilde{G}_n(\nu_1,\ldots,\nu_{2n}),\\[0.3cm]&
 \begin{split}
 \tilde{G}(\nu_1,\ldots,\nu_{2n})=\int_0^{\beta}d\tau_1&\ldots\int_0^{\beta}d\tau_{2n}\;e^{i(\nu_1\tau_1+\ldots-\nu_{2n}\tau_{2n})}\times \\&\times G_n(\tau_1,\ldots,\tau_{2n}),
 \end{split}
 \end{split}
\end{equation}
where $\nu_i\!=\!\frac{\pi}{\beta}(2n_i+1)$ are fermionic Matsubara-frequencies. 
The calculation of the Fourier-coefficients can be simplified by means of the following considerations. One uses Eq. (\ref{equ:ngreentimetranslation}) and performs the substitutions $\tau_i=\tau_i'-\tau_{2n}, i=1,\ldots,2n-1$.
Next, one can shift the integration intervals of $\tau_1',\ldots,\tau_{2n-1}'$ from $[-\tau_{2n},\beta-\tau_{2n}]$ to $[0,\beta]$ due to the anti-periodicity condition (\ref{equ:ngreenantiperiodic}). Hence, the $\tau_{2n}$ integration in Eq. (\ref{equ:nparticleinversefourier}) can be performed analytically and leads (beside a factor $\beta$) to energy conservation $\nu_1-\nu_2+\ldots+\nu_{2n-1}-\nu_{2n}=0$.
Therefore, it is sufficient to consider a $(2n-1)$-frequency object
\begin{equation}
 \label{equ:nminus1greenfunction}
 \begin{split}
 \tilde{G}_{\bar{n}}(\nu_1,\ldots,&\nu_{2n-1})=\int_0^{\beta}d\tau_1\ldots\int_0^{\beta}d\tau_{2n-1}\times \\ \times &e^{i(\nu_1\tau_1+\ldots-\nu_{2n-1}\tau_{2n-1})}G_n(\tau_1,\ldots,\tau_{2n-1},0),
 \end{split}
\end{equation}
related to the full $2n$-frequency Green's function via
\begin{equation}
 \label{equ:ngreenminus1}
 \tilde{G}_n(\nu_1,\ldots,\nu_{2n})=\beta\delta_{(\nu_1+\ldots+\nu_{2n-1})\nu_{2n}}\tilde{G}_{\bar{n}}(\nu_1,\ldots,\nu_{2n-1}). 
\end{equation}

\section{Spin-diagonalization}
\label{app:spindiag}
In this appendix, we summarize the spin-dependence of the three (ir)reducible channels (i.e., $ph,\overline{ph}$ and $pp$) and give a derivation of the corresponding Bethe-Salpeter-equations for the SU(2)-symmetric case. The formalism presented here is similar to that of Ref. [\onlinecite{bickers}]. 

\subsubsection{The longitudinal (horizontal) channel $\Gamma_{ph}$}
\begin{figure}[t]
 \centering
 \includegraphics[width=0.48\textwidth]{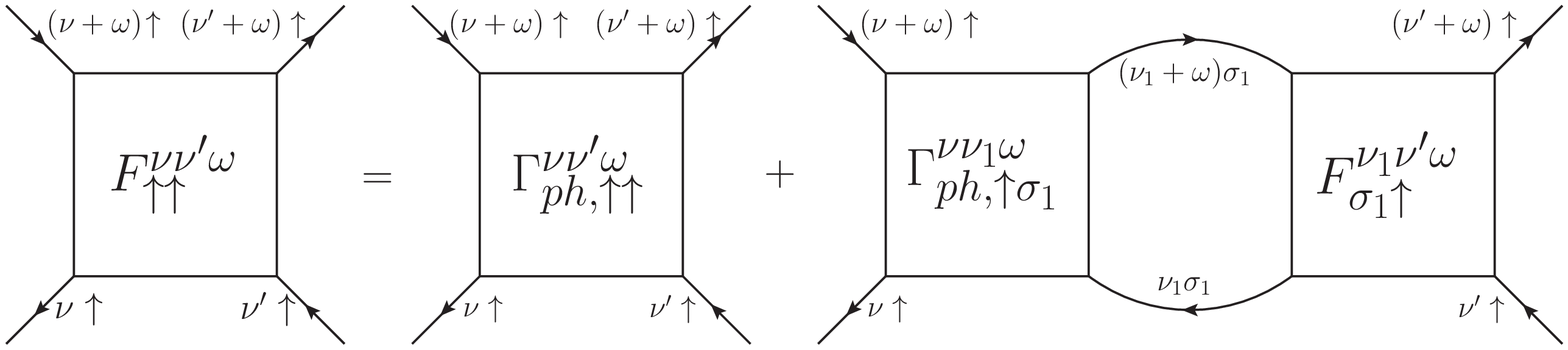}\\[0.3cm]
 \includegraphics[width=0.48\textwidth]{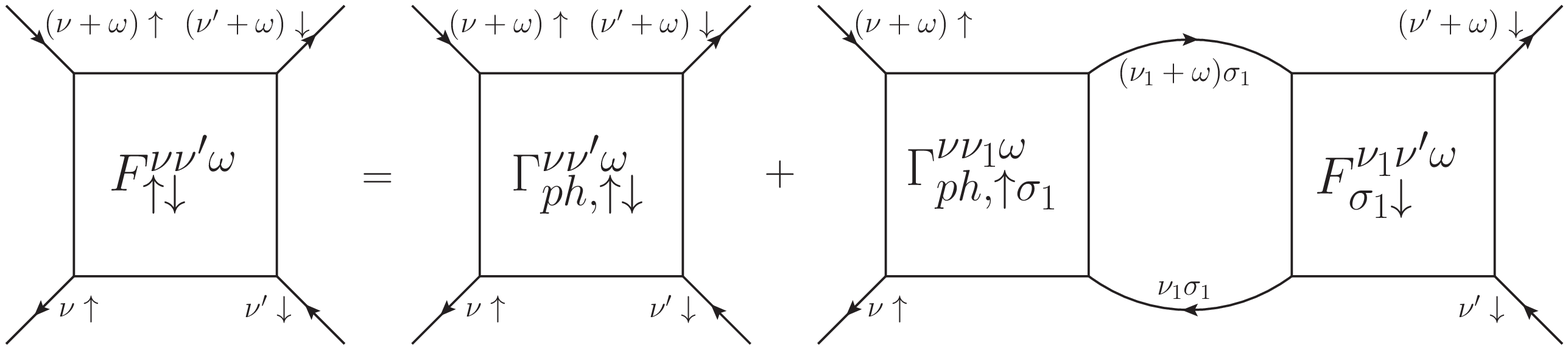}\\[0.3cm]
 \includegraphics[width=0.48\textwidth]{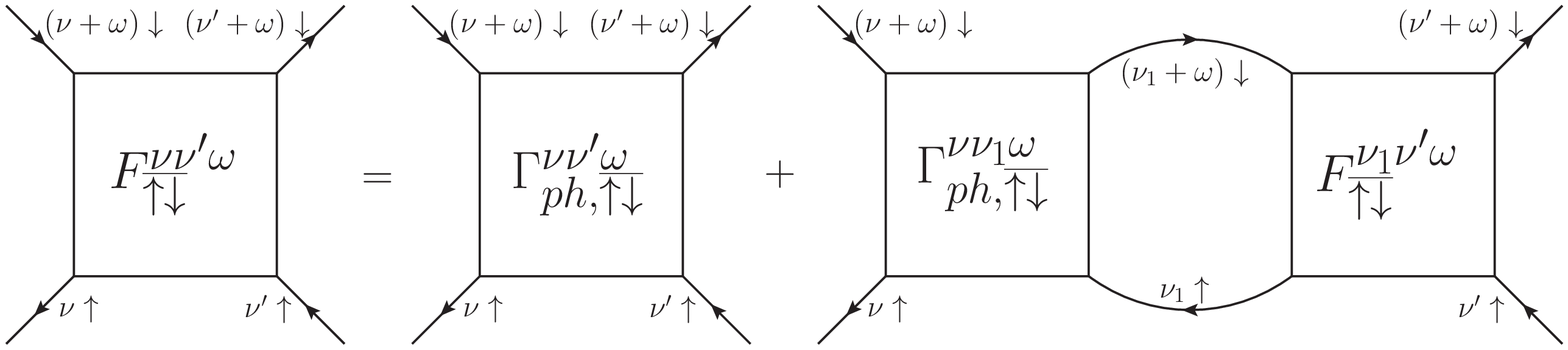}
 \caption{Bethe-Salpeter equations in the longitudinal channel.}
 \label{fig:gamma_ph}
\end{figure}

We start with the Bethe-Salpeter equations for the three independent spin-combinations $\Gamma_{ph,\uparrow\uparrow}$, $\Gamma_{ph,\uparrow\downarrow}$ and $\Gamma_{ph,\overline{\uparrow\downarrow}}$.  Diagrammatically they take the form shown in Fig. \ref{fig:gamma_ph}. 
Algebraically they read as
{\small
\begin{subequations}
\label{equ:bethesalpeterlong}
\begin{equation}
\label{equ:bethesalpeterlongupup}
F_{\uparrow\uparrow}^{\nu\nu'\omega}=\Gamma_{ph,\uparrow\uparrow}^{\nu\nu'\omega}+\frac{1}{\beta}\sum_{\nu_1\sigma_1}\Gamma_{ph,\uparrow\sigma_1}^{\nu\nu_1\omega}G(\nu_1)G(\nu_1+\omega)F_{\sigma_1\uparrow}^{\nu_1\nu'\omega}
\end{equation}
\begin{equation}
\label{equ:bethesalpeterlongupdown}
F_{\uparrow\downarrow}^{\nu\nu'\omega}=\Gamma_{ph,\uparrow\downarrow}^{\nu\nu'\omega}+\frac{1}{\beta}\sum_{\nu_1\sigma_1}\Gamma_{ph,\uparrow\sigma_1}^{\nu\nu_1\omega}G(\nu_1)G(\nu_1+\omega)F_{\sigma_1\downarrow}^{\nu_1\nu'\omega}
\end{equation}
\begin{equation}
\label{equ:bethesalpeterlongupdownbar}
F_{\overline{\uparrow\downarrow}}^{\nu\nu'\omega}=\Gamma_{ph,\overline{\uparrow\downarrow}}^{\nu\nu'\omega}+\frac{1}{\beta}\sum_{\nu_1}\Gamma_{ph,\overline{\uparrow\downarrow}}^{\nu\nu_1\omega}G(\nu_1)G(\nu_1+\omega)F_{\overline{\uparrow\downarrow}}^{\nu_1\nu'\omega}.
\end{equation}
\end{subequations}
}
It is easy to verify the plus-sign in front of the second summand on the right hand side of these equations by comparison with 2$^{\text{nd}}$-order perturbation theory: The corresponding perturbative contribution shown in Fig. \ref{fig:vertex_perturb}, upper left diagram, exhibits a plus-sign (see also Eq. (\ref{equ:secondorderupup1})).\\
One can see that Eqs. (\ref{equ:bethesalpeterlongupup}) and ({\ref{equ:bethesalpeterlongupdown}) are coupled, while Eq. (\ref{equ:bethesalpeterlongupdownbar}) contains only $\Gamma_{ph,\overline{\uparrow\downarrow}}^{\nu\nu'\omega}$. Anyway, we will postpone the calculation of this vertex function to the transversal particle-hole case since  
$\Gamma_{ph,\overline{\uparrow\downarrow}}$ is related to $\Gamma_{\overline{ph},\uparrow\downarrow}$ by the crossing relation Eq. (\ref{equ:crossinggammageneralph}) which reads as
\begin{equation}
\label{equ:bethesalpeterlongcrossing}
\Gamma_{ph,\overline{\uparrow\downarrow}}^{\nu\nu'\omega}=-\Gamma_{\overline{ph},\uparrow\downarrow}^{\nu(\nu+\omega)(\nu'-\nu)}
\end{equation}
for this specific case. \\ 
By hands of SU(2)-symmetry the two other equations can be decoupled analytically considering the sum and the difference of Eqs. (\ref{equ:bethesalpeterlongupup}) and (\ref{equ:bethesalpeterlongupdown}), respectively:
\begin{subequations}
\label{equ:defdensitymagnetic}
\begin{equation}
\label{equ_defdensity}
F_{d(ensity)}^{\nu\nu'\omega}:=F_{\uparrow\uparrow}^{\nu\nu'\omega}+F_{\uparrow\downarrow}^{\nu\nu'\omega},
\end{equation}
\begin{equation}
\label{equ:defmagnetic}
F_{m(agnetic)}^{\nu\nu'\omega}:=F_{\uparrow\uparrow}^{\nu\nu'\omega}-F_{\uparrow\downarrow}^{\nu\nu'\omega},
\end{equation}
\end{subequations}
which correspond to Eqs. (\ref{equ:channeldefdensity}) and (\ref{equ:channeldefmagnetic}) for the $\Gamma$'s. The two decoupled equations for the density and magnetic channel are
\begin{subequations}
\label{equ:bethesalpeterdensitymagnetic}
\begin{equation}
\label{equ:bethesalpeterdensity}
F_{d}^{\nu\nu'\omega}=\Gamma_{d}^{\nu\nu'\omega}+\frac{1}{\beta}\sum_{\nu_1}\Gamma_{d}^{\nu\nu_1\omega}G(\nu_1)G(\nu_1+\omega)F_{d}^{\nu_1\nu'\omega},
\end{equation}
\begin{equation}
\label{equ:bethesalpetermagnetic}
F_{m}^{\nu\nu'\omega}=\Gamma_{m}^{\nu\nu'\omega}+\frac{1}{\beta}\sum_{\nu_1}\Gamma_{m}^{\nu\nu_1\omega}G(\nu_1)G(\nu_1+\omega)F_{m}^{\nu_1\nu'\omega}.
\end{equation}
\end{subequations}
These equations can be solved for the $\Gamma$'s by an inversion of the matrix $(\mathds{1}+\frac{1}{\beta}GGF)^{\nu\nu'\omega}$ in the $\nu\nu'$-space, i.e.,
\begin{equation}
 \label{equ:invertchidivchi0}
 \Gamma_{d,m}^{\nu\nu'\omega}=\sum_{\nu_1}F_{d,m}^{\nu\nu_1\omega}\bigr[(\mathds{1}+\frac{1}{\beta}GGF_{d,m})^{-1}\bigl]^{\nu_1\nu'\omega}.
\end{equation}
Considering the definition of $\chi$ in Eq. (\ref{equ:chi_decomposition}) one can write the quantity which is inverted as $\chi_{d,m}^{\nu\nu'\omega}/\chi_0^{\nu\nu'\omega}$. \\
For the sake of completeness, we want to rewrite this equation into the form which was used for extracting the $\Gamma$'s shown in this paper. Defining $\chi_d^{\nu\nu'\omega}$ and $\chi_{m}^{\nu\nu'\omega}$ and combining Eq. (\ref{equ:chi_decomposition}) with Eqs. (\ref{equ:bethesalpeterdensitymagnetic}) one finds the corresponding Bethe-Salpeter-equations for the $\chi$'s:
\begin{equation}
\label{equ:bethesalpeterdensitymagneticchi}
\chi_{d,m}^{\nu\nu'\omega}=\chi_0^{\nu\nu'\omega}-\frac{1}{\beta^2}\sum_{\nu_1\nu_2}\chi_0^{\nu\nu_1\omega}\Gamma_{d,m}^{\nu_1\nu_2\omega}\chi_{d,m}^{\nu_2\nu'\omega},
\end{equation}
Solving these equations for $\Gamma_d^{\nu\nu'\omega}$ and $\Gamma_m^{\nu\nu'\omega}$ yields
\begin{equation}
 \label{equ:invertchi}
 \Gamma_{d,m}^{\nu\nu'\omega}=\beta^2(\chi_{d,m}^{-1}-\chi_0^{-1})^{\nu\nu'\omega}.
\end{equation}

\subsubsection{The transverse (vertical) channel}  
The Bethe-Salpeter equations for the three different spin-combinations shown diagrammatically in Fig. \ref{fig:gamma_phbar} read as
\begin{figure}[t]
 \centering
 \includegraphics[width=0.48\textwidth]{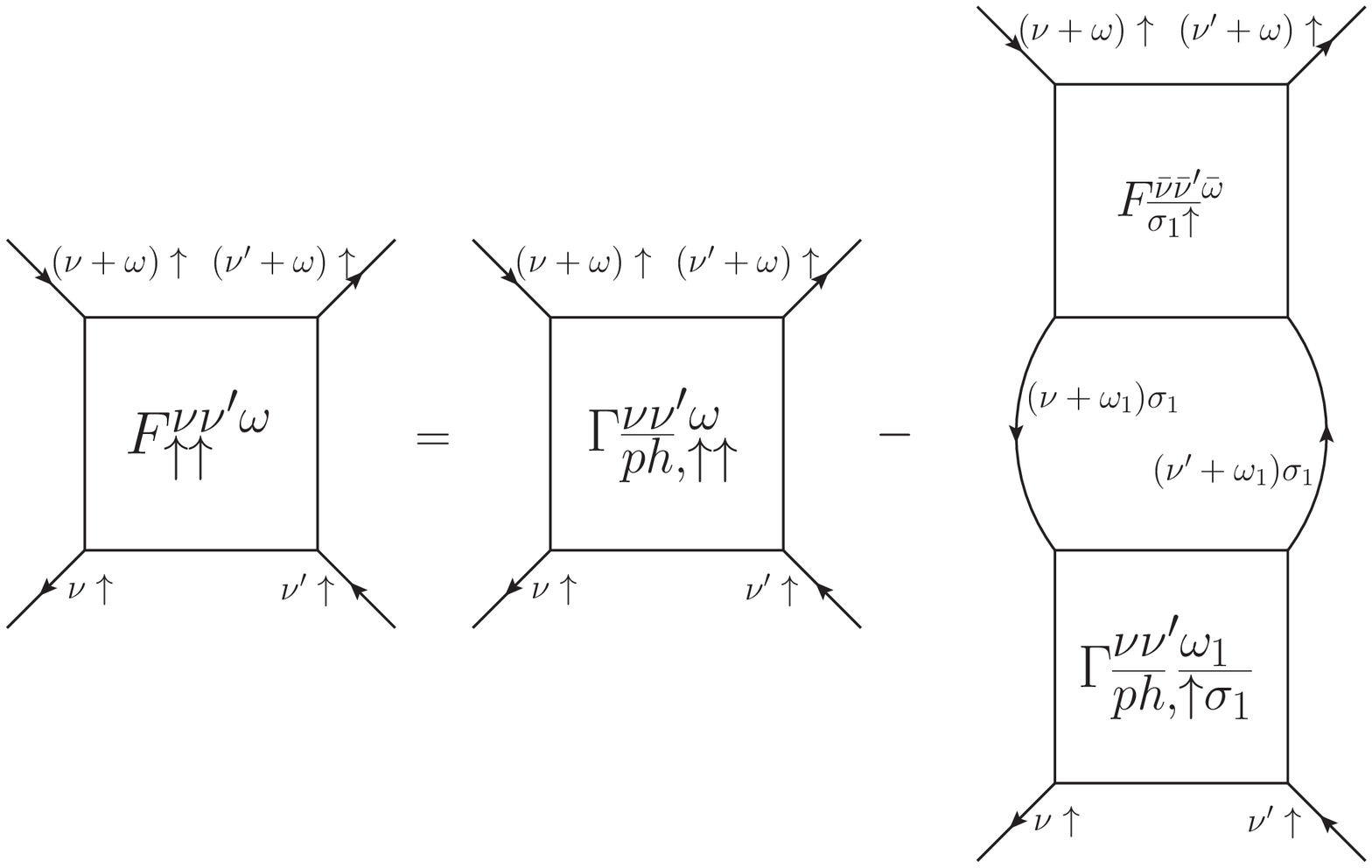}\\[0.3cm]
 \includegraphics[width=0.48\textwidth]{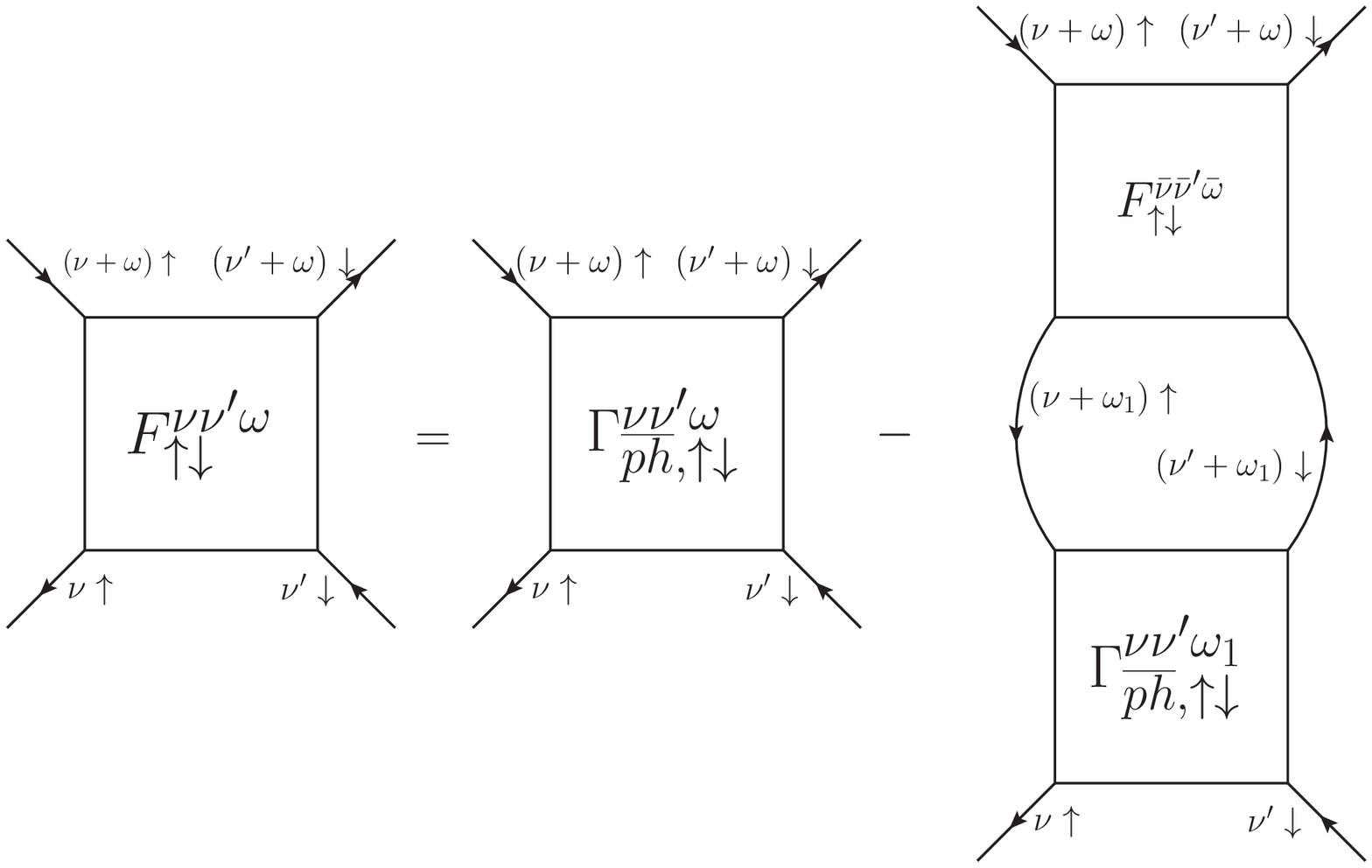}\\[0.3cm]
 \includegraphics[width=0.48\textwidth]{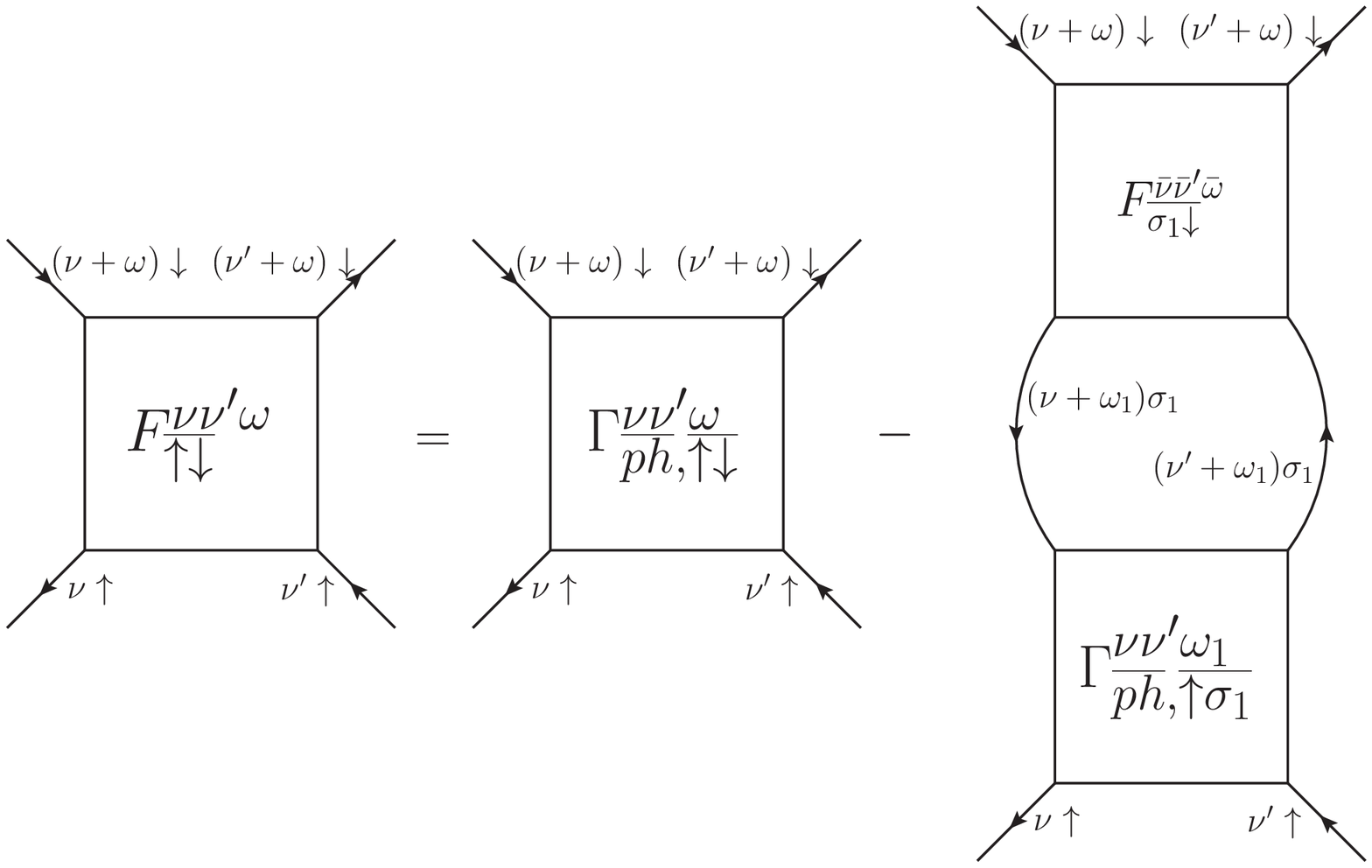}
 \caption{Bethe-Salpeter equations in the transverse channel with $\bar{\nu}=\nu+\omega_1$, $\bar{\nu}'=\nu'+\omega_1$, $\bar{\omega}=\omega-\omega_1$.}
 \label{fig:gamma_phbar}
\end{figure}
{\small
\begin{subequations}
\label{equ:bethesalpetertrans}
\begin{equation}
\label{equ:bethesalpetertransupup}
\begin{split}
F_{\uparrow\uparrow}^{\nu\nu'\omega}=\Gamma_{\overline{ph},\uparrow\uparrow}^{\nu\nu'\omega}-\frac{1}{\beta}\sum_{\omega_1\sigma_1}&\Gamma_{\overline{ph},\overline{\uparrow\sigma_1}}^{\nu\nu'\omega_1}G(\nu+\omega_1)G(\nu'+\omega_1)\times\\&\times F_{\overline{\sigma_1\uparrow}}^{(\nu+\omega_1)(\nu'+\omega_1)(\omega-\omega_1)}
\end{split}
\end{equation}
\begin{equation}
\label{equ:bethesalpetertransupdown}
\begin{split}
F_{\uparrow\downarrow}^{\nu\nu'\omega}=\Gamma_{\overline{ph},\uparrow\downarrow}^{\nu\nu'\omega}-\frac{1}{\beta}\sum_{\omega_1}&\Gamma_{\overline{ph},\uparrow\downarrow}^{\nu\nu'\omega_1}G(\nu+\omega_1)G(\nu'+\omega_1)\times\\&\times F_{\uparrow\downarrow}^{(\nu+\omega_1)(\nu'+\omega_1)(\omega-\omega_1)}
\end{split}
\end{equation}
\begin{equation}
\label{equ:bethesalpetertransupdownbar}
\begin{split}
F_{\overline{\uparrow\downarrow}}^{\nu\nu'\omega}=\Gamma_{\overline{ph},\overline{\uparrow\downarrow}}^{\nu\nu'\omega}-\frac{1}{\beta}\sum_{\omega_1\sigma_1}&\Gamma_{\overline{ph},\overline{\uparrow\sigma_1}}^{\nu\nu'\omega_1}G(\nu+\omega_1)G(\nu'+\omega_1)\times\\&\times F_{\overline{\sigma_1\downarrow}}^{(\nu+\omega_1)(\nu'+\omega_1)(\omega-\omega_1)}.
\end{split}
\end{equation}
\end{subequations}
}
As in the longitudinal channel the minus-sign in front of the reducible part of these equations can be inferred from comparison with 2$^{\text{nd}}$ order perturbation-theory (vertical diagrams in Fig. \ref{fig:vertex_perturb} as well as Eqs. (\ref{equ:secondorderupup2}) and (\ref{equ:secondorderupdown1})). \\
One can see that Eqs. (\ref{equ:bethesalpetertransupup}) and (\ref{equ:bethesalpetertransupdownbar}) are not independent, i.e., in the transverse channel the $\uparrow\uparrow$- and the $\overline{\uparrow\downarrow}$-vertex are coupled in the same way as it was the case for $\Gamma_{ph,\uparrow\uparrow}$ and $\Gamma_{ph,\uparrow\downarrow}$ in the longitudinal channel (see Eqs. (\ref{equ:bethesalpeterlongupup}) and (\ref{equ:bethesalpeterlongupdown})). This is not surprising since these functions are connected via the crossing relations
\begin{subequations}
\label{equ:crossingphphbar}
\begin{equation}
\label{equ:crossingphphbarupup}
\Gamma_{\overline{ph},\uparrow\uparrow}^{\nu\nu'\omega}=-\Gamma_{ph,\uparrow\uparrow}^{\nu(\nu+\omega)(\nu'-\nu)}
\end{equation}
\begin{equation}
\label{equ:crossingphphbarapdown}
\Gamma_{\overline{ph},\overline{\uparrow\downarrow}}^{\nu\nu'\omega}=-\Gamma_{ph,\uparrow\downarrow}^{\nu(\nu+\omega)(\nu'-\nu)}.
\end{equation}
\end{subequations}
Therefore the only ``new'' (independent) quantity in the transverse (vertical) channel is $\Gamma_{\overline{ph},\uparrow\downarrow}$ (Eq. (\ref{equ:bethesalpetertransupdown})) which corresponds to $\Gamma_{ph,\overline{\uparrow\downarrow}}$  via the crossing relation Eq. (\ref{equ:bethesalpeterlongcrossing}). Hence, in the following we will discuss only Eq. (\ref{equ:bethesalpetertransupdown}) in more detail: First of all we can perform the transformation $\omega_1=\nu_1-\nu$ of the summed index yielding
\begin{equation}
\label{equ:bethesalpetertransupdownfermionicsum}
\begin{split}
F_{\uparrow\downarrow}^{\nu\nu'\omega}=\Gamma_{\overline{ph},\uparrow\downarrow}^{\nu\nu'\omega}-\frac{1}{\beta}\sum_{\nu_1}&\Gamma_{\overline{ph},\uparrow\downarrow}^{\nu\nu'(\nu_1-\nu)}G(\nu_1)G(\nu_1+\nu'-\nu)\times\\&\times F_{\uparrow\downarrow}^{\nu_1(\nu_1+\nu'-\nu)(\omega-\nu_1+\nu)}.
\end{split}
\end{equation}  
In the next step we introduce the transformation $\nu\!\rightarrow\!\nu$, $\nu'\!\rightarrow\!\nu\!+\!\omega$ and $\omega\!\rightarrow\!\nu'\!-\!\nu$ and make use of the SU(2)-symmetry relation (\ref{equ:chiFrotation}) $F_{\uparrow\downarrow}^{\nu(\nu+\omega)(\nu'-\nu)}\!=\!-(F_{\uparrow\uparrow}^{\nu\nu'\omega}\!-\!F_{\uparrow\downarrow}^{\nu\nu'\omega})\!=\!-F_m^{\nu\nu'\omega}$. Furthermore we define $\tilde{\Gamma}^{\nu\nu'\omega}\!=\!-\Gamma_{\overline{ph},\uparrow\downarrow}^{\nu(\nu+\omega)(\nu'-\nu)}$. Hence Eq. (\ref{equ:bethesalpetertransupdownfermionicsum}) reads
\begin{equation}
\label{equ:bethesalpetertransupdownfrequencytrans}
F_{m}^{\nu\nu'\omega}=\tilde{\Gamma}^{\nu\nu'\omega}+\frac{1}{\beta}\sum_{\nu_1}\tilde{\Gamma}^{\nu\nu_1\omega}G(\nu_1)G(\nu_1+\omega)F_m^{\nu_1\nu'\omega}.
\end{equation}
This is exactly the same equation we derived for $\Gamma_m^{\nu\nu'\omega}$ (Eq. \ref{equ:bethesalpetermagnetic}) which means that 
\begin{equation}
\label{equ:bethesalpetertransmagnetic}
\tilde{\Gamma}^{\nu\nu'\omega}=\Gamma_m^{\nu\nu'\omega}.
\end{equation}
Together with the definition of $\tilde{\Gamma}$ this yields
\begin{equation}
\label{equ:bethesalpetertransmagnetic1}
\Gamma_{\overline{ph},\uparrow\downarrow}^{\nu\nu'\omega}=-\Gamma_m^{\nu(\nu+\omega)(\nu'-\nu)}.
\end{equation}
Hence, the transverse channel does not provide any ``new'' information (in the SU(2)-symmetric case), and $\Gamma_m$ and $\Gamma_d$ are, in fact, the only two independent functions for the two irreducible particle-hole channels. 

\subsubsection{The particle-particle channel}
The particle-particle channel is completely independent of the two particle-hole channels and fulfills a crossing relation itself (Eq. (\ref{equ:crossinggammageneralpp})). The Bethe-Salpeter equations for the three possible spin-combinations shown diagrammatically in Fig. \ref{fig:gamma_pp} read as 
\begin{figure}[t]
\centering
\includegraphics[width=0.48\textwidth]{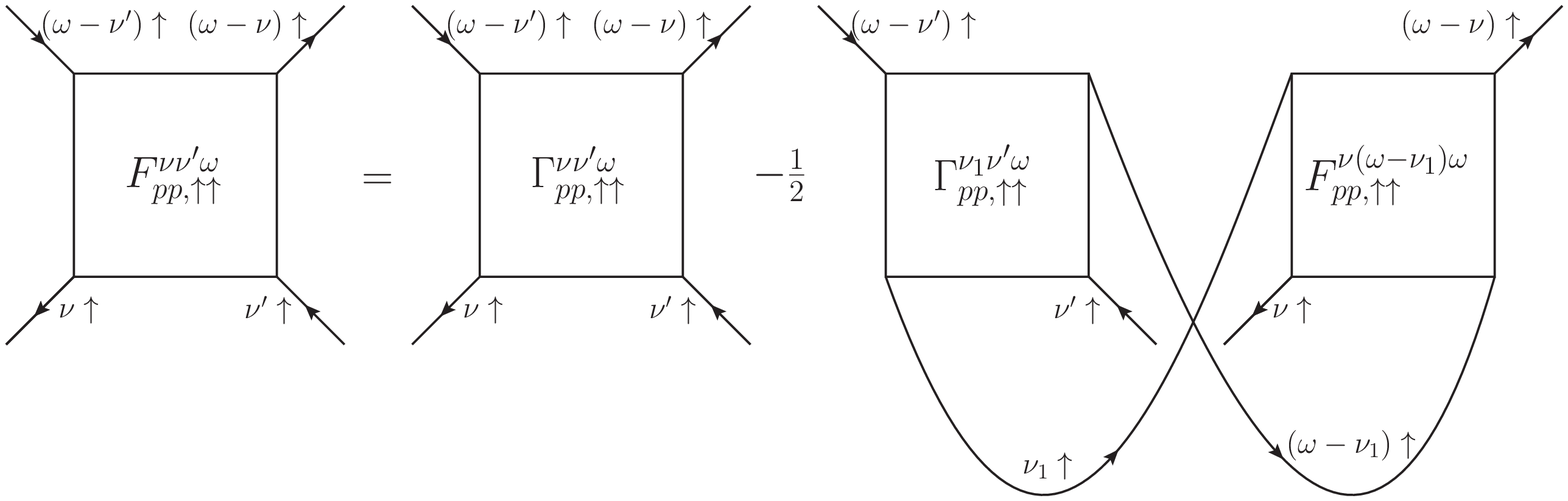}\\[0.3cm]
\includegraphics[width=0.48\textwidth]{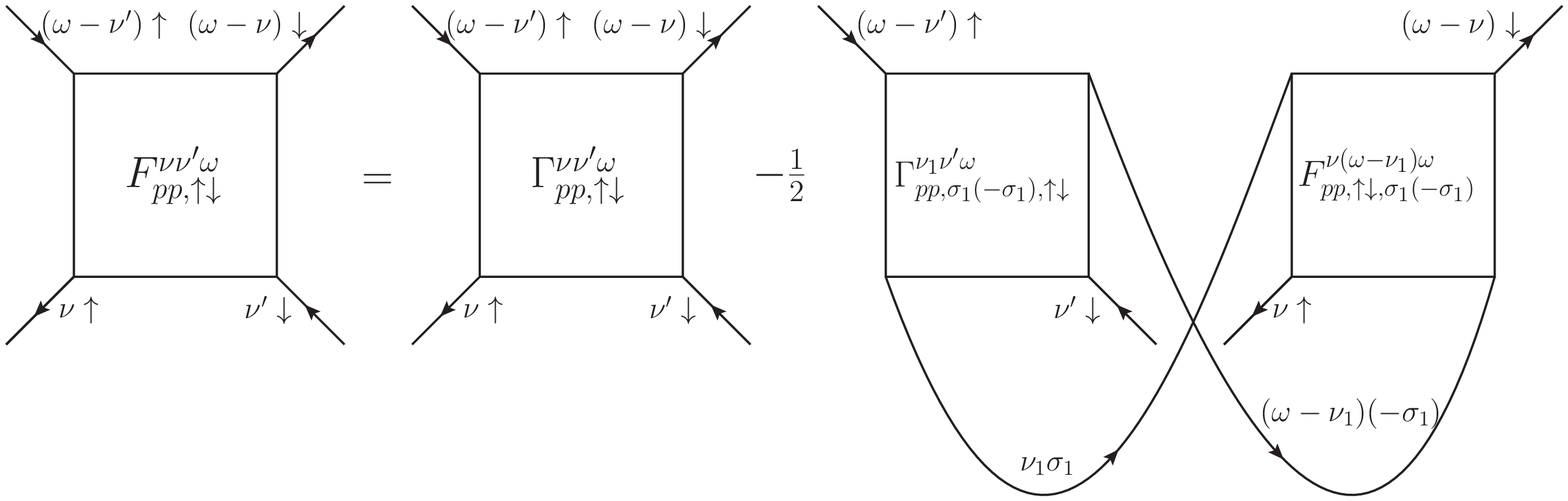}\\[0.3cm]
\includegraphics[width=0.48\textwidth]{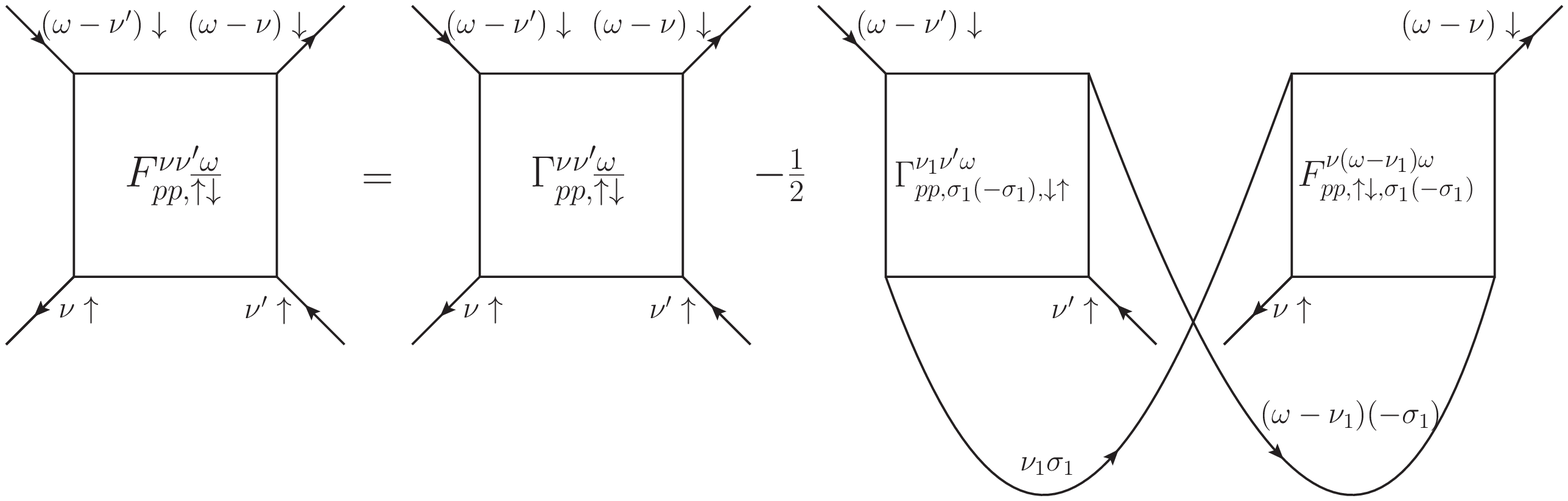}
\caption{Bethe-Salpeter equations in the particle-particle channel.}
\label{fig:gamma_pp}
\end{figure}
\begin{subequations}
\label{equ:bethesalpeterpp}
\begin{equation}
\label{equ:bethesalpeterppupup}
F_{pp,\uparrow\uparrow}^{\nu\nu'\omega}=\Gamma_{pp,\uparrow\uparrow}^{\nu\nu'\omega}-\frac{1}{2}\frac{1}{\beta}\sum_{\nu_1}\Gamma_{pp,\uparrow\uparrow}^{\nu_1\nu'\omega}G(\nu_1)G(\omega-\nu_1)F_{pp,\uparrow\uparrow}^{\nu(\omega-\nu_1)\omega}
\end{equation}
\begin{equation}
\label{equ:bethesalpeterppupdown}
\begin{split}
F_{pp,\uparrow\downarrow}^{\nu\nu'\omega}=\Gamma_{pp,\uparrow\downarrow}^{\nu\nu'\omega}-\frac{1}{2}\frac{1}{\beta}\sum_{\nu_1\sigma_1}&\Gamma_{pp,\sigma_1(-\sigma_1),\uparrow\downarrow}^{\nu_1\nu'\omega}G(\nu_1)G(\omega-\nu_1)\times\\&\times F_{pp,\uparrow\downarrow,\sigma_1(-\sigma_1)}^{\nu(\omega-\nu_1)\omega}
\end{split}
\end{equation}
\begin{equation}
\label{equ:bethesalpeterppupdownbar}
\begin{split}
F_{pp,\overline{\uparrow\downarrow}}^{\nu\nu'\omega}=\Gamma_{pp,\overline{\uparrow\downarrow}}^{\nu\nu'\omega}-\frac{1}{2}\frac{1}{\beta}\sum_{\nu_1\sigma_1}&\Gamma_{pp,\sigma_1(-\sigma_1),\downarrow\uparrow}^{\nu_1\nu'\omega}G(\nu_1)G(\omega-\nu_1)\times\\&\times F_{pp,\uparrow\downarrow,\sigma_1(-\sigma_1)}^{\nu(\omega-\nu_1)\omega}.
\end{split}
\end{equation}
\end{subequations}
The factor $\frac{1}{2}$ appearing in these equations is needed in order to avoid double-counting since we are dealing with indistinguishable particles (see e.g., [\onlinecite{bickers}]). The minus-sign in the reducible part again can be inferred from comparison with 2$^{\text{nd}}$-order perturbation theory. \\
We see that in the particle-particle channel the $\uparrow\uparrow$-vertex is completely independent from the two other spin-combinations, while $\Gamma_{pp,\uparrow\downarrow}$ and $\Gamma_{pp,\overline{\uparrow\downarrow}}$ are not. 
Since they are coupled in the same way as $\Gamma_{ph,\uparrow\uparrow}$ and $\Gamma_{ph,\uparrow\downarrow}$ they can be decoupled introducing the linear combinations
\begin{subequations}
\label{equ:deftripletsinglet}
\begin{equation}
\label{equ:defsinglet}
F_{s(inglet)}^{\nu\nu'\omega}:=F_{pp,\uparrow\downarrow}^{\nu\nu'\omega}-F_{pp,\overline{\uparrow\downarrow}}^{\nu\nu'\omega},
\end{equation}
\begin{equation}
\label{equ:deftriplet}
F_{t(riplet)}^{\nu\nu'\omega}:=F_{pp,\uparrow\downarrow}^{\nu\nu'\omega}+F_{pp,\overline{\uparrow\downarrow}}^{\nu\nu'\omega},
\end{equation}
\end{subequations}
which correspond to Eqs. (\ref{equ:channeldefsinglet}) and (\ref{equ:channeldeftriplet}) for the $\Gamma$'s in complete analogy to the definition of the density and magnetic channel in Eqs. (\ref{equ:defdensitymagnetic}).  By adding and subtracting Eqs. (\ref{equ:bethesalpeterppupdown}) and (\ref{equ:bethesalpeterppupdownbar}) one gets the Bethe-Salpeter equation for the singlet and the triplet channel
\begin{subequations}
\label{equ:bethesalpetertripletsinglet}
\begin{equation}
\label{equ:bethesalpetersinglet}
F_{s}^{\nu\nu'\omega}=\Gamma_{s}^{\nu\nu'\omega}-\frac{1}{2}\frac{1}{\beta}\sum_{\nu_1}\Gamma_{s}^{\nu_1\nu'\omega}G(\nu_1)G(\omega-\nu_1)F_{s}^{\nu(\omega-\nu_1)\omega},
\end{equation}
\begin{equation}
\label{equ:bethesalpetertriplet}
F_{t}^{\nu\nu'\omega}=\Gamma_{t}^{\nu\nu'\omega}-\frac{1}{2}\frac{1}{\beta}\sum_{\nu_1}\Gamma_{t}^{\nu_1\nu'\omega}G(\nu_1)G(\omega-\nu_1)F_{t}^{\nu(\omega-\nu_1)\omega}
\end{equation}
\end{subequations}
Writing the crossing relation (\ref{equ:crossingphnotationF}) in particle-particle notation yields
\begin{equation}
 \label{equ:crossingFpp}
 \begin{split}
 &F_{pp,\uparrow\uparrow}^{\nu\nu'\omega}=F_{ph,\uparrow\uparrow}^{\nu\nu'(\omega-\nu-\nu')}=-F_{ph,\uparrow\uparrow}^{\nu(\omega-\nu')(\nu'-\nu)}=-F_{pp,\uparrow\uparrow}^{\nu(\omega-\nu')\omega}\\
 &F_{pp,\overline{\uparrow\downarrow}}^{\nu\nu'\omega}=F_{ph,\overline{\uparrow\downarrow}}^{\nu\nu'(\omega-\nu-\nu')}=-F_{ph,\uparrow\downarrow}^{\nu(\omega-\nu')(\nu'-\nu)}=-F_{pp,\uparrow\downarrow}^{\nu(\omega-\nu')\omega}.
 \end{split}
\end{equation}
Applying these relations to the definitions of singlet- and triplet-channel gives
\begin{equation}
 \label{equ:crossingFtripletsinglet}
 \begin{split}
 &\Gamma_s^{\nu(\omega-\nu')\omega}=\Gamma_s^{\nu\nu'\omega}\\
 &\Gamma_t^{\nu(\omega-\nu')\omega}=-\Gamma_t^{\nu\nu'\omega}.
 \end{split}
\end{equation}
Inserting the crossing-relations for the $\uparrow\uparrow$- the singlet- and the triplet-vertex in Eqs. (\ref{equ:bethesalpeterppupup}) and (\ref{equ:bethesalpetertripletsinglet}) yields again the standard matrix multiplication-form of the Bethe-Salpeter equations. Furthermore, combining these equations with the definition of the susceptibility in Eq. (\ref{equ:chi_decomposition}) yields the corresponding Bethe-Salpeter equations for the generalized susceptibilities $\chi$ which read as
\begin{subequations}
\label{equ:bethesalpetertripletsingletchi}
\begin{equation}
\label{equ:bethesalpetersingletchi}
\chi_{s}^{\nu\nu'\omega}=-\chi_{0,pp}^{\nu\nu'\omega}-\frac{1}{2}\frac{1}{\beta^2}\sum_{\nu_1\nu_2}(\chi_{0,pp}^{\nu\nu_1\omega}-\chi_{s}^{\nu\nu_1\omega})\Gamma_{t}^{\nu_1\nu_2\omega}\chi_{0,pp}^{\nu_2\nu'\omega},
\end{equation}
\begin{equation}
\label{equ:bethesalpetertripletchi}
\chi_{t}^{\nu\nu'\omega}=\chi_{0,pp}^{\nu\nu'\omega}-\frac{1}{2}\frac{1}{\beta^2}\sum_{\nu_1\nu_2}(\chi_{0,pp}^{\nu\nu_1\omega}+\chi_t^{\nu\nu_1\omega})\Gamma_{t}^{\nu_1\nu_2\omega}\chi_{0,pp}^{\nu_2\nu'\omega},
\end{equation}
\end{subequations}
where $\chi_s$ and $\chi_t$ are defined analogously to the $F$'s in Eqs. (\ref{equ:defsinglet}) and (\ref{equ:deftriplet}).\\
Solving Eqs. (\ref{equ:bethesalpetertripletsingletchi}) for $\Gamma_s^{\nu\nu'\omega}$ and $\Gamma_t^{\nu\nu'\omega}$ yields
\begin{equation}
 \label{equ:invertchipp}
 \begin{split}
 &\Gamma_{s}^{\nu\nu'\omega}=\beta^2\bigl[4(\chi_s-\chi_{0,pp})^{-1}+2\chi_{0,pp}^{-1}\bigr]^{\nu\nu'\omega}\\
 &\Gamma_{t}^{\nu\nu'\omega}=\beta^2\bigl[4(\chi_t+\chi_{0,pp})^{-1}-2\chi_{0,pp}^{-1}\bigr]^{\nu\nu'\omega}.
 \end{split}
\end{equation}
Considering the crossing relations (\ref{equ:crossingFpp}) and the SU(2) symmetry (Eq. \ref{equ:chiFrotation}) one can express the singlet and the triplet channel in the following way
\begin{equation}
 \label{equ:singtripsu2def}
 \begin{split}
 &F_s^{\nu\nu'\omega}=-F_{pp,\uparrow\uparrow}^{\nu\nu'\omega}+2F_{pp,\uparrow\downarrow}^{\nu\nu'\omega}\\
 &F_t^{\nu\nu'\omega}=F_{pp,\uparrow\uparrow}^{\nu\nu'\omega}.
 \end{split}
\end{equation}
This means that in the SU(2)-symmetric case there are only two independent irreducible particle-particle vertices namely, $\Gamma_s$ and $\Gamma_t$ or $\Gamma_{\uparrow\downarrow}$ and $\Gamma_{\uparrow\uparrow}$.\\
However, this is to be expected since in the particle-particle case the $\uparrow\downarrow$ and $\overline{\uparrow\downarrow}$ are connected via the crossing relation (\ref{equ:crossingFpp}). 
Because of that there is another possibility to decouple the $\uparrow\downarrow$ from the $\overline{\uparrow\downarrow}$-channel. Using the crossing relation (\ref{equ:crossingFpp}) we can eliminate $\Gamma_{pp,\overline{\uparrow\downarrow}}^{\nu\nu'\omega}$ from Eq. (\ref{equ:bethesalpeterppupdown}) and obtain an equation containing $\Gamma_{pp,\uparrow\downarrow}^{\nu\nu'\omega}$ only
\begin{equation}
\label{equ:bethesalpeterppupdownfinal}
F_{pp,\uparrow\downarrow}^{\nu\nu'\omega}=
\Gamma_{pp,\uparrow\downarrow}^{\nu\nu'\omega}-\frac{1}{\beta}\sum_{\nu_1}\Gamma_{pp,\uparrow\downarrow}^{\nu_1\nu'\omega}G(\nu_1)G(\omega-\nu_1)F_{pp,\uparrow\downarrow}^{\nu(\omega-\nu_1)\omega}.
\end{equation}
Note that the factor $\frac{1}{2}$ and the spin-summation have disappeared in this equation. Physically this result can be understood in the following way: The factor $\frac{1}{2}$ was introduced in the particle-particle channel to avoid double-counting of diagrams since the two particles are indistinguishable. This clearly holds for the $\uparrow\uparrow$-case. However, in the $\uparrow\downarrow$-case the spin can be fixed (i.e., no spin-summation in the Bethe-Salpeter equation) and hence, the two particles are now distinguishable by their spin. \\
Finally we write Eq. (\ref{equ:bethesalpeterppupdownfinal}) in terms of the corresponding susceptibility $\chi_{pp,\uparrow\downarrow}^{\nu\nu'\omega}$
\begin{equation}
\label{equ:bethesalpeterppupdownchi}
\chi_{pp,\uparrow\downarrow}^{\nu\nu'\omega}=
-\frac{1}{\beta^2}\sum_{\nu_1\nu_2}(\chi_{0,pp}^{\nu\nu_1\omega}-\chi_{pp,\uparrow\downarrow}^{\nu(\omega-\nu_1)\omega})
\Gamma_{pp,\uparrow\downarrow}^{\nu_1\nu_2\omega}\chi_{0,pp}^{\nu_2\nu'\omega}.
\end{equation}
In contrast to Eqs. (\ref{equ:bethesalpetertripletsingletchi}) this equation does not have the form of a matrix-multiplication since it contains $\chi_{pp,\uparrow\downarrow}^{\nu(\omega-\nu_1)\omega}$ instead $\chi_{pp,\uparrow\downarrow}^{\nu\nu_1\omega}$ inside the sum. Nevertheless, it is possible to rewrite it by means of the substitution $\nu'\!\rightarrow\!\omega\!-\!\nu'$ and the transformation $\nu_2\!\rightarrow\!\omega\!-\!\nu_2$ of the summation variable $\nu_2$. Considering that $\chi_{0,pp}^{(\omega-\nu_2)(\omega-\nu')\omega}\!=\!\chi_{0,pp}^{\nu_2\nu'\omega}$ one gets
\begin{equation}
\label{equ:bethesalpeterppupdownchinewfrequency}
\chi_{pp,\uparrow\downarrow}^{\nu(\omega-\nu')\omega}=
-\frac{1}{\beta^2}\sum_{\nu_1\nu_2}(\chi_{0,pp}^{\nu\nu_1\omega}-\chi_{pp,\uparrow\downarrow}^{\nu(\omega-\nu_1)\omega})
\Gamma_{pp,\uparrow\downarrow}^{\nu_1(\omega-\nu_2)\omega}\chi_{0,pp}^{\nu_2\nu'\omega}.
\end{equation}
With the definition $\tilde{\chi}_{pp,\uparrow\downarrow}^{\nu\nu'\omega}\!=\!\chi_{pp,\uparrow\downarrow}^{\nu(\omega-\nu')\omega}$ (and the same for the $\Gamma$'s) one gets the Bethe-Salpeter equation (\ref{equ:bethesalpeterppupdownchinewfrequency}) in the usual form of a matrix multiplication
\begin{equation}
\label{equ:bethesalpeterppupdownchitilde}
\tilde{\chi}_{pp,\uparrow\downarrow}^{\nu\nu'\omega}=
-\frac{1}{\beta^2}\sum_{\nu_1\nu_2}(\chi_{0,pp}^{\nu\nu_1\omega}-\tilde{\chi}_{pp,\uparrow\downarrow}^{\nu\nu_1\omega})
\tilde{\Gamma}_{pp,\uparrow\downarrow}^{\nu_1\nu_2\omega}\chi_{0,pp}^{\nu_2\nu'\omega},
\end{equation}
It can be solved for $\tilde{\Gamma}$ yielding
\begin{equation}
 \label{equ:invertchippupdown}
 \tilde{\Gamma}_{pp,\uparrow\downarrow}^{\nu\nu'\omega}=\beta^2\bigl[(\tilde{\chi}_{pp,\uparrow\downarrow}-\chi_{0,pp})^{-1}+\chi_{0,pp}^{-1}\bigr]^{\nu\nu'\omega}.
\end{equation}

\section{Parquet equations}
\label{app:parquet}
In this section we give the explicit form of the parquet Eq. (\ref{equ:parquet}) taking into their frequency dependence in terms of the density, magnetic, singlet and triplet channel introduced in the previous section. In order to simplify the notation we use the definition of reducible vertex $\Phi$
\begin{equation}
 \label{equ:defreducible}
 \Phi_r^{\nu\nu'\omega}=F_r^{\nu\nu'\omega}-\Gamma_r^{\nu\nu'\omega},\quad r=d,m,s,t
\end{equation}
Hence, the parquet equations read
\begin{align}
 \label{equ:parquetchannels1}
 \Lambda_d^{\nu\nu'\omega}=\Gamma_d^{\nu\nu'\omega}
   &+\frac{1}{2}\Phi_d^{\nu(\nu+\omega)(\nu'-\nu)}+\frac{3}{2}\Phi_m^{\nu(\nu+\omega)(\nu'-\nu)}-\notag\\
   &-\frac{1}{2}\Phi_s^{\nu\nu'(\nu+\nu'+\omega)}-\frac{3}{2}\Phi_t^{\nu\nu'(\nu+\nu'+\omega)}
  \\[0.3cm]
  \label{equ:parquetchannels2}
  \Lambda_m^{\nu\nu'\omega}=\Gamma_m^{\nu\nu'\omega}
  &+\frac{1}{2}\Phi_d^{\nu(\nu+\omega)(\nu'-\nu)}-\frac{1}{2}\Phi_m^{\nu(\nu+\omega)(\nu'-\nu)}+\notag\\
  &+\frac{1}{2}\Phi_s^{\nu\nu'(\nu+\nu'+\omega)}-\frac{1}{2}\Phi_t^{\nu\nu'(\nu+\nu'+\omega)}
  \\[0.3cm]
  \label{equ:parquetchannels3}
  \Lambda_s^{\nu\nu'\omega}=\Gamma_s^{\nu\nu'\omega}
   &-\frac{1}{2}\Phi_d^{\nu\nu'(\omega-\nu-\nu')}+\frac{3}{2}\Phi_m^{\nu\nu'(\omega-\nu-\nu')}-\notag\\
   &-\frac{1}{2}\Phi_d^{\nu(\omega-\nu')(\nu'-\nu)}+\frac{3}{2}\Phi_m^{\nu(\omega-\nu')(\nu'-\nu)}
  \\[0.3cm]
  \label{equ:parquetchannels4}
  \Lambda_t^{\nu\nu'\omega}=\Gamma_t^{\nu\nu'\omega}
   &-\frac{1}{2}\Phi_d^{\nu\nu'(\omega-\nu-\nu')}-\frac{1}{2}\Phi_m^{\nu\nu'(\omega-\nu-\nu')}+\notag\\
   &+\frac{1}{2}\Phi_d^{\nu(\omega-\nu')(\nu'-\nu)}+\frac{1}{2}\Phi_m^{\nu(\omega-\nu')(\nu'-\nu)}.
\end{align}
For the $\Lambda_s$ and $\Lambda_t$ particle-particle notation was adopted. Since at the level of $\Lambda$ no dependency on an irreducible channel ($ph$, $\overline{ph}$ or $pp$) is present $\Lambda_s$ and $\Lambda_t$ can be expressed in terms of the $\Lambda_d$ and $\Lambda_m$
\begin{equation}
 \label{equ:parquetchanneldependence}
 \begin{split}
  &\Lambda_s^{\nu\nu'\omega}=\frac{1}{2}\Lambda_d^{\nu\nu(\omega-\nu-\nu')}-\frac{3}{2}\Lambda_m^{\nu\nu'(\omega-\nu-\nu')}\\
  &\Lambda_t^{\nu\nu'\omega}=\frac{1}{2}\Lambda_d^{\nu\nu(\omega-\nu-\nu')}+\frac{1}{2}\Lambda_m^{\nu\nu'(\omega-\nu-\nu')}.
 \end{split}
\end{equation}

\section{Symmetries}
\label{app:symmetries}
In this appendix, we summarize for convenience some symmetry properties of one-particle Green's function $G$ and the generalized susceptibility $\chi$.

\subsection{Time-Reversal Symmetry}
\label{app:timereversal}
A system without spin-orbit coupling is invariant under time-reversal if its Hamiltonian $\hat{\mathcal{H}}$ (assumed to be time-independent) is a real function of the momentum operator $\hat{p}$ and the position operator $\hat{x}$. This usually holds in absence of an external magnetic field. It can be shown that one can always find real eigenfunctions $\psi(\vec{r})$ in this case and and analogously, the $n$-particle Green's function is a purely real function of the (imaginary) times $\tau_i$
\begin{equation}
 \label{equ:timereversalngreenreal}
 G_n^*(\tau_1,\ldots,\tau_{2n})=G_n(\tau_1,\ldots,\tau_{2n}).
\end{equation}
This property of the $n$-particle Green's function can be easily proven by passing on to its functional integral representation\cite{timereversal}.\\
As the AIM defined in Eq. (\ref{equ:defanderson}) complies with all above-mentioned conditions, the imaginary-times $n$-particle Green's functions are real. Hence, one can derive the following relations for the one- and the two-particle Green's functions (i.e., the generalized susceptibility) of this model in frequency-space
\begin{subequations}
 \label{equ:timereversalfrequencycond}
 \begin{equation}
  \label{equ:timereversalfrequencycond1}  
   G^*(\nu)=G(-\nu)
 \end{equation}
 \begin{equation}
  \label{equ:timereversalfrequencycond2}
  \chi_{\sigma\sigma'}^{\nu\nu'\omega}=\chi_{\sigma'\sigma}^{\nu'\nu\omega}.
 \end{equation}
\end{subequations}
Let us also give an equation relating the generalized susceptibility $\chi$ to its complex conjugate
\begin{equation}
 \label{equ:chicomplex}
 (\chi_{\sigma\sigma'}^{\nu\nu'\omega})^*=\chi_{\sigma'\sigma}^{(-\nu')(-\nu)(-\omega)}=\chi_{\sigma\sigma'}^{(-\nu)(-\nu')(-\omega)}.
\end{equation}

\subsection{Crossing Symmetry}
\label{app:crossingsymmetry}
This symmetry is simply a consequence of the Pauli-principle, i.e., exchanging two identical fermions leads to a minus-sign in the wave function. Considering Eq. (\ref{equ:fouriertransdefph}) the exchange of annihilation operators in the time-ordered matrix element yields a minus-sign and leads to an exchange of the corresponding frequencies $\nu'$ and $\nu+\omega$. Taking into account additional $\chi_0$-contributions one gets the following crossing relations for $\chi$, $F$ and $\Lambda$ in particle-hole notation
\begin{subequations}
 \label{equ:crossingphnotation}
 \begin{equation}
  \label{equ:crossingphnotationchi}
  \chi_{\overline{\sigma\sigma'}}^{\nu\nu'\omega}-\delta_{\sigma\sigma'}\chi_0^{\nu(\nu+\omega)(\nu'-\nu)}=-\chi_{\sigma\sigma'}^
  {\nu(\nu+\omega)(\nu'-\nu)}+\chi_0^{\nu\nu'\omega},
 \end{equation}
 \begin{equation}
  \label{equ:crossingphnotationF}
  F_{\overline{\sigma\sigma'}}^{\nu\nu'\omega}=-F_{\sigma\sigma'}^{\nu(\nu+\omega)(\nu'-\nu)},
 \end{equation}
 \begin{equation}
  \label{equ:crossingphnotationlambda}
  \Lambda_{\overline{\sigma\sigma'}}^{\nu\nu'\omega}=-\Lambda_{\sigma\sigma'}^{\nu(\nu+\omega)(\nu'-\nu)}.
 \end{equation}
\end{subequations}

\subsection{SU(2) Symmetry}
\label{su2symmetry}
If the Hamiltonian of the system does not contain terms breaking rotation symmetry (e.g., a magnetic field), the $\chi$'s and the $F$'s satisfy some specific relations.\\ 
Every matrix-element hast to fulfill spin-conservation, e.g., $G_{\uparrow\downarrow}=0$.
The one-particle Green's function is independent of the spin, i.e., $G_{\uparrow\uparrow}=G_{\downarrow\downarrow}\equiv G$. At the two particle level similarly $\chi_{\uparrow\uparrow}=\chi_{\downarrow\downarrow}$ and $\chi_{\uparrow\downarrow}=\chi_{\downarrow\uparrow}$ hold. These relations can be easily proven by rotating all spins through an angle $\pi$ about the $x$-  or $y$-axis. Furthermore, performing a rotation through an angle $\frac{\pi}{2}$, i.e., rotating a spin in $z$-direction into the $xy$-plane, yields
\begin{align}
 \label{equ:chichirotation}
 &\chi^{\nu\nu'\omega}_{\sigma\sigma}=\chi^{\nu\nu'\omega}_{\sigma(-\sigma)}-\chi^{\nu(\nu+\omega)(\nu'-\nu)}_{\sigma(-\sigma)}+\chi^{\nu\nu'\omega}_{0}\\
 \label{equ:chiFrotation}
 &F^{\nu\nu'\omega}_{\sigma\sigma}=F^{\nu\nu'\omega}_{\sigma(-\sigma)}-F^{\nu(\nu+\omega)(\nu'-\nu)}_{\sigma(-\sigma)}.
\end{align}

\subsection{Mapping onto the Attractive model}
\label{app:particleholesymmetry}

The usual partial particle-hole transformations which map repulsive onto attractive Hubbard interactions are defined for lattice systems\cite{particle-hole-transform}. Obviously, one can find an equivalent (but local) transformation for the corresponding AIM as will be shown in the following. \\
Starting point is the Hamiltonian of the AIM (Eq. (\ref{equ:defanderson})) containing also the chemical potential term $-\mu(\hat{n}_{\uparrow}+\hat{n}_{\downarrow})$ since we consider a grand canonical ensemble
\begin{equation}
\begin{split}
\label{equ:defandersonmu}
\hat{\mathcal{H}}=\sum_{\ell\sigma}\varepsilon_{\ell}\hat{a}^{\dagger}_{\ell\sigma}\hat{a}_{\ell\sigma}&+\sum_{\ell\sigma}V_{\ell}(\hat{c}^{\dagger}_{\sigma}\hat{a}_{\ell\sigma}+\hat{a}^{\dagger}_{\ell\sigma}\hat{c}_{\sigma})+\\&+U\hat{n}_{\uparrow}\hat{n}_{\downarrow}-\mu(\hat{n}_{\uparrow}+\hat{n}_{\downarrow}).
\end{split}
\end{equation}
The sum over $\ell$ (bath sites) ranges from 2 to N, $\ell=1$ denotes the impurity (i.e., $\hat{a}_{1\sigma}^{(\dagger)}=\hat{c}_{\sigma}^{(\dagger)}$). The (partial) particle-hole transformation we are considering is defined by the unitary operator $\hat{\mathcal{W}}$
\begin{equation}
\label{equ:defparticleholetransform}
\hat{\mathcal{W}}=(\hat{a}_{\text{N}\downarrow}^{\dagger}-\hat{a}_{\text{N}\downarrow})\ldots(\hat{a}_{2\downarrow}^{\dagger}-\hat{a}_{2\downarrow})(\hat{c}_{\downarrow}^{\dagger}+\hat{c}_{\downarrow}).
\end{equation}
The action of the transformation $\hat{\mathcal{W}}$ on the creation and annihilation operators is given by 
\begin{subequations}
\label{equ:particleholecatransform}
\begin{equation}
\begin{split}
\label{equ:particleholectransform}
&\hat{\mathcal{W}}^{\dagger}(\hat{c}_{\downarrow}^{\dagger},\hat{c}_{\downarrow})\hat{\mathcal{W}}=(-1)^{(\text{N}-1)}(\hat{c}_{\downarrow},\hat{c}_{\downarrow}^{\dagger})
\\&
\hat{\mathcal{W}}^{\dagger}(\hat{c}_{\uparrow}^{\dagger},\hat{c}_{\uparrow})\hat{\mathcal{W}}=(\hat{c}_{\uparrow}^{\dagger},\hat{c}_{\uparrow})
\end{split}
\end{equation}
\begin{equation}
\label{equ:particleholeatransform}
\begin{split}
&\hat{\mathcal{W}}^{\dagger}(\hat{a}_{\ell\downarrow}^{\dagger},\hat{a}_{\ell\downarrow})\hat{\mathcal{W}}=(-1)^{\text{N}}(\hat{a}_{\ell\downarrow},\hat{a}_{\ell\downarrow}^{\dagger})
\\&
\hat{\mathcal{W}}^{\dagger}(\hat{a}_{\ell\uparrow}^{\dagger},\hat{a}_{\ell\uparrow})\hat{\mathcal{W}}=(-1)^{\text{N}}(\hat{a}_{\ell\uparrow}^{\dagger},\hat{a}_{\ell\uparrow}).
\end{split}
\end{equation}
\end{subequations}
This means that for $\sigma=\downarrow$ the annihilation and creation operators are interchanged, while the $\uparrow$-operators are not modified by the transformation $\hat{\mathcal{W}}$ (despite a phase-factor $(-1)^{\text{N}}$). Therefore $\hat{\mathcal{W}}$ is coined {\sl partial} particle-hole transformation.\\ 
The transformation of the AIM-Hamiltonian given in Eq. (\ref{equ:defandersonmu}) yields
{\small
\begin{equation}
\label{equ:particleholetransformhamilt}
\begin{split}
\hat{\mathcal{W}}^{\dagger}\hat{\mathcal{H}}\hat{\mathcal{W}}=&\sum_{\sigma,\ell=2}^{\text{N}}\left[\varepsilon_\ell\hat{a}_{\ell\uparrow}^{\dagger}\hat{a}_{\ell\uparrow}-\varepsilon_{\ell}\hat{a}_{\ell\downarrow}^{\dagger}\hat{a}_{\ell\downarrow}+V_{\ell}(\hat{c}^{\dagger}_{\sigma}\hat{a}_{\ell\sigma}+\hat{a}^{\dagger}_{\ell\sigma}\hat{c}_{\sigma})\right]-\\&-U\hat{n}_{\uparrow}\hat{n}_{\downarrow}-[(\mu-U)\hat{n}_{\uparrow}-\mu\hat{n}_{\downarrow})-\mu+\sum_{\ell=2}^{\text{N}}\varepsilon_{\ell}.
\end{split}
\end{equation}}
We are now restricting ourselves to an even number of bath sides, i.e., N has to be odd. Furthermore, we assume that the bath levels are distributed symmetrically around $0$, i.e., $\varepsilon_{\ell}=-\varepsilon_{\ell+\frac{\text{N}}{2}}$ for $\ell=2..\frac{\text{N}}{2}+1$. In addition, the hybridization between the bath and the impurity should be the same for positive and the corresponding negative bath-energies which means that $V_{\ell}=V_{\ell+\frac{N}{2}}$ for $\ell=2..\frac{\text{N}}{2}+1$. Hence, the negative energy-sector of the bath is completely equivalent to the positive one. Performing the index-transformation $\ell\leftrightarrow(\ell+\frac{\text{N}}{2})$ for the $\downarrow$-spins in Eq. (\ref{equ:particleholetransformhamilt}) changes the minus-sign in front of $\varepsilon_{\ell}\hat{a}_{i\downarrow}^{\dagger}\hat{a}_{\ell\downarrow}$ back into a plus-sign as in the original Hamiltonian. \\
Furthermore, choosing the chemical potential as
\begin{equation}
\label{equ:muhalffilling}
\mu=\frac{U}{2},
\end{equation}
in Eq. (\ref{equ:particleholetransformhamilt}) one retrieves the same structure of the original Hamiltonian whereas only the sign of $U$ has changed (the constant contribution $-\mu+\sum_{\ell=2}^{\text{N}}\varepsilon_{\ell}$ can be neglected). \\
This way it has been shown how the transformation $\hat{\mathcal{W}}$ maps the repulsive AIM Hamiltonian ($U>0, \mu=\frac{U}{2}$) on the attractive one ($-U,\mu=-\frac{U}{2}$) provided that the additional conditions for N, $\varepsilon_{\ell}$ and $V_{\ell}$ are fulfilled, which is the case for the particle-hole symmetric AIM associated to the DMFT solution of the half-filled Hubbard model considered here.\\
Next, we discuss some symmetry-relations for the $n$-particle Green's functions $G_{n,\sigma_1,\ldots,\sigma_{2n}}^U$ and the generalized susceptibility $\chi_{U,\sigma\sigma'}^{\nu\nu'\omega}$. The additional index $U$ indicates whether the quantity under consideration is calculated for repulsive ($U$ or $+U$) or for the corresponding attractive ($-U$) model. \\ 
First, applying the particle-hole transformation to the {\sl one-particle Green's function} with spin-$\uparrow$ lets this function unchanged, since $\hat{\mathcal{W}}$ only acts on the $\downarrow$-creation- and annihilation operators, i.e.,
\begin{equation}
 \label{equ:particlehole1particleupup}
 G_{\uparrow\uparrow}^U(\tau_1,\tau_2)=G_{\uparrow\uparrow}^{(-U)}(\tau_1,\tau_2).
\end{equation}
For the Green's function with spin-$\downarrow$ $\hat{c}_{\downarrow}$ and $\hat{c}^{\dagger}_{\downarrow}$ change their role which leads to an exchange of $\tau_1$ and $\tau_2$ as well as to an additional minus-sign
\begin{equation}
 \label{equ:particlehole1particleupdown}
 G_{\downarrow\downarrow}^U(\tau_1,\tau_2)=-G_{\downarrow\downarrow}^{(-U)}(\tau_2,\tau_1).
\end{equation}
For the SU(2) symmetric case $G_{\uparrow\uparrow}=G_{\downarrow\downarrow}\equiv G$ one can combine relations (\ref{equ:particlehole1particleupup}) and (\ref{equ:particlehole1particleupdown}) and gets
\begin{equation}
 \label{equ:particlehole1particlecombined}
 G^U(\tau_1,\tau_2)=-G^U(\tau_2,\tau_1).
\end{equation}
which means in Fourier-space
\begin{equation}
\label{equ:particleholegreenfourier}
G^{*}(\nu)=-G(\nu),
\end{equation}
expressing the fact that in the particle-hole symmetric case the one-particle Green's function is purely imaginary.\\ 
Taking the limit $\tau_2\rightarrow\tau_1+$ (i.e. $\tau_2\rightarrow\tau_1$ and $\tau_2>\tau_1$) in Eq.  (\ref{equ:particlehole1particlecombined}) leads to the result that the average density at the impurity $\langle\hat{n}\rangle=n=1$, which means that the system is ''half-filled`` in the particle-hole symmetric case. \\
Next, we consider the {\sl two-particle Green's function}, i.e., the generalized susceptibility. As in the one-particle-case 
\begin{equation}
\label{equ:particlehole2particleupup}
G_{2,\uparrow\uparrow\uparrow\uparrow}^U(\tau_1,\tau_2,\tau_3,\tau_4)=G_{2,\uparrow\uparrow\uparrow\uparrow}^{(-U)}(\tau_1,\tau_2,\tau_3,\tau_4).
\end{equation}
The two-particle Green's function containing only $\downarrow$-spins transforms under $\hat{\mathcal{W}}$ as follows:
\begin{equation}
\label{equ:particlehole2particledowndown}
G_{2,\downarrow\downarrow\downarrow\downarrow}^U(\tau_1,\tau_2,\tau_3,\tau_4)=G_{2,\downarrow\downarrow\downarrow\downarrow}^{(-U)}(\tau_4,\tau_3,\tau_2,\tau_1). 
\end{equation}
Combining eqs. (\ref{equ:particlehole2particleupup}) and (\ref{equ:particlehole2particledowndown}) and using again SU(2)-symmetry yields
\begin{equation}
\label{equ:particlehole2particlecombined}
G_{2,\uparrow\uparrow\uparrow\uparrow}^{U}(\tau_1,\tau_2,\tau_3,\tau_4)=G_{2,\uparrow\uparrow\uparrow\uparrow}^U(\tau_4,\tau_3,\tau_2,\tau_1).
\end{equation}
In Fourier-space this relation states that the two-particle Green functions are purely real and the same holds true also for the susceptibilities
\begin{equation}
\label{equ:particleholesuscupupfourier}
\left(\chi_{\sigma\sigma'}^{\nu\nu'\omega}\right)^*=\chi_{\sigma\sigma'}^{\nu\nu'\omega}.
\end{equation}
Furthermore we want to study how the $\uparrow\downarrow$-function transforms under the particle-hole-transformation. In the corresponding matrix element only the operators corresponding to the times $\tau_3$ and $\tau_4$ carry $\downarrow$-spins and therefore $\hat{\mathcal{W}}$ acts only on them
\begin{equation}
\label{equ:particlehole2particleupdown}
G_{2,\uparrow\uparrow\downarrow\downarrow}^U(\tau_1,\tau_2,\tau_3,\tau_4)=-G_{2,\uparrow\uparrow\downarrow\downarrow}^{(-U)}(\tau_1,\tau_2,\tau_4,\tau_3).
\end{equation}
In Fourier-space this is equivalent to the transformations $(\nu'+\omega)\rightarrow(-\nu')$ and $\nu'\rightarrow(-\nu'-\omega)$, i.e.,
\begin{equation}
\label{equ:particleholesuscupdownfourier} 
\chi_{U,\uparrow\downarrow}^{\nu\nu'\omega}-\chi_0^{\nu(\nu+\omega)(\nu'-\nu)}=-\chi_{(-U),\uparrow\downarrow}^{\nu(-\nu'-\omega)\omega}+\chi_0^{\nu(\nu+\omega)(-\nu-\nu'-\omega)}.
\end{equation}
Using SU(2) symmetry on the left hand side of this equations yields
\begin{equation}
\label{equ:particleholesuscupdownsu2}
-\chi_{U,m}^{\nu(\nu+\omega)(\nu'-\nu)}=-\chi_{(-U),\uparrow\downarrow}^{\nu(-\nu'-\omega)\omega}+\chi_0^{\nu(\nu+\omega)(-\nu-\nu'-\omega)}.
\end{equation}
Performing the frequency transformation $\nu'\rightarrow\nu-\omega$ and $\omega\rightarrow\nu'-\nu$ and transforming the right hand side to the particle-particle notation gives
\begin{equation}
\label{equ:particlehole2ndfrequencyshift}
\chi_{U,m}^{\nu\nu'(-\omega)}=\chi_{(-U),pp,\uparrow\downarrow}^{\nu(\omega-\nu')\omega}-\chi_{0,pp}^{\nu\nu'\omega}.
\end{equation}
This equation can be interpreted as follows: The inversion $\chi_{U,m}^{\nu\nu'(-\omega)}$ yields $\Gamma_m^{\nu\nu'(-\omega)}$ as discussed in Sec. \ref{app:spindiag}. The inversion of the quantity on the right hand side of Eq. (\ref{equ:particlehole2ndfrequencyshift}) gives the irreducible $\uparrow\downarrow$-vertex in the particle-particle channel, i.e., $\Gamma_{(-U),pp,\uparrow\downarrow}^{\nu(\omega-\nu')\omega}$ (see Eq. \ref{equ:invertchippupdown}). Hence, 
\begin{equation}
 \label{equ:particleholegamma}
 \Gamma_m^{\nu\nu'(-\omega)}=\Gamma_{(-U),pp,\uparrow\downarrow}^{\nu(\omega-\nu')\omega},
\end{equation}
which is also shown diagrammatically in Sec. \ref{Sec:DMFTneg}.\\
If one performs the sum over $\nu$ and $\nu'$ in Eq. (\ref{equ:particlehole2ndfrequencyshift}) one sees that fluctuations of the spin for the repulsive model are mapped on fluctuations of an electron-pair for the attractive case. This is consistent with the well-known fact that for a lattice model the anti-ferromagnetic instability for $U>0$ corresponds to the superconducting instability in the attractive model.


\begin{thebibliography}{}
\bibitem{noteDOS} Such a choice for the standard deviation of the DOS ($0.5$) for the cubic lattice makes our DMFT results comparable with the corresponding ones for a square lattice with $t=0.25$ and a Bethe lattice of semibandwidth $W/2=1$.
\bibitem{note_reduc} This can be easily seen by the following argument: Consider a diagram that is reducible in one of the two particle-hole channels as that shown in Fig. \ref{fig:phreduciblegeneric}. It is evident that the subdiagram $A$ can be separated from $B$ by cutting the Green function lines $a$ and $b$. Let us assume that the diagram is reducible in the (14)-(23) channel as well, i.e. it can be split into two parts one containing the outer legs 1 and 4 the other one containing 2 and 3 by cutting two internal lines. This would be only possible if the subdiagrams $A$ and $B$ could be split into two parts by cutting only one internal line. But this is {\sl impossible}, since it would lead to subdiagrams with three outer legs which violates particle-conservation. Therefore, it is proven that each diagram is either fully irreducible or reducible in exactly one channel.
\bibitem{spinreducibility} While this is obvious for $\Gamma_d$, $\Gamma_s$ and $\Gamma_t$, this is even true for $\Gamma_m$. In fact, $\Gamma_m$ contains diagrams reducible in the transverse $ph$-channel, and, due to the SU(2)- and the crossing-symmetry they coincide with the corresponding magnetic ones.
\bibitem{note_error} The error bars have been estimated by comparing the results for the summations over a finite frequency range with the exact DMFT result, which can be obtained directly  for $n_\uparrow n_\downarrow$ and $\chi_{d,m}^{loc}$ from the ED impurity solver.
\end{thebibliography}
\end{document}